\documentclass{aa}  

\usepackage{xcolor}
\usepackage{graphicx}
\usepackage[varg]{txfonts}
\usepackage{isotope}
\usepackage{siunitx}
\usepackage{ulem}
\usepackage{bm}
\usepackage{natbib}
\usepackage{url}
\usepackage{xspace}
\usepackage{placeins}
\usepackage{capt-of}
\bibpunct{(}{)}{;}{a}{}{,}

\newcommand{\teff}{$T_{\rm eff}$}

\begin{document}

\title{NLTE spectroscopic analysis of the $\isotope[3]{He}$ anomaly in subluminous B-type stars}

\author{D. Schneider\inst{1}, A. Irrgang\inst{1}, U. Heber\inst{1}, M. F. Nieva\inst{2}, N. Przybilla\inst{2}
       }

\institute{ Dr. Karl Remeis-Observatory \& ECAP, Astronomical Institute,
               Friedrich-Alexander University Erlangen-Nuremberg, 
Sternwartstr. 7,
               96049 Bamberg, Germany\\ \email{david.schneider@fau.de}
         \and Institut f\"{u}r Astro- und Teilchenphysik, Universit\"{a}t Innsbruck, Technikerstr. 25/8, 6020 Innsbruck, Austria 
         }

   \date{Received 6 April 2018 / Accepted 6 July 2018}

        \abstract
        {Several B-type main-sequence stars show chemical peculiarities. A particularly striking class are the $\isotope[3]{He}$ stars, which exhibit a remarkable enrichment of $\isotope[3]{He}$ with respect to $\isotope[4]{He}$. This isotopic anomaly has also been found in blue horizontal branch (BHB) and subdwarf B (sdB) stars, which are helium-core burning stars of the extreme horizontal branch. Recent surveys uncovered 11 $\isotope[3]{He}$ sdBs. The $\isotope[3]{He}$ anomaly is not due to thermonuclear processes, but caused by atomic diffusion in the stellar atmosphere. Using a hybrid local/non-local thermodynamic equilibrium (LTE/NLTE) approach for B-type stars, we analyzed high-quality spectra of two known $\isotope[3]{He}$ BHBs and nine known $\isotope[3]{He}$ sdBs to determine their isotopic helium abundances and $\isotope[4]{He}$/$\isotope[3]{He}$ abundance ratios. We redetermined their atmospheric parameters and analyzed selected $\ion{He}{i}$ lines, including $\lambda$\SI{4922}{\angstrom} and $\lambda$\SI{6678}{\angstrom}, which are very sensitive to $\isotope[4]{He}$/$\isotope[3]{He}$. Most of the $\isotope[3]{He}$ sdBs cluster in a narrow temperature strip between \SI{26000}{\kelvin} and \SI{30000}{\kelvin} and are helium deficient in accordance with previous LTE analyses. BD+48$^\circ$ 2721 is reclassified as a BHB star because of its low temperature (\teff$=$\,\SI{20700}{\kelvin}). Whereas $\isotope[4]{He}$ is almost absent ($\isotope[4]{He}$/$\isotope[3]{He}<0.25$) in most of the known $\isotope[3]{He}$ stars, other sample stars show abundance ratios up to $\isotope[4]{He}$/$\isotope[3]{He}\sim2.51$. A search for $\isotope[3]{He}$ stars among 26 candidate sdBs from the ESO SPY survey led to the discovery of two new $\isotope[3]{He}$ sdB stars (HE 0929-0424 and HE 1047-0436). The observed helium line profiles of all BHBs and of three sdBs are not matched by chemically homogeneous atmospheres, but hint at vertical helium stratification. This phenomenon has been seen in other peculiar B-type stars, but is found for the first time for sdBs. We estimate helium to increase from the outer to the inner atmosphere by factors ranging from 1.4 (SB 290) up to 8.0 (BD+48$^\circ$ 2721).   
        }

\authorrunning{D. Schneider et al.}

\maketitle


\section{Introduction}\label{Introduction}
The chemical composition of a large fraction of stars is similar to that of the Sun. However, abundance anomalies can be observed throughout many parts of the Hertzsprung-Russell diagram \citep{Michaud_1991}. Some abundance anomalies may be traced back to thermonuclear burning reaching the stellar surface. Possible causes may be strong mass loss in Wolf-Rayet stars \citep{Langer_2012}, internal mixing in PG 1159 stars \citep{Werner_2006}, or mass transfer in binaries, for example, in dwarf carbon stars \citep{Heber_1993, Green_2000}. In many cases, however, the abundance anomalies result from atomic transport, that is, diffusion processes occurring in the stellar atmosphere (\citealt{Greenstein_1967}; see \citealt{Michaud_Atomic_Diffusion_in_Stars_2015} for a detailed review). For instance, there is no doubt that the surface abundances of white dwarfs are caused by atomic transport processes. Moreover, abundance anomalies are also observed on the main sequence (MS) for B, A, and F-type stars (see, e.g., \citealt{Smith_1996}) and for evolved stars on the horizontal branch (HB), such as blue horizontal branch (BHB, \teff\,$\gtrsim$ \SI{12000}{\kelvin}) or extreme horizontal branch (EHB, \teff\,$\gtrsim$ \SI{22000}{\kelvin}) stars (see \citealt{Heber_2009,Heber_2016} for reviews).\\
The latter particularly contain hot subluminous B stars (sdBs), which have similar colors and spectral characteristics as B-type MS stars, but are much less luminous and are considered to burn helium in their cores. These rather compact objects ($R_{\text{sdB}}\sim$ 0.1-0.3 $R_{\text{\sun}}$) have very thin hydrogen envelopes ($M{_\text{env}}\sim$ 0.01\,$M_{\text{\sun}}$) and total masses of $M_{\text{sdB}}\sim$ 0.5\,$M_{\text{\sun}}$. They show effective temperatures between $\sim$\,\SI{22000}{\kelvin} and $\sim$\,\SI{40000}{\kelvin} with high surface gravities of $\log{(g)}\sim$ 5.0\,-\,6.0.\\
In a simplistic atmospheric diffusion model, the equilibrium abundance of a particular element is set by a balance between gravitational settling and radiative levitation, since the radiation pressure experienced by an ion depends on its abundance. However, such simple atomic diffusion models predict that the atmospheres of all chemically peculiar B-type MS and EHB stars should be depleted in helium to such low abundances on timescales much shorter than the evolutionary one that no helium spectral lines are predicted at all by atmospheric models in the optical spectra of these stars. This is at odds with observations (see \citealt{Fontaine_1997} for a review).\\
\begin{table}
\captionof{table}{Transitions and isotopic shifts $\Delta\lambda:=\lambda (\isotope[3]{He})-\lambda (\isotope[4]{He})$ of selected $\ion{He}{i}$ lines in the near-ultraviolet, optical, and near-infrared spectral range up to principal quantum number $n=8$.}\label{summary of isotopic shifts of neutral helium lines}
\begin{tabular}{cccc}
\hline\hline
Transition & $\lambda$ ($\isotope[4]{He}$) & $\lambda$ ($\isotope[3]{He}$) & $\Delta\lambda$\\
 & [\si{\angstrom}] & [\si{\angstrom}] & [\si{\angstrom}]\\
\hline
\vspace*{0.5pt}
1s\,2s\,$^3$S$_1$--1s\,2p\,$^3$P$^0_2$ & 10\,830.340 & 10\,831.658 & 1.318\\
\vspace*{0.5pt}
1s\,2s\,$^3$S$_1$--1s\,3p\,$^3$P$^0_2$ & 3888.649 & 3888.862 & 0.213\\
\vspace*{0.5pt}
1s\,2s\,$^3$S$_1$--1s\,4p\,$^3$P$^0_2$ & 3187.745 & 3187.903 & 0.158\\
\vspace*{0.5pt}
1s\,2s\,$^3$S$_1$--1s\,5p\,$^3$P$^0_2$ & 2945.104 & 2945.246 & 0.142\\
\vspace*{0.5pt}
1s\,2s\,$^3$S$_1$--1s\,6p\,$^3$P$^0_2$ & 2829.081 & 2829.216 & 0.135\\
\vspace*{0.5pt}
1s\,2s\,$^3$S$_1$--1s\,7p\,$^3$P$^0_2$ & 2763.803 & 2763.934 & 0.131\\
\vspace*{0.5pt}
1s\,2s\,$^3$S$_1$--1s\,8p\,$^3$P$^0_2$ & 2723.192 & 2723.320 & 0.128\\
\hline
\vspace*{0.5pt}
1s\,2s\,$^1$S$_0$--1s\,3p\,$^1$P$^0_1$ & 5015.678 & 5015.890 & 0.212\\
\vspace*{0.5pt}
1s\,2s\,$^1$S$_0$--1s\,4p\,$^1$P$^0_1$ & 3964.729 & 3964.912 & 0.183\\
\vspace*{0.5pt}
1s\,2s\,$^1$S$_0$--1s\,5p\,$^1$P$^0_1$ & 3613.642 & 3613.812 & 0.170\\
\vspace*{0.5pt}
1s\,2s\,$^1$S$_0$--1s\,6p\,$^1$P$^0_1$ & 3447.589 & 3447.753 & 0.164\\
\vspace*{0.5pt}
1s\,2s\,$^1$S$_0$--1s\,7p\,$^1$P$^0_1$ & 3354.555 & 3354.715 & 0.160\\
\vspace*{0.5pt}
1s\,2s\,$^1$S$_0$--1s\,8p\,$^1$P$^0_1$ & 3296.773 & 3296.930 & 0.157\\
\hline
\vspace*{0.5pt}
1s\,2p\,$^1$P$^0_1$--1s\,3s\,$^1$S$_0$ & 7281.351 & 7281.904 & 0.553\\
\vspace*{0.5pt}
1s\,2p\,$^1$P$^0_1$--1s\,3d\,$^1$D$_2$ & 6678.152 & 6678.654 & 0.502\\
\vspace*{0.5pt}
1s\,2p\,$^1$P$^0_1$--1s\,4s\,$^1$S$_0$ & 5047.739 & 5048.078 & 0.339\\
\vspace*{0.5pt}
1s\,2p\,$^1$P$^0_1$--1s\,4d\,$^1$D$_2$ & 4921.931 & 4922.262 & 0.331\\
\vspace*{0.5pt}
1s\,2p\,$^1$P$^0_1$--1s\,5s\,$^1$S$_0$ & 4437.553 & 4437.841 & 0.288\\
\vspace*{0.5pt}
1s\,2p\,$^1$P$^0_1$--1s\,5d$^1$D$_2$ & 4387.929 & 4388.213 & 0.284\\
\vspace*{0.5pt}
1s\,2p\,$^1$P$^0_1$--1s\,6s\,$^1$S$_0$ & 4168.971 & 4169.237 & 0.266\\
\vspace*{0.5pt}
1s\,2p\,$^1$P$^0_1$--1s\,6d\,$^1$D$_2$ & 4143.759 & 4144.023 & 0.264\\
\vspace*{0.5pt}
1s\,2p\,$^1$P$^0_1$--1s\,7s\,$^1$S$_0$ & 4023.980 & 4024.233 & 0.253\\
\vspace*{0.5pt}
1s\,2p\,$^1$P$^0_1$--1s\,7d\,$^1$D$_2$ & 4009.257 & 4009.509 & 0.252\\
\vspace*{0.5pt}
1s\,2p\,$^1$P$^0_1$--1s\,8s\,$^1$S$_0$ & 3935.945 & 3936.192 & 0.247\\
\vspace*{0.5pt}
1s\,2p\,$^1$P$^0_1$--1s\,8d\,$^1$D$_2$ & 3926.544 & 3926.790 & 0.246\\
\hline
\vspace*{0.5pt}
1s\,2p\,$^3$P$^0_1$--1s\,3s\,$^3$S$_1$ & 7065.215 & 7065.205 & -0.010\\
\vspace*{0.5pt}
1s\,2p\,$^3$P$^0_1$--1s\,3d\,$^3$D$_1$ & 5875.625 & 5875.669 & 0.044\\
\vspace*{0.5pt}
1s\,2p\,$^3$P$^0_2$--1s\,4s\,$^3$S$_1$ & 4713.139 & 4713.208 & 0.069\\
\vspace*{0.5pt}
1s\,2p\,$^3$P$^0_2$--1s\,4d\,$^3$D$_1$ & 4471.470 & 4471.544 & 0.074\\
\vspace*{0.5pt}
1s\,2p\,$^3$P$^0_2$--1s\,5s\,$^3$S$_1$ & 4120.811 & 4120.887 & 0.076\\
\vspace*{0.5pt}
1s\,2p\,$^3$P$^0_2$--1s\,5d\,$^3$D$_1$ & 4026.184 & 4026.262 & 0.078\\
\vspace*{0.5pt}
1s\,2p\,$^3$P$^0_2$--1s\,6s\,$^3$S$_1$ & 3867.472 & 3867.550 & 0.078\\
\vspace*{0.5pt}
1s\,2p\,$^3$P$^0_2$--1s\,6d\,$^3$D$_1$ & 3819.602 & 3819.680 & 0.078\\
\vspace*{0.5pt}
1s\,2p\,$^3$P$^0_2$--1s\,7s\,$^3$S$_1$ & 3732.863 & 3732.942 & 0.079\\
\vspace*{0.5pt}
1s\,2p\,$^3$P$^0_2$--1s\,7d\,$^3$D$_1$ & 3704.995 & 3705.074 & 0.079\\
\vspace*{0.5pt}
1s\,2p\,$^3$P$^0_2$--1s\,8s\,$^3$S$_1$ & 3651.982 & 3652.060 & 0.078\\
\vspace*{0.5pt}
1s\,2p\,$^3$P$^0_2$--1s\,8d\,$^3$D$_1$ & 3634.231 & 3634.310 & 0.079\\
\hline
\end{tabular}
\tablefoot{The listed transitions are implemented in NLTE in the $\isotope[3]{He}$ and $\isotope[4]{He}$ model atoms used for this study (see Sect. \ref{Model Atmospheres and Synthetic Spectra} for details). All wavelengths for the $\isotope[4]{He}$ and $\isotope[3]{He}$ components of the listed $\ion{He}{i}$ lines were extracted from the \textit{Atomic Spectra Database} of NIST (\url{https://physics.nist.gov/PhysRefData/ASD/lines_form.html}). For each individual line, only one of the transitions with the highest relative intensity according to NIST is listed.  
}
\end{table}\noindent
The existence of $\isotope[3]{He}$ isotope enhancement in helium-weak B-type MS stars with \SI{14000}{\kelvin} $\la$ \teff\,$\la$ \SI{21000}{\kelvin} \citep{Sargent_1961, Hartoog_1979b} as well as in BHB \citep{Hartoog_1979a} and sdB stars with \SI{27000}{\kelvin} $\la$ \teff\,$\la$ \SI{31000}{\kelvin} \citep{Heber_1987, Geier_2013a} is also difficult to reconcile with the simplistic diffusion model because of the general weakness of the radiative acceleration of helium. In principle, the $\isotope[4]{He}$/$\isotope[3]{He}$ abundance ratio decreases with time since the more massive $\isotope[4]{He}$ settles more quickly than $\isotope[3]{He}$ \citep{Michaud_2011, Michaud_Atomic_Diffusion_in_Stars_2015}. Unfortunately, the time needed to obtain the observed $\isotope[3]{He}$ overabundances is too long compared to the stellar lifetime \citep{Vauclair_1974, Michaud_1979, Michaud_2011}. That is why other diffusion models such as the diffusion mass-loss model for MS stars \citep{Vauclair_1975} in combination with stellar fractionated winds \citep{Babel_1996}, the light-induced drift \citep{Atutov_1986, LeBlanc_1993}, meridional circulations \citep{Quievy_2009, Michaud_2008, Michaud_2011}, or thermohaline mixing were developed.\\
\citet{Hartoog_1979a} discovered the first blue horizontal branch star (Feige 86) to show the $\isotope[3]{He}$ anomaly. \citet{Heber_1987} classified two additional BHB stars (PHL 25 and PHL 382\footnote{There is no uniform classification of PHL 382. First, the star was classified as a BHB star by \citet{Heber_1987}. However, it might also be an evolved low-mass star that has left the He-core burning phase (post-BHB star). Finally, it could also be an MS star, as suggested by \citet{Kilkenny_1990} and \citet{Dufton_1993} because of its low surface gravity. For the work at hand, we initially make use of the classification of \citet{Heber_1987} and call PHL 382 a BHB star. This classification is further discussed in Sect. \ref{Effective Temperatures, Surface Gravities and Helium Content}.}) as $\isotope[3]{He}$ stars and discovered that the rotating sdB star SB 290 showed the same anomaly as well. Later, \citet{Edelmann_1997, Edelmann_1999, Edelmann_2001} found another three sdBs (Feige 36, BD+48$^\circ$ 2721, PG 0133+114) in which $\isotope[3]{He}$ is enriched in the atmosphere. \citet{Geier_2013a} added another seven $\isotope[3]{He}$ sdBs. However, $\isotope[3]{He}$ stars are rare among sdBs. \citet{Heber_2004} estimated that less than 5\%\ of the sdB stars show this anomaly, whereas \citet{Geier_2013a} estimated a higher fraction of 18\%.\\
Most of the $\isotope[3]{He}$ BHBs and sdBs do not show periodic radial velocity (RV) variations. Consequently, there is no evidence that binary evolution facilitated the photospheric $\isotope[3]{He}$ enrichment in $\isotope[3]{He}$ BHBs and sdBs. However, three $\isotope[3]{He}$ sdB stars are known to be close binaries (Feige 36, PG 1519+640, and PG 0133+114). While Feige 36 has a RV semi-amplitude of K$\,=\,$\SI{134.6}{\kilo\metre\per\second} \citep{Saffer_1998} and a period of P$\,=\,0.35386\pm0.00014$\,d \citep{Moran_1999}, PG 1519+640 has K$\,=\,36.7\pm1.2$\,\si{\kilo\metre\per\second} and P$\,=\,0.539\pm0.003$\,d \citep{Rueda_2003b, Edelmann_2004, Copperwheat_2011}. PG 0133+114 exhibits K$\,=\,82.0\pm0.3$\,\si{\kilo\metre\per\second} and P$\,=\,1.23787\pm0.00003$\,d \citep{Rueda_2003a, Edelmann_2005}.\\ 
Identifying the possible diffusion processes occurring in the stellar atmosphere of $\isotope[3]{He}$ B-type stars is indispensable for the detailed understanding of their evolution, and empirical information on the photospheric $\isotope[4]{He}$/$\isotope[3]{He}$ abundance ratios are required to constrain theoretical concepts.  $\isotope[3]{He}$ is usually identified by precisely measuring the small isotopic shifts of the $\ion{He}{i}$ absorption lines in the optical part of the spectrum. The isotopic shifts with respect to the $\isotope[4]{He}$ isotope vary from line to line. Physically, two effects are important, as elaborated by \citet{Hughes_1930}: {\sc i}) a shift of term energies that affects all terms, $\Delta E = (\Delta\mu/m)E$, where $\Delta\mu$ is the difference between the reduced masses of the two isotopes, $m$ is the electron mass, and $E$ is the $\isotope[4]{He}$ term energy; {\sc ii}) a specific shift that depends on the wave functions, which becomes non-zero for a two-electron system such as $\ion{He}{i}$ only for the $P$ terms. While the reduced mass effect leads to an overall reduction of the term energies for $^3$He compared to $^4$He, the specific shift reduces the singlet $P$ term energies and increases the triplet ones. Table \ref{summary of isotopic shifts of neutral helium lines} lists all neutral helium line transitions in the near-ultraviolet, optical, and near-infrared spectral range up to principal quantum number $n=8$. While both effects cancel out for the $^3P$ series to some extent (all corresponding transitions have isotopic shifts of $|\Delta\lambda|\la$\,\SI{0.1}{\angstrom}), larger shifts occur for the $^1S$, $^3S$, and $^1P$ series ($|\Delta\lambda|\gtrsim$\,\SI{0.13}{\angstrom}). For example, $\isotope[3]{He}$\;{\sc i} \SI{5875}{\angstrom} is shifted only slightly ($\sim$\,0.04\,\si{\angstrom}) toward redder wavelengths and can therefore be used as a reference line, whereas the strongest isotopic line shifts in the optical amongst others are observed for $\isotope[3]{He}$\;{\sc i} \SI{7281}{\angstrom}, $\isotope[3]{He}$\;{\sc i} \SI{6678}{\angstrom}, and $\isotope[3]{He}$\;{\sc i} \SI{4922}{\angstrom}; they are $\sim$\,0.55\,\si{\angstrom}, $\sim$\,0.50\,\si{\angstrom}, and $\sim$\,0.33\,\si{\angstrom}, respectively (see Table \ref{summary of isotopic shifts of neutral helium lines} and \citealt{Fred_1951}). In order to precisely measure the position of the lines in optical spectra, high signal-to-noise ratios (S/N) are desirable. In general, the RV of the particular $\isotope[3]{He}$ star has to be well determined before the wavelengths of the observed helium lines can be interpreted  because it strongly influences the measured isotopic line shifts and hence the abundance ratio $\isotope[4]{He}$/$\isotope[3]{He}$. RVs can be measured best from the sharp metal lines of the slowly rotating BHB and sdB stars.\\ 
Here, we present the final results of a quantitative spectral analysis performed in order to determine the $\isotope[3]{He}$ and $\isotope[4]{He}$ isotopic abundances and the $\isotope[4]{He}$/$\isotope[3]{He}$ abundance ratios of 13 subluminous B-type stars. We also determined effective temperatures and surface gravities as well as projected rotation velocities from high-resolution spectra using Kurucz ATLAS12 model atmospheres and allowing for NLTE effects for the synthesis of the helium line spectrum using DETAIL and SURFACE. We used updated model spectra that are discussed in detail in Sect. \ref{Model Atmospheres and Synthetic Spectra}.\\
In order to search for possibly unclassified $\isotope[3]{He}$ hot subdwarf B stars, we selected candidates from a sample of 76 sdBs from the hot subdwarf list of the ESO Supernova Ia Progenitor Survey (ESO SPY; \citealt{Napiwotzki_2001}).\\
In Sect. \ref{Target Sample and Data} an overview of the target sample and the spectroscopic data is given. Section \ref{Methods} describes the model atmospheres and spectrum synthesis, the model grids, the spectroscopic analysis technique, and the general effect of $^4$He/$^3$He on helium line formation. The resulting atmospheric parameters, isotopic helium abundances, and abundance ratios are reported in Sects. \ref{Effective Temperatures and Surface Gravities} and \ref{isotope helium abundances}. Evidence for vertical helium stratification in the $\isotope[3]{He}$ BHB and in three of the $\isotope[3]{He}$ sdB program stars is presented in Sect. \ref{Helium Line Profile Anomalies}. Section \ref{Sensitivity study} provides a sensitivity study based on mock spectra in order to verify and interpret the results. The paper ends with a summary of the most important results and gives an outlook on future work.

\section{Target sample and data}\label{Target Sample and Data}
The original target sample consisted of two known $\isotope[3]{He}$ BHB (PHL 25, PHL 382; \citealt{Heber_1987}) and 13 sdB stars. The latter consisted of eight known $\isotope[3]{He}$ stars from \citet{Geier_2013a}: EC 03263-6403, EC 03591-3232, EC 12234-2607, EC 14338-1445, Feige 38, BD+48$^\circ$ 2721, PG 1519+640, and PG 1710+490. Furthermore, the sdB sample included the rotating star SB 290 \citep{Geier_SB290_2013}, which was first discovered to show $\isotope[3]{He}$ by \citet{Heber_1987}, Feige 36 \citep{Edelmann_1997, Edelmann_1999}, and PG 0133+114 \citep{Edelmann_2001}. For comparison, two well-studied prototypical sdBs, HD 4539 and CD-35$^\circ$ 15910, were also included.\\
High-resolution spectra were obtained with five different Echelle spectrographs (see Table \ref{summary of analyzed spectra} in Appendix \ref{Target Sample and Data appendix}). While most stars were observed with the FEROS spectrograph ($R\sim 48\,000$, 3530-9200\,\si{\angstrom}; \citealt{Kaufer_1999}) mounted at the ESO/MPG \SI{2.2}{\metre} telescope in La Silla, Chile, two spectra of BD+48$^\circ$ 2721 and PG 1710+490 were obtained using the FOCES spectrograph ($R\sim 40\,000$, 3800-7000\,\si{\angstrom}; \citealt{Pfeiffer_1998}) mounted at the CAHA \SI{2.2}{\metre} telescope. Both the FEROS and the FOCES spectra were reduced with the MIDAS (\textit{Munich Image Data Analysis System}) package. The BHB star PHL 25 was observed with the HRS spectrograph ($R\sim60\,000$, 3660-9950\,\si{\angstrom}; \citealt{Tull_1998}) mounted at the Hobby-Eberly telescope located at the McDonald Observatory and Feige 36 using the HIRES spectrograph ($R\sim36\,000$, 4270-6720\,\si{\angstrom}; \citealt{Vogt_1994}) on the Keck \SI{10}{\metre} telescope. The FOCES spectra of PG 1519+640 and PG 0133+114 used in the RV study of \citet{Edelmann_2005} were insufficient for a quantitative spectral analysis of the stars, therefore we could not investigate the $\isotope[3]{He}$ anomaly for both stars. The spectra of HE 0929-0424, HE 1047-0436, HE 2156-3927, and HE 2322-0617 were taken with the UVES spectrograph ($R\sim 18\,500$, 3290-6640\,\si{\angstrom}; \citealt{Dekker_2000}) at the ESO VLT in the framework of the ESO SPY project \citep{Napiwotzki_2001}. These stars are discussed in Sect. \ref{Two 3He sdB Stars from the ESO SPY Project}.\\
We RV-corrected single spectra of all program stars and coadded them in order to achieve a higher S/N, which is essential for the analysis to be performed.

\section{Methods}\label{Methods}
\subsection{Model atmospheres and synthetic spectra}\label{Model Atmospheres and Synthetic Spectra}
\begin{figure*}
\begin{minipage}[b]{0.5\linewidth}
\resizebox{\hsize}{!}{\includegraphics{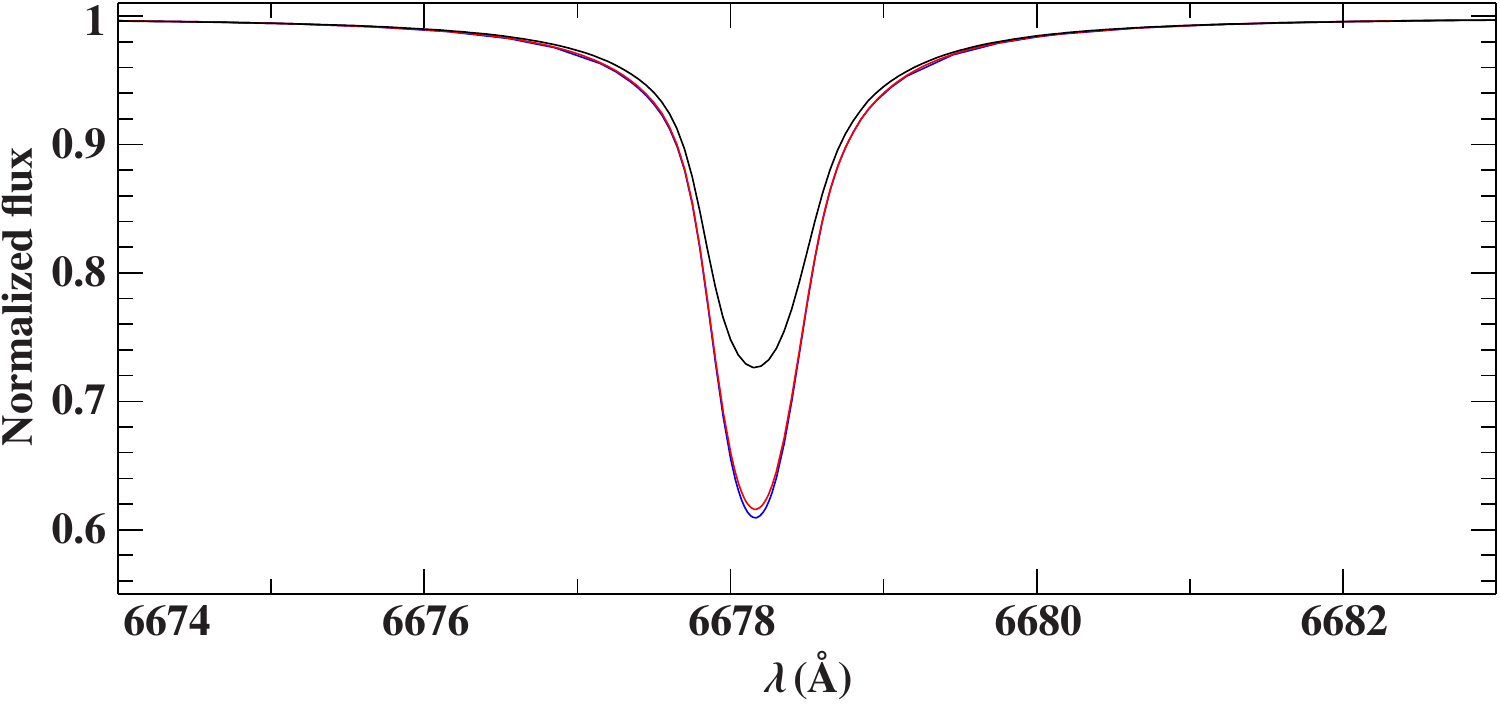}}
\centering
\end{minipage}\hfill
\begin{minipage}[b]{0.5\linewidth}
\resizebox{\hsize}{!}{\includegraphics{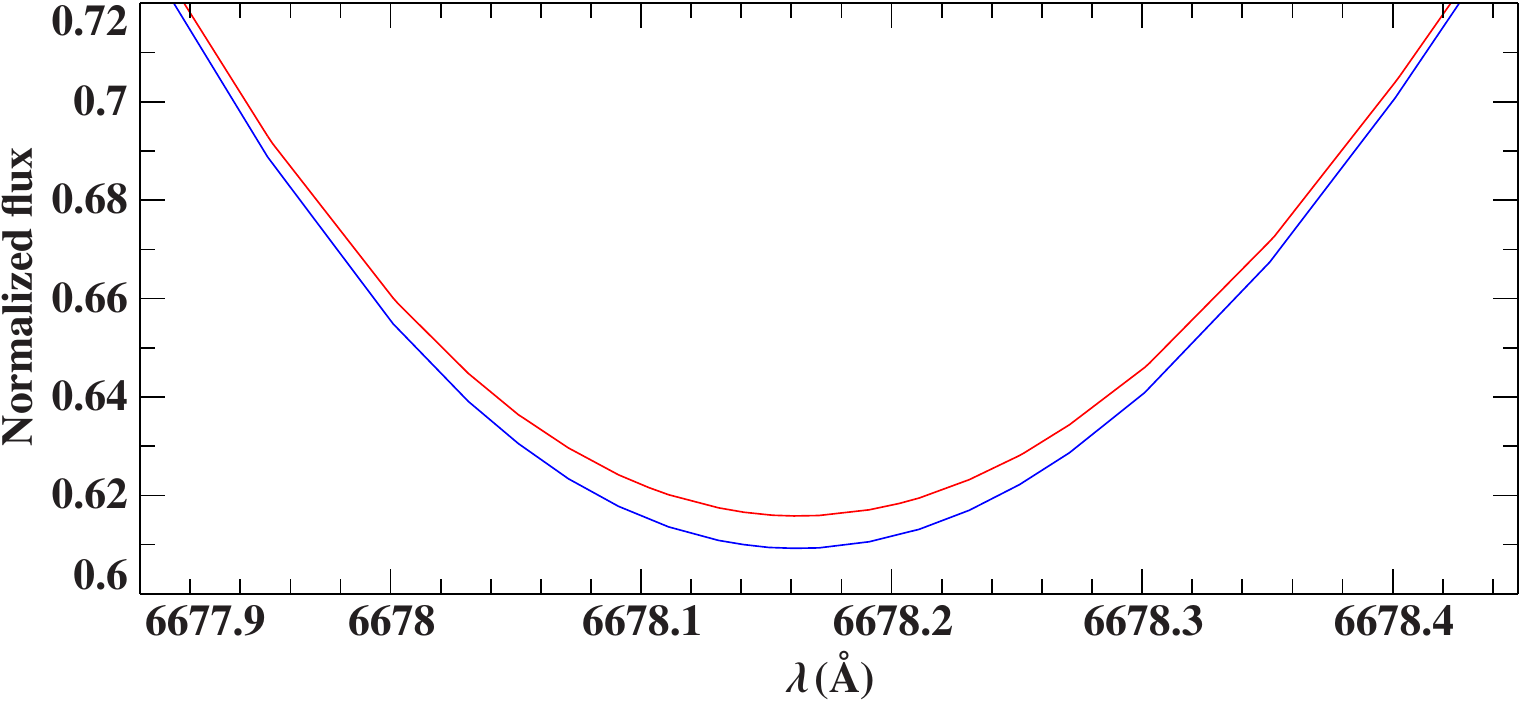}}
\centering
\end{minipage}\hfill
\captionof{figure}{\textit{Left-hand panel}: Model spectra for $\ion{He}{i}$ \SI{6678}{\angstrom} calculated in LTE (black curve) and NLTE (red and blue curve) for $T_{\text{eff}}=$ \SI{28000}{\kelvin}, $\log{(g)}=5.60$, and $\log{(y)}=-2.00$. $\ion{He}{i}$ \SI{6678}{\angstrom} is clearly strongly affected by NLTE effects. While the red model is based on `old' Stark broadening tables for hydrogen and $\ion{He}{i}$ \citep{Dimitrijevic_1990}, the blue model uses `new' broadening tables from \citet{Beauchamp_1997}. \textit{Right-hand panel}: Same as the left-hand panel, but zoomed-in. The difference between the `old' (red curve) and `new' (blue curve) model is marginal, but sufficient in order to explain small deviations in atmospheric parameter determination (see Sect. \ref{Model Atmospheres and Synthetic Spectra} for details).}\label{comparison He6678 LTE_NLTE}
\end{figure*}\noindent
We calculated model spectra making use of a detailed $\isotope[3]{He}$ model atom that has been applied recently by \citet{Maza_2014}. This model atom is identical to that of $\isotope[4]{He}$ by \citet{Przybilla_2005}, except that isotopic line shifts as measured by \citet{Fred_1951} are taken into account. In this model atom, all 29 singlet and triplet terms for principal quantum number $n$\,$\leq$\,5 as well as 6 additional `superlevels' plus the $\ion{He}{ii}$ ground state are considered individually in the non-LTE calculations. Concerning `superlevels', the terms for $n$\,=\,6 to 8 are grouped individually into one level for singlets and triplets. All in all, a total of 162 explicitly treated (multiplet) line transitions plus photoionizations from all levels, free-free radiative processes, and all interconnecting bound-bound and bound-free collisions are considered. Atomic data from \textit{\textup{ab initio}} calculations are employed where available, supplemented by approximations in all other cases, as given by \citet{Przybilla_2005}. The use of identical oscillator strengths, photoionization cross sections, etc. for $^3$He and $^4$He is motivated by the rather small effects of the isotopic shifts on the term structure and wave functions, which impact the atomic data on much smaller scales than the intrinsic uncertainties attributed to them, a few percent at best for ab initio\textup{} oscillator strengths to an order of magnitude for approximate data.\\
\citet{Auer_1973} showed that NLTE effects are quite small for helium lines in the blue-violet region of the spectrum, whereas the red lines, in particular $\ion{He}{i}$ \SI{6678}{\angstrom}, are considerably strengthened by departures from LTE (see the left-hand panel of Fig. \ref{comparison He6678 LTE_NLTE}). The $\ion{He}{i}$ \SI{5875}{\angstrom} line is also strengthened, but less pronounced. Because the modeling of these two lines is crucial for our spectroscopic analysis, it is most important to model their line profiles as precisely as possible accounting for NLTE strengthening. Therefore, we used the hybrid local thermodynamic/non-local thermodynamic equilibrium (LTE/NLTE) approach for B-type stars \citep{Przybilla_2006a, Przybilla_2006b, Przybilla_2011, Nieva_2007, Nieva_2008}. This approach is based on the three generic codes ATLAS12 \citep{Kurucz_1996}, DETAIL, and SURFACE (\citealt{Giddings_1981, Butler_1985}, extended and updated). ATLAS12 model atmospheres were computed in LTE, whereby plane-parallel geometry, chemical homogeneity, and hydrostatic as well as radiative equilibrium were assumed. These metal-rich and line-blanketed model atmospheres were based on the mean metallicity for hot subdwarf B stars determined by \citet{Naslim_2013}. Non-LTE occupation number densities for hydrogen and helium as well as for the included metals were determined with DETAIL, solving the coupled radiative transfer and statistical equilibrium equations. Realistic line-broadening functions were used within SURFACE in order to synthesize the full emergent flux spectrum. When solving the statistical equilibrium equations and the radiative transfer for $\isotope[3]{He}$ and $\isotope[4]{He}$, both isotopes were treated simultaneously since all of their spectral lines overlap. In addition to hydrogen ($\ion{H}{i}$) and helium ($\ion{He}{i/ii}$), the calculated synthetic spectra included spectral lines of $\ion{C}{ii/iii}$, $\ion{N}{ii}$, $\ion{O}{ii}$, $\ion{Ne}{i/ii}$, $\ion{Mg}{ii}$, $\ion{Al}{iii}$, $\ion{Si}{ii/iii/iv}$, $\ion{S}{ii/iii}$, $\ion{Ar}{ii}$, and $\ion{Fe}{ii/iii}$ (see Table \ref{summary of model atoms used}), which were used to precisely measure radial as well as projected rotation velocities.\\
This hybrid LTE/NLTE approach has been applied to hot subdwarf B stars before by \citet{Przybilla_2005b}, \citet{Geier_2007}, and \citet{Latour_2016}. It was also used for the preliminary results presented in \citet{Schneider_2017}. However, hotter BHB stars are analyzed with a hybrid LTE/NLTE technique for the first time here.\footnote{\citet{Przybilla_2010} provided a hybrid LTE/NLTE study of the fast halo BHB star SDSSJ153935.67$+$023909.8 in order to constrain the mass of the Galactic dark matter halo. By making use of the same hybrid approach, \citet{Marino_2014} analyzed HB/BHB stars in the globular cluster NGC 2808. However, all previously investigated BHBs are cooler than those in our sample, which are PHL 25 and PHL 382.}\\
\begin{table}
\caption{Model atoms for non-LTE calculations. Updated and corrected models as described by \citet{Nieva_2012} are marked with `$^a$'.}\label{summary of model atoms used}
\centering
\begin{tabular}{cc}
\hline\hline
Ion & Model atom\\
\hline
$\ion{H}{i}$ & \citet{Przybilla_2004}\\
$\ion{He}{i/ii}$ & \citet{Przybilla_2005}\\
$\ion{C}{ii/iii}$ & \citet{Nieva_2006, Nieva_2008}\\
$\ion{N}{ii}$ & \citet{Przybilla_2001a}\tablefootmark{a}\\
$\ion{O}{ii}$ & \citet{Becker_1988}\tablefootmark{a}\\
$\ion{Ne}{i/ii}$ & \citet{Morel_2008}\tablefootmark{a}\\
$\ion{Mg}{ii}$ & \citet{Przybilla_2001b}\\
$\ion{Al}{iii}$ & Przybilla (in prep.)\\
$\ion{Si}{ii/iii/iv}$ & Przybilla \& Butler (in prep.)\\
$\ion{S}{ii/iii}$ & \citet{Vrancken_1996}\tablefootmark{a}\\
$\ion{Ar}{ii}$ & Butler (in prep.)\\
$\ion{Fe}{ii/iii}$ & \citet{Becker_1998}, \citet{Morel_2006}\tablefootmark{a}\\
\hline
\end{tabular}
\end{table}\noindent
\begin{table}
\caption{Model grid used for the quantitative spectral analysis of sdBs.}\label{model grid used for the quantitative spectral analysis of sdBs}
\centering
\begin{tabular}{ccc}
\hline\hline
Parameter & Grid Size & Step Size\\
\hline
\teff & \SI{20000}{\kelvin} - \SI{35000}{\kelvin} & \SI{1000}{\kelvin}\\
$\log{(g)}$ & 5.0 - 6.0 & 0.2 \\
$\log{n(\text{\isotope[4]{He}})}$ & $^a$ & 0.2\\
$\log{n(\text{\isotope[3]{He}})}$ & $^a$ & 0.2\\
\hline
\end{tabular}
\tablefoot{
\tablefoottext{a}{Depending on the individual star.}
}
\end{table}\noindent
ATLAS12, DETAIL, and SURFACE were modified to account for level dissolution of the $\ion{H}{i}$ and $\ion{He}{ii}$ levels as described by \citet{Hubeny_1994}. Moreover, we used updated Stark broadening tables for hydrogen and $\ion{He}{i}$ according to \citet{Tremblay_2009} and \citet{Beauchamp_1997}, respectively. The latter were used for all synthesized $\ion{He}{i}$ lines, if the particular parameter space (effective temperature, surface gravity, helium abundance) and therefore the respective atmospheric electron densities of the helium line formation depths were included in these tables. Otherwise, Stark broadening tables for $\ion{He}{i}$ according to \citet{Dimitrijevic_1990} were used. As an example, the right-hand panel of Fig. \ref{comparison He6678 LTE_NLTE} displays the influence of the new broadening tables on $\ion{He}{i}$ \SI{6678}{\angstrom}. The influence is marginal, but sufficient in order to explain small deviations in atmospheric parameter determination compared to the preliminary results presented in \citet{Schneider_2017}.

\subsection{Model grid}\label{Model Grid}
We determined the particular atmospheric parameters through a grid of model spectra in a four-dimensional parameter space. A multi-dimensional mesh spanned by effective temperature, surface gravity, and isotopic helium abundances was calculated for the analysis of the sdBs (see Table \ref{model grid used for the quantitative spectral analysis of sdBs}). Arbitrary parameter combinations within this mesh were approximated by linear interpolation between the calculated synthetic spectra. A detailed description on how the individual model spectra were calculated can be found in \citet{Irrgang_2014b}. Since the parameter regime for BHB stars (PHL 25, and PHL 382) is not covered by the calculated grid for sdBs, we used small tailored meshes with the same step sizes as given in Table \ref{model grid used for the quantitative spectral analysis of sdBs} in order to investigate them.

\subsection{Spectroscopic analysis technique}\label{Spectroscopic Analysis Technique}
The spectral analysis made use of the objective, $\chi^2$-based spectroscopic approach as described by \citet{Irrgang_2014b}. This approach uses the whole spectrum at once. The entire wavelength range of the model spectrum, including all synthesized metal lines, can be adjusted to the real spectrum according to the Doppler effect in order to precisely determine both the radial, $v_{\text{rad}}$, and the projected rotation velocity, $v\sin{i}$. We set both macroturbulence $\zeta$ and microturbulence $\xi$ to zero, since there is no indication for additional line-broadening due to these effects in sdB stars \citep{Geier_2012}. In total, six free parameters were fitted simultaneously: \teff, $\log{(g)}$, $\log{n(^3\text{He})}$, $\log{n(^4\text{He})}$, $v_{\text{rad}}$, and $v\sin{i}$.\\
\begin{figure*}
\begin{minipage}[b]{0.5\linewidth}
\resizebox{\hsize}{!}{\includegraphics{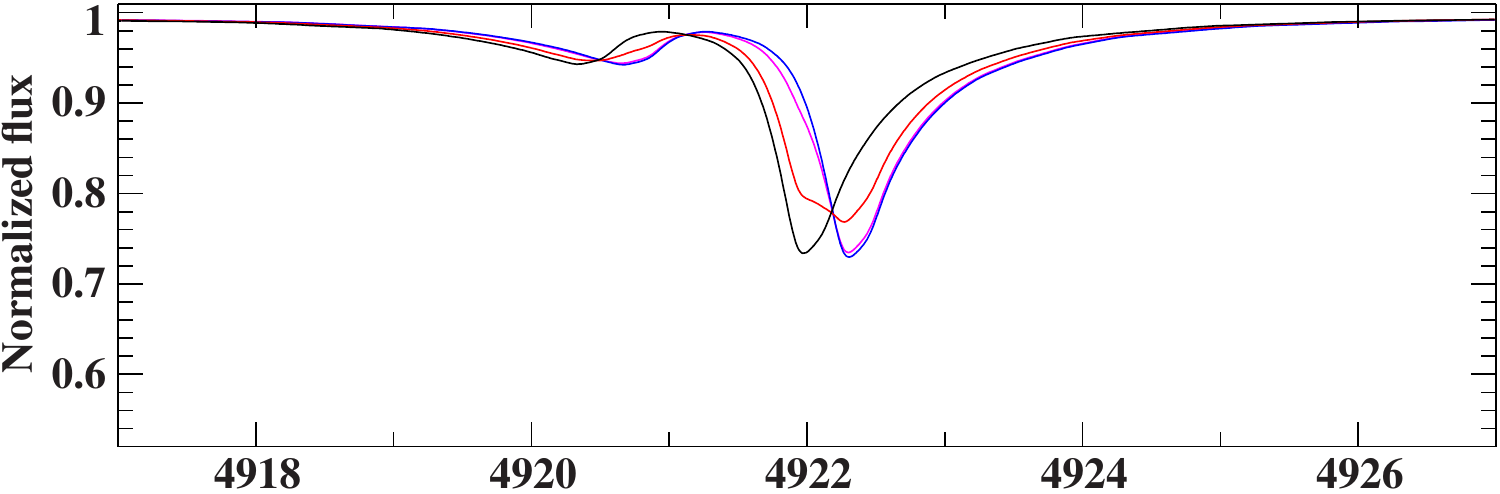}}
\resizebox{\hsize}{!}{\includegraphics{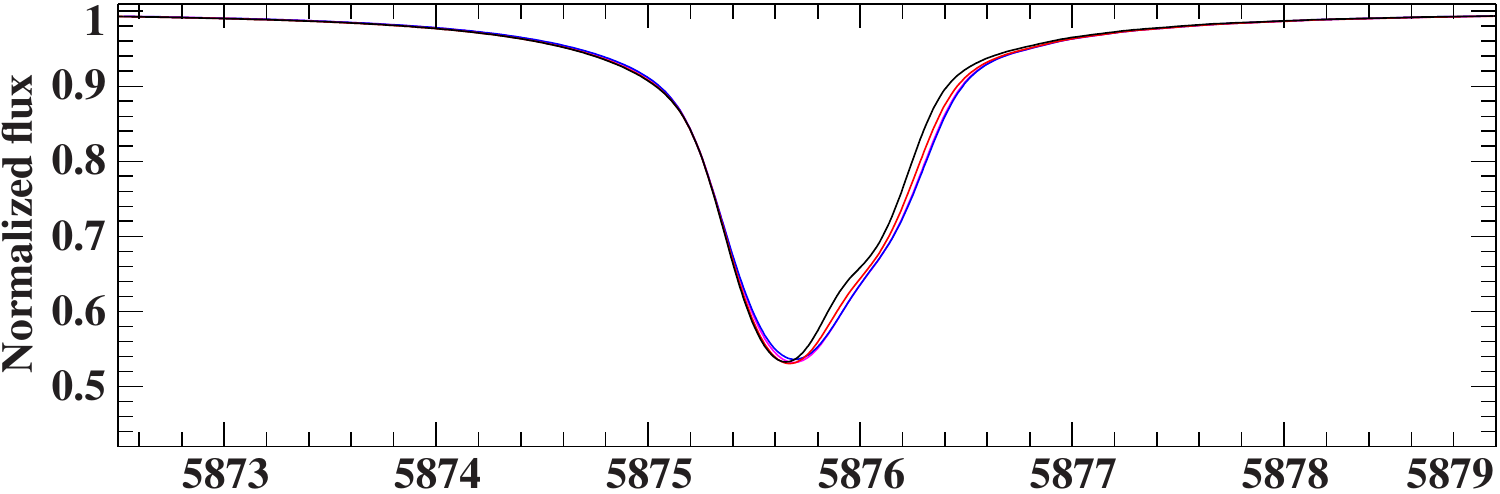}}
\resizebox{\hsize}{!}{\includegraphics{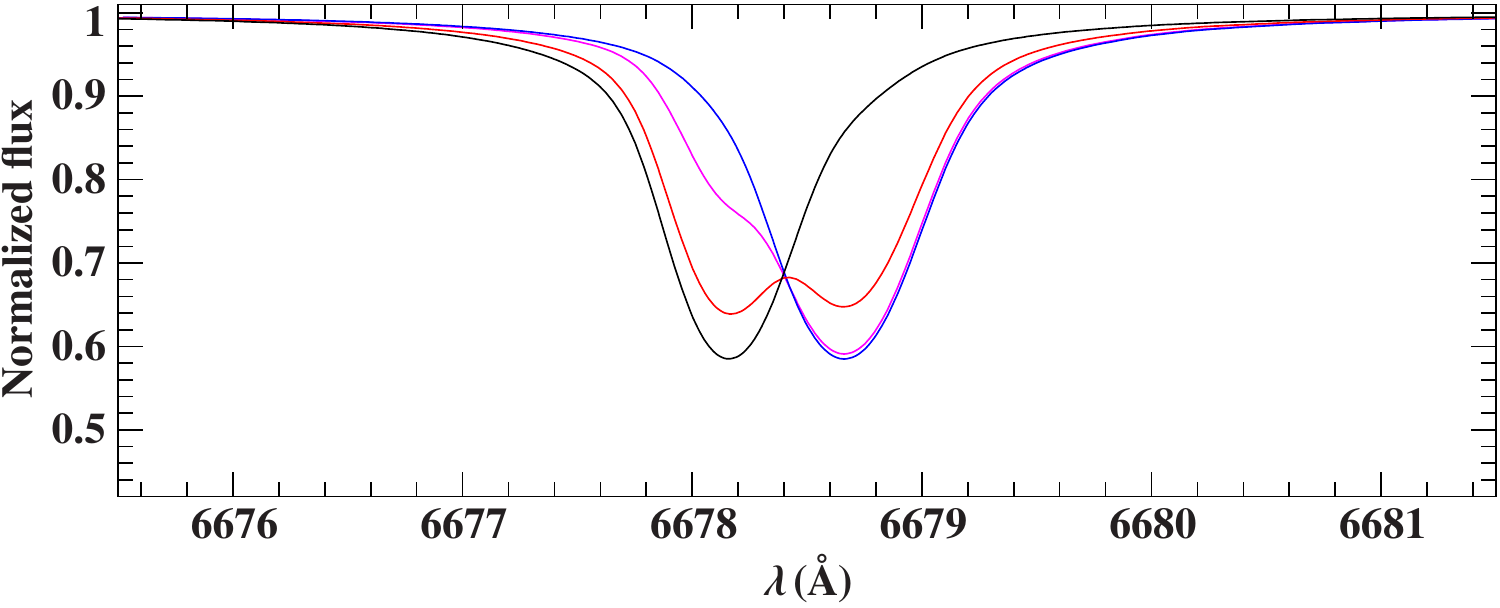}}
\centering
\end{minipage}\hfill
\begin{minipage}[b]{0.5\linewidth}
\resizebox{\hsize}{!}{\includegraphics{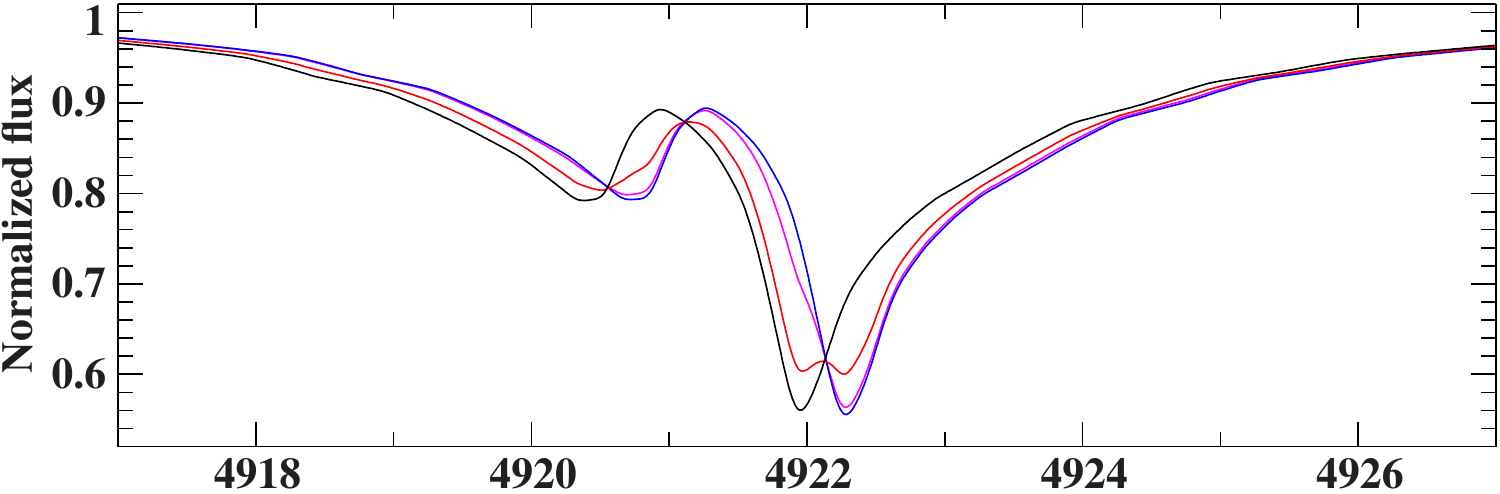}}
\resizebox{\hsize}{!}{\includegraphics{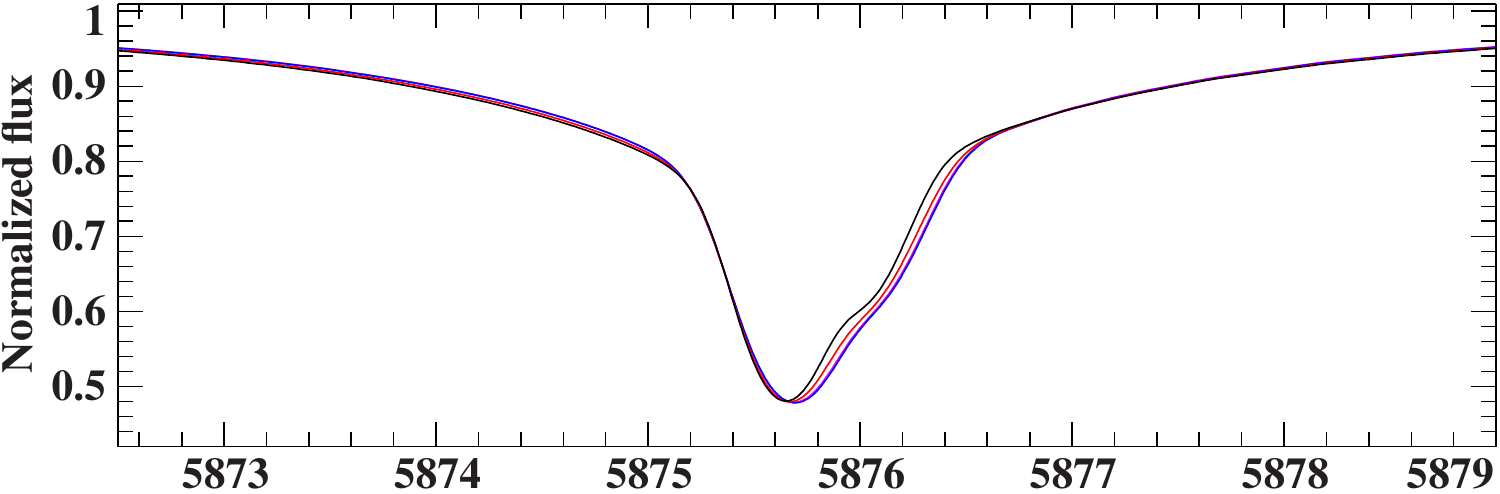}}
\resizebox{\hsize}{!}{\includegraphics{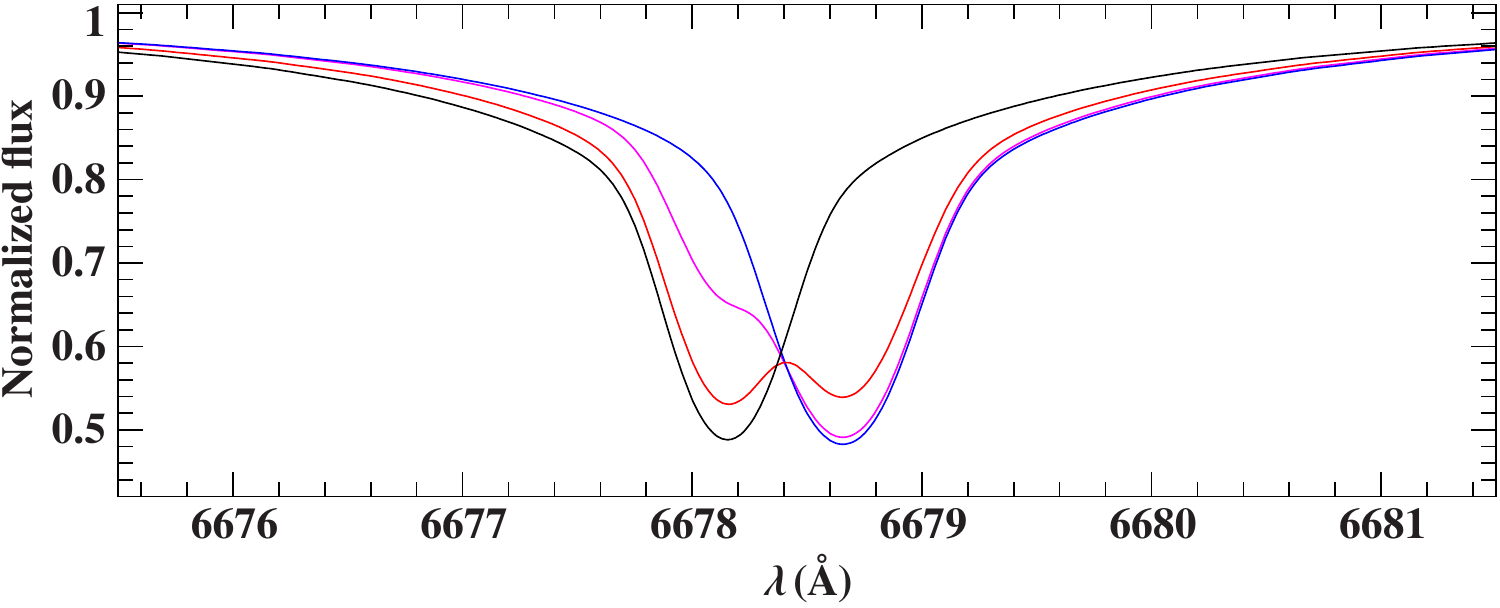}}
\centering
\end{minipage}\hfill
\caption{\textit{Left-hand panels}: Folded ($R=50,000$) model spectra showing $\ion{He}{i}$ \SI{4922}{\angstrom}, $\ion{He}{i}$ \SI{5875}{\angstrom}, and $\ion{He}{i}$ \SI{6678}{\angstrom} for fixed effective temperature \teff$=$ \SI{28000}{\kelvin}, fixed surface gravity $\log{(g)}=5.60$, fixed total helium abundance $\log{n(\isotope[4]{He}+\isotope[3]{He})}\sim-2.00$, but for four different combinations of $^3$He and $^4$He abundances. The corresponding isotopic helium abundances are $\log{n(^3\text{He})}=-4.00$ and $\log{n(^4\text{He})}=-2.00$ (black curve), $\log{n(^3\text{He})}=-2.30$ and $\log{n(^4\text{He})}=-2.30$ (red curve), and $\log{n(^3\text{He})}=-2.00$ and $\log{n(^4\text{He})}=-4.00$ (blue curve), respectively. In addition, a model spectrum for $\log{n(^3\text{He})}=-2.05$ and $\log{n(^4\text{He})}=-3.05$ is shown in magenta. \textit{Right-hand panels}: Same as the left-hand panels, but for a fixed total helium abundance of $\log{n(\isotope[4]{He}+\isotope[3]{He})}\sim-1.00$.}\label{3He/4He comparison for helium lines}
\end{figure*}\noindent
We used gradient methods such as \textit{mpfit} in combination with non-gradient minimization algorithms such as \textit{powell} and \textit{simplex} in order to find the global minimum of the individual $\chi^2$ landscape, which was generally well-behaved such that the best fit was found after a relatively small number of steps. We also determined $1\sigma$ ($\approx$ 68\%) single parameter confidence intervals for all derived quantities.\\
Features that were not properly included in the models used were excluded from fitting. Such features could be cosmics, normalization problems, interstellar or telluric lines, dead or hot pixels, reduction artifacts, noise or non-overlapping diffraction orders at the end of the individual spectra, or photospheric lines with inappropriate or inaccurate atomic data.

\subsection{Influence of the isotopic abundance ratio and total helium abundance on the shape of helium lines}\label{3He model atom and its influence on spectral line formation}
The different isotopic line shifts strongly influence the helium line shapes and depend on both the $^4$He/$^3$He isotopic abundance ratio and on the total helium abundance. The dependence on the $^4$He/$^3$He ratio is demonstrated in the left-hand panels of Fig. \ref{3He/4He comparison for helium lines}, where we display different synthetic profiles for $\ion{He}{i}$ \SI{4922}{\angstrom} (top panel), $\ion{He}{i}$ \SI{5875}{\angstrom} (middle panel), and $\ion{He}{i}$ \SI{6678}{\angstrom} (bottom panel) calculated for \teff$=$ \SI{28000}{\kelvin} and $\log{(g)}=5.60$, that is, typical for an sdB star, but for a fixed total helium abundance of $\log{n(\isotope[4]{He}+\isotope[3]{He})}\sim-2.00$. Because of the small isotopic shift, the shape of $\ion{He}{i}$ \SI{5875}{\angstrom} varies very little when the abundance ratio is varied between $^4$He/$^3$He\,$=$\,1/100 and 100. However, the shapes of the profiles of $\ion{He}{i}$ \SI{4922}{\angstrom} and $\ion{He}{i}$ \SI{6678}{\angstrom} change significantly. The effect of the isotopic anomaly is most pronounced for a ratio of unity, for which the $\ion{He}{i}$ \SI{6678}{\angstrom} line develops a hump absorption profile. For $^4$He/$^3$He as high as 100, or as low as 1/100, the distortion of the line profiles by the minority isotope is invisible to the eye. Hence, we may be able to determine the isotopic ratio from $\ion{He}{i}$ \SI{4922}{\angstrom} and $\ion{He}{i}$ \SI{6678}{\angstrom} if it lies within this range. At $^4$He/$^3$He\,$=$\,1/10 (see the magenta line in Fig. \ref{3He/4He comparison for helium lines}), the line asymmetry remains detectable for $\ion{He}{i}$ \SI{6678}{\angstrom} but not for $\ion{He}{i}$ \SI{4922}{\angstrom} because of the smaller isotopic line shift, which means that $\ion{He}{i}$ \SI{6678}{\angstrom} is the more sensitive diagnostic tool. Increasing the total helium abundance generally results in stronger absorption lines with much broader line wings. This is shown in the right-hand panels of Fig. \ref{3He/4He comparison for helium lines}, where the same profiles are displayed as in the left-hand panels, but for a total helium abundance of $\log{n(\isotope[4]{He}+\isotope[3]{He})}\sim-1.00$. Interestingly, the hump absorption profile for an abundance ratio of unity can now be identified for $\ion{He}{i}$ \SI{4922}{\angstrom} as well.


\section{Atmospheric parameters and projected rotation velocities}\label{Effective Temperatures and Surface Gravities}
\begin{figure}
\begin{center}
\resizebox{\hsize}{!}{\includegraphics{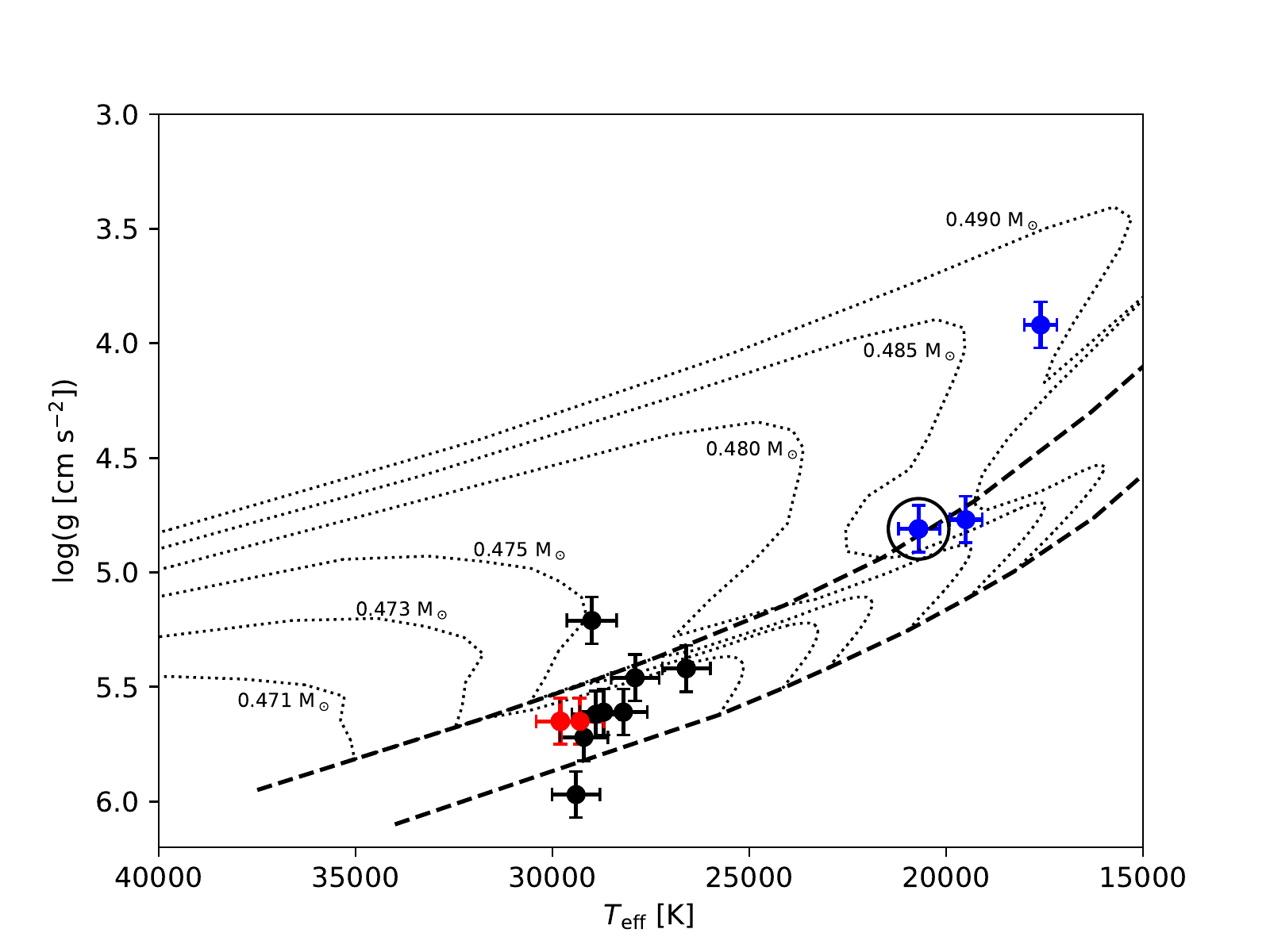}}
\caption{\teff-$\log{(g)}$ diagram of the program stars. While known $\isotope[3]{He}$ sdBs from our sample are marked in black, the two $\isotope[3]{He}$-enriched ESO SPY sdBs (HE 0929-0424, HE 1047-0436; see Sect. \ref{Two 3He sdB Stars from the ESO SPY Project}) are marked in red. The three blue dots at the cool end of the sequence represent PHL 25, PHL 382, and BD+48$^\circ$ 2721, highlighted with a solid circle (see Sect. \ref{Results from NLTE versus LTE analyses} for details). The zero-age (ZAHB) and terminal-age horizontal branch (TAHB) as well as evolutionary tracks for different stellar masses but solar metallicity according to \citet{Dorman_1993} are also shown with dashed and dotted lines, respectively.}\label{NLTE_Teff_logg_diagram_1.pdf}
\end{center}
\end{figure}\noindent
Effective temperatures and surface gravities of the program stars were determined by fitting the calculated model spectra to the hydrogen and helium lines of the observed spectra listed in Table \ref{summary of analyzed spectra}. In addition, sharp metal lines were used for a precise determination of $v_{\text{rad}}$ and $v\sin{i}$. Preliminary results have been reported by \citet{Schneider_2017}, which were based on standard ATLAS12 model atmospheres and a DETAIL/SURFACE spectrum synthesis, that is, without the modifications described in Sect. \ref{Model Atmospheres and Synthetic Spectra}.\\
\begin{table*}
\small
\caption{Effective temperatures, surface gravities, and helium abundances of the analyzed subluminous B-type stars compared to literature values. In both cases the same high-resolution spectra were analyzed, but different approaches were used (see Sect. \ref{Results from NLTE versus LTE analyses} for details). Isotopic helium abundances and abundance ratios according to our analysis are also given.}\label{briefly summarized 4He/3He table hybrid LTE/NLTE 1}
\centering
\begin{tabular}{cccccccc}
\hline\hline
Object & \teff & $\log{(g[\si{\cm\per\square\second}])}$ & $\log{n(\isotope[4]{He}+\isotope[3]{He})}$ & $\log{n(\text{\isotope[4]{He}})}$ & $\log{n(\text{\isotope[3]{He}})}$ & $\frac{n(\text{\isotope[4]{He}})}{n(\text{\isotope[3]{He}})}$& References\\
 & [\si{\kelvin}] &  &  &  &  & \\
\hline
HD 4539 & $23\,200\pm100$ & $5.20\pm0.01$ & $-2.27\pm0.24$ & $-2.27\pm0.01$ & $<-4.89$\tablefootmark{a} & $>410^a$ & [1]\\
CD-35$^\circ$ 15910 & $27\,200\pm100$ & $5.39\pm0.01$ & $-2.94\pm0.23$ & $-2.94\pm0.01$ & $<-4.89$\tablefootmark{a} & $>80^a$ & [1]\\
& $27\,000\pm1100$ & $5.32\pm0.14$ & $-2.90\pm0.10$ &  &  &  & [2]\\
\hline
PHL 25\tablefootmark{b} & $19\,500\pm100$ & $4.77^{+0.01}_{-0.02}$ & $-2.42\pm0.18$ & $-2.75\pm0.03$\tablefootmark{b} & $-2.69\pm0.03$\tablefootmark{b} & $0.87:$ & [1]\\
PHL 382\tablefootmark{b} & $17\,600\pm100$ & $3.92\pm0.01$ & $-2.54\pm0.20$ & $-3.25\pm0.05$\tablefootmark{b} & $-2.63\pm0.02$\tablefootmark{b} & $0.24:$ & [1]\\
BD+48$^\circ$ 2721\tablefootmark{b} & $20\,700^{+100}_{-200}$ & $4.81\pm0.02$ & $-2.51^{+0.27}_{-0.30}$ & $-3.34^{+0.09}_{-0.11}$\tablefootmark{b} & $-2.57^{+0.09}_{-0.11}$\tablefootmark{b} & $0.17:$ & [1]\\
& $24\,800\pm1100$ & $5.38\pm0.14$ & $-2.23\pm0.10$ &  &  &  & [2]\\
\hline
SB 290\tablefootmark{b}\tablefootmark{c} & $26\,600\pm100$ & $5.42\pm0.01$ & $-2.69\pm0.22$ & $-3.73^{+0.12}_{-0.11}$\tablefootmark{b} & $-2.73\pm0.02$\tablefootmark{b} & $0.10:$ & [1]\\
EC 03263-6403 & $29\,000\pm200$ & $5.21\pm0.02$ & $-2.84\pm0.24$ & $-4.75^{+0.29}_{-0.32}$ & $-2.85^{+0.03}_{-0.02}$ & $0.01\pm0.01$ & [1]\\
& $29\,300\pm1100$ & $5.48\pm0.14$ & $-2.51\pm0.10$ &  &  &  & [2]\\
EC 03591-3232\tablefootmark{b} & $28\,700\pm100$ & $5.61\pm0.01$ & $-2.09\pm0.23$ & $-3.51^{+0.14}_{-0.19}$\tablefootmark{b} & $-2.11\pm0.01$\tablefootmark{b} & $0.04:$ & [1]\\
& $28\,000\pm1100$ & $5.55\pm0.14$ & $-2.03\pm0.10$ &  &  &  & [2]\\
EC 12234-2607\tablefootmark{b} & $28\,900\pm100$ & $5.62\pm0.02$ & $-1.53^{+0.20}_{-0.19}$ & $-1.65^{+0.03}_{-0.02}$\tablefootmark{b} & $-2.14\pm0.05$\tablefootmark{b} & $3.09:$ & [1]\\
& $28\,000\pm1100$ & $5.58\pm0.14$ & $-1.58\pm0.10$ &  &  &  & [2]\\
EC 14338-1445 & $27\,900\pm100$ & $5.46^{+0.01}_{-0.02}$ & $-3.01\pm0.21$ & $-3.75^{+0.07}_{-0.08}$ & $-3.10\pm0.03$ & $0.22\pm0.09$ & [1]\\
& $27\,700\pm1100$ & $5.54\pm0.14$ & $-2.82\pm0.10$ &  &  &  & [2]\\
Feige 38 & $28\,200\pm100$ & $5.61\pm0.01$ & $-2.70\pm0.21$ & $-3.48^{+0.08}_{-0.11}$ & $-2.78\pm0.02$ & $0.20^{+0.08}_{-0.09}$ & [1]\\
PG 1710+490 & $29\,200\pm100$ & $5.72\pm0.02$ & $-2.66\pm0.22$ & $-3.67^{+0.05}_{-0.04}$ & $-2.70\pm0.01$ & $0.11\pm0.04$ & [1]\\
Feige 36\tablefootmark{d} & $29\,400\pm100$ & $5.97\pm0.01$ & $-2.18^{+0.18}_{-0.19}$ & $-2.49^{+0.04}_{-0.06}$ & $-2.48^{+0.04}_{-0.06}$ & $0.98^{+0.35}_{-0.38}$ & [1]\\
& $29\,800\pm100$ & $5.97\pm0.02$ & $-2.17\pm0.02$ &  &  &  & [3]\\
\hline
HE 0929-0424\tablefootmark{d} & $29\,300\pm100$ & $5.65\pm0.01$ & $-1.95\pm0.19$ & $-2.10\pm0.03$ & $-2.50\pm0.06$ & $2.51\pm0.91$ & [1]\\
& $29\,602\pm529$ & $5.69\pm0.07$ & $-2.01\pm0.07$ &  &  &  & [4]\\
HE 1047-0436\tablefootmark{d} & $29\,800\pm100$ & $5.65\pm0.01$ & $-2.44\pm0.18$ & $-2.76\pm0.04$ & $-2.72\pm0.03$ & $0.91\pm0.32$ & [1]\\
& $30\,280\pm529$ & $5.71\pm0.07$ & $-2.35\pm0.07$ &  &  &  & [4]\\
\hline
\end{tabular}
\tablefoot{
We give $1\sigma$ statistical single parameter errors for \teff, $\log{(g)}$, $\log{n(\text{\isotope[4]{He}})}$, and $\log{n(\text{\isotope[3]{He}})}$. The systematic uncertainties are $\pm2$\% in \teff, and $\pm0.10$ for $\log{(g)}$, $\log{n(\text{\isotope[4]{He}})}$, and $\log{n(\text{\isotope[3]{He}})}$ in all cases. In the case of $\log{n(\isotope[4]{He}+\isotope[3]{He})}$ and $n(\text{\isotope[4]{He}})/n(\text{\isotope[3]{He}}),$ the given uncertainties result from the statistical and systematic errors on $\log{n(\text{\isotope[3]{He}})}$ and $\log{n(\text{\isotope[4]{He}})}$, for which Gaussian error propagation was used.\\
\tablefoottext{a}{The solar value of $\log{n(\text{\isotope[3]{He}})}=-4.89$ was adopted.}\tablefoottext{b}{Anomalous helium line profiles (see Sect. \ref{Helium Line Profile Anomalies}). Therefore, the isotopic abundance ratio $n(\text{\isotope[4]{He}})/n(\text{\isotope[3]{He}})$ is uncertain as indicated by a colon.}\tablefoottext{c}{Rotating star.}\tablefoottext{d}{RV-variable star.}
}
\tablebib{
(1) This work; (2) \citet{Geier_2013a}; (3) \citet{Edelmann_1999}; (4) \citet{Lisker_2005}.   
}
\end{table*}
\subsection{Effective temperatures, surface gravities, and helium content}\label{Effective Temperatures, Surface Gravities and Helium Content}
Table \ref{briefly summarized 4He/3He table hybrid LTE/NLTE 1} lists the resulting effective temperatures, surface gravities, and helium abundances. Because statistical uncertainties are small, the error budget is dominated by systematic uncertainties. As suggested in the study of \citet{Lisker_2005}, we added systematic uncertainties of $\pm$2\% for the effective temperatures and $\pm$0.1 for $\log{(g)}$. Given uncertainties on $\log{n(\isotope[4]{He}+\isotope[3]{He})}$ in Table \ref{briefly summarized 4He/3He table hybrid LTE/NLTE 1} result from Gaussian error propagation, for which we used $\pm$0.1 as systematic errors for $\log{n(\text{\isotope[3]{He}})}$ and $\log{n(\text{\isotope[4]{He}})}$, respectively. All stars are helium-deficient compared to the solar helium abundance, and except for EC 12234-2607 show typical abundances of -2.00\,$\lesssim$\,$\log{n(\isotope[4]{He}+\isotope[3]{He})}$\,$\lesssim$\,-3.00. Fig. \ref{NLTE_Teff_logg_diagram_1.pdf} shows the \teff-$\log{(g)}$ diagram, where we compare our results to predictions of evolutionary models for the horizontal branch and beyond assuming solar metallicity and canonical masses between 0.471\,$M_{\text{\sun}}$ and 0.490\,$M_{\text{\sun}}$ \citep{Dorman_1993}. Most of the analyzed $\isotope[3]{He}$ stars lie within the EHB band, as expected. EC 03263-6403 has already evolved beyond the helium-core burning phase as the star lies above the canonical HB. The same is true for the post-BHB star PHL 382. Consequently, EC 03263-6403 and PHL 382 should have evolved off the HB and therefore should not be burning helium at their cores. Moreover, it is worthwhile to note that Feige 36 lies below the ZAEHB, which may be explained by a lower-than-canonical mass of the star (see \citealt{Kupfer_2015}). Most importantly, however, all $\isotope[3]{He}$ stars cluster in a narrow temperature strip between $\sim$\,\SI{26000}{\kelvin} and $\sim$\,\SI{30000}{\kelvin}, very similar to the results of \citet{Geier_2013a}.
\begin{figure}
\begin{center}
\resizebox{\hsize}{!}{\includegraphics{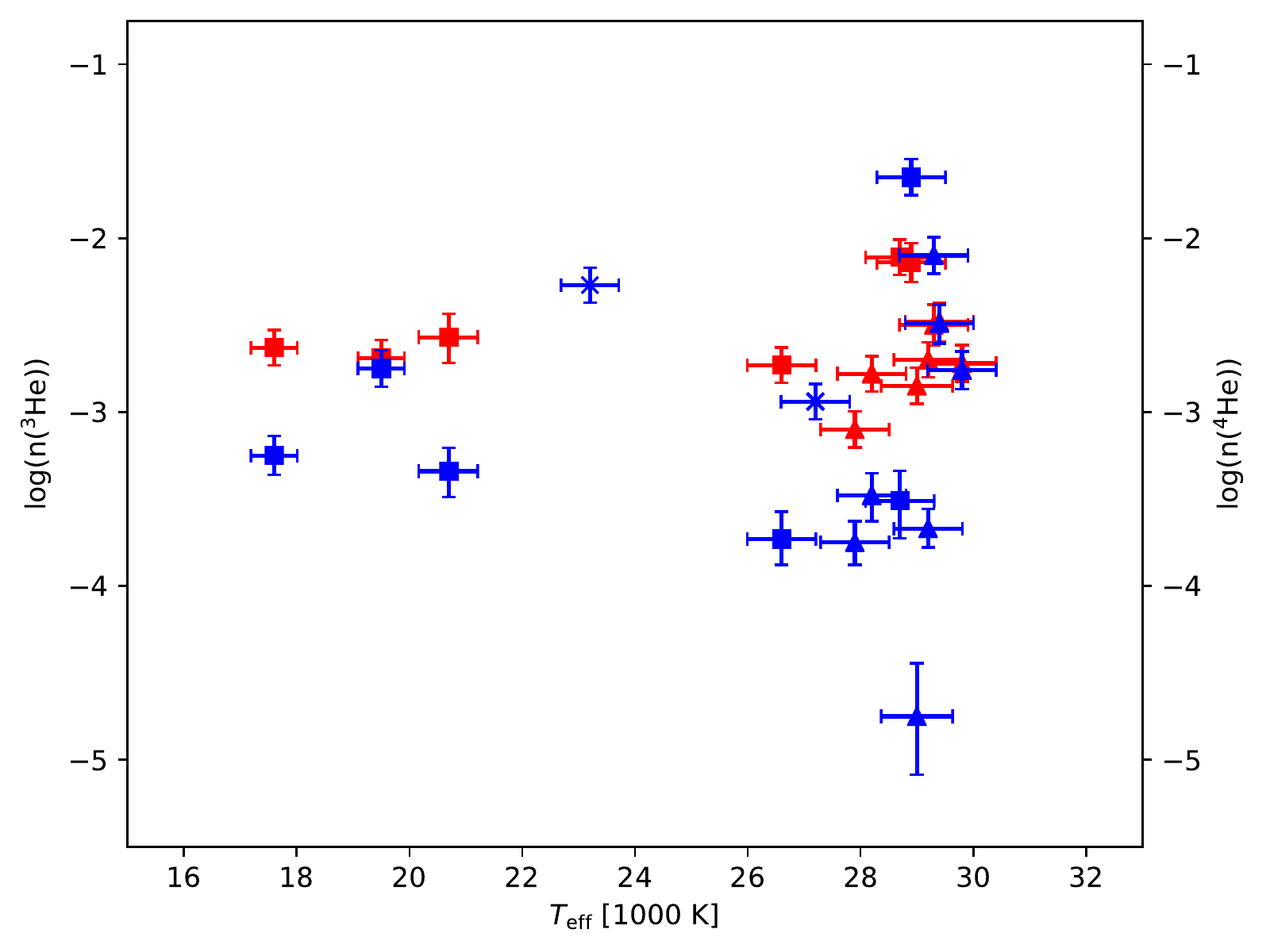}}
\end{center}
\captionof{figure}{$\isotope[3]{He}$ (red) and $\isotope[4]{He}$ (blue) abundances of the analyzed stars plotted against effective temperature. Stars showing anomalous helium line profiles (see Sect. \ref{Helium Line Profile Anomalies}) are marked with squares. Triangles represent $\isotope[3]{He}$ stars for which we were able to match the helium line profiles. In addition, both He-normal comparison stars (HD 4539 and CD-35$^\circ$ 15910) are marked with crosses. Their $\isotope[3]{He}$ upper limits (see Table \ref{briefly summarized 4He/3He table hybrid LTE/NLTE 1}) are not displayed.}\label{NLTE_abundances_plot.pdf}
\end{figure}\noindent
\subsection{Results from NLTE versus LTE analyses}\label{Results from NLTE versus LTE analyses}
Nine of the $\isotope[3]{He}$ sdB stars have been analyzed from the same high-resolution spectra used here (see \citealt{Edelmann_1999}, \citealt{Lisker_2005}, and \citealt{Geier_2013a} for details). All published analyses made use of the grid of metal line-blanketed LTE model atmospheres and spectral synthesis described by \citet{Heber_2000}. Atmospheric parameters were determined by fitting the observed hydrogen Balmer and helium lines to grids of synthetic spectra in a similar way to the analysis presented here. Previous analyses also relied on $\chi^2$ minimization, but used the codes of \citet{Napiwotzki_1999} or its variant SPAS (\citealt{Hirsch_2009}; see, e.g., \citealt{Copperwheat_2011} for a detailed description) and a solar or supersolar metallicity.\\
Hence, a comparison of our results to the published ones allows us to test systematic effects, that is, the cumulative impact of departures from LTE, different metal contents, and the different analysis strategies used (objective $\chi^2$ -based spectroscopic approach vs. SPAS). In Table \ref{briefly summarized 4He/3He table hybrid LTE/NLTE 1} we compare our results to the published ones for the relevant stars.\\
No systematic differences can be identified between the two approaches for effective temperatures and surface gravities. We therefore conclude that the cumulative effect of departures from LTE, different metal contents, and the different analysis strategies is minor. All values agree to within the given uncertainties, except for BD+48$^\circ$ 2721 and EC 03263-6403. In the case of EC 03263-6403, we derived a lower surface gravity of $\log{(g)}=5.21\pm0.11$ compared to literature values ($5.48\pm0.14$). However, we derived a drastically lower \teff\, by $\sim$\,\SI{4000}{\kelvin} and $\log{(g)}$ by $\sim$\,0.57 for BD+48$^\circ$ 2721, although we used the same FOCES spectrum as \citet{Geier_2013a}. At \teff$=20\,700\pm600$\,\si{\kelvin} and $\log{(g)}=4.81\pm0.11$, the atmospheric parameters of BD+48$^\circ$ 2721 are quite similar to those of the BHB star PHL 25. Because of this similarity, BD+48$^\circ$ 2721 should no longer be considered an sdB, but should be reclassified as a BHB star.
\subsection{Rotational broadening}\label{Rotational Broadening}
Subluminous B stars are slow rotators, unless they are spun up by a compact companion \citep{Geier_2010, Geier_2012}. Projected rotational velocities have been reported for all $\isotope[3]{He}$ sdB stars to be low, that is, close to the detection limit, which is on the order of the typical spectral resolution element of the instrumental profile ($\sim$\,5-8\,\si{\kilo\metre\per\second}). Our analysis of the metal line profiles confirms the slow rotation, which implies that rotation is irrelevant for modeling the helium line profiles to determine the isotopic abundances and the abundance ratios. However, there are two exceptions. Although apparently single, SB 290 is known to be a rapid rotator. \citet{Geier_SB290_2013} derived the projected rotation velocity from metal lines to be $v\sin{i}=48.0\pm2.0$\,\si{\kilo\metre\per\second}. They noted, however, that the observed helium lines require a higher rotational broadening of $v\sin{i}=58.0\pm1.0$\,\si{\kilo\metre\per\second} to be matched by synthetic spectra. We confirm this discrepancy and discuss it in Sect. \ref{Helium Line Profile Anomalies}. PHL 382 also shows significant rotation, but at a lower $v\sin{i}=12.9\pm0.1$\,\si{\kilo\metre\per\second} than SB 290.

\section{Isotopic helium abundances}\label{isotope helium abundances}
\begin{figure*}
\begin{minipage}[b]{0.5\linewidth}
\resizebox{\hsize}{!}{\includegraphics{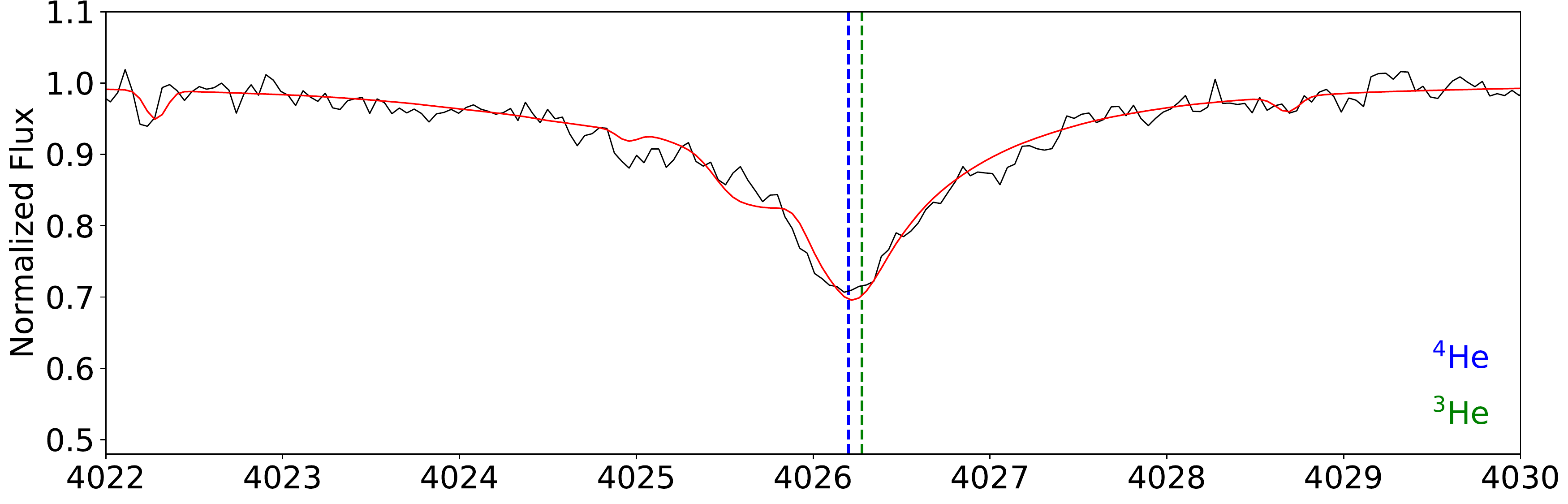}}
\centering
        \end{minipage}\hfill
\begin{minipage}[b]{0.5\linewidth}
\resizebox{\hsize}{!}{\includegraphics{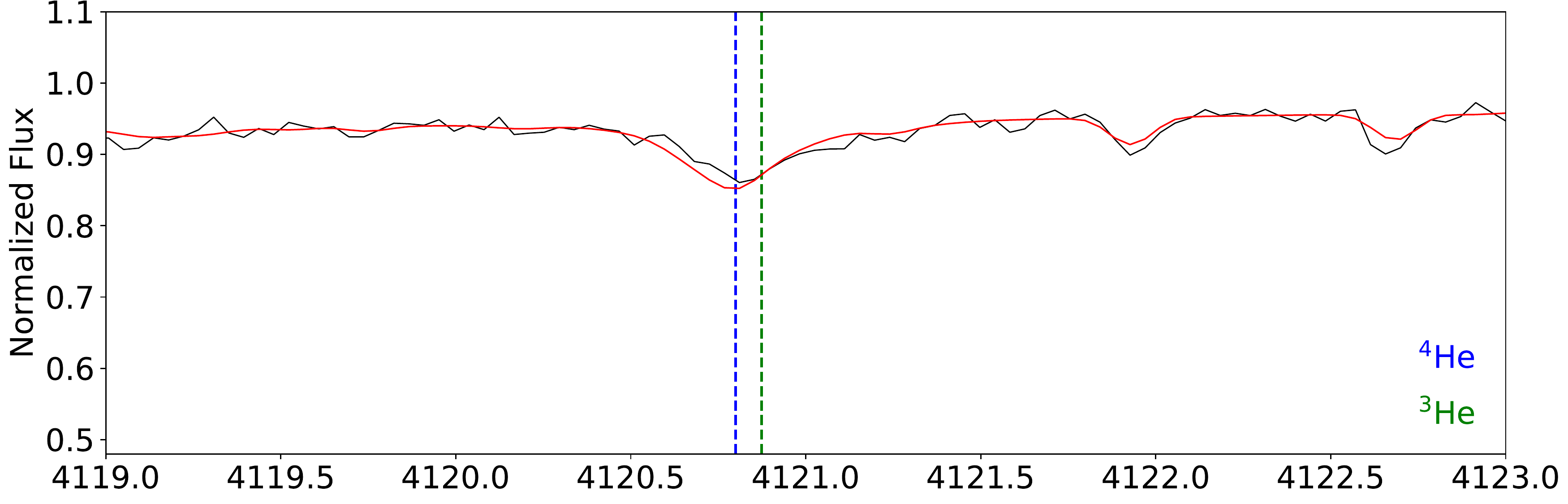}}
        \centering
        \end{minipage}\hfill    
\begin{minipage}[b]{0.5\linewidth}
\resizebox{\hsize}{!}{\includegraphics{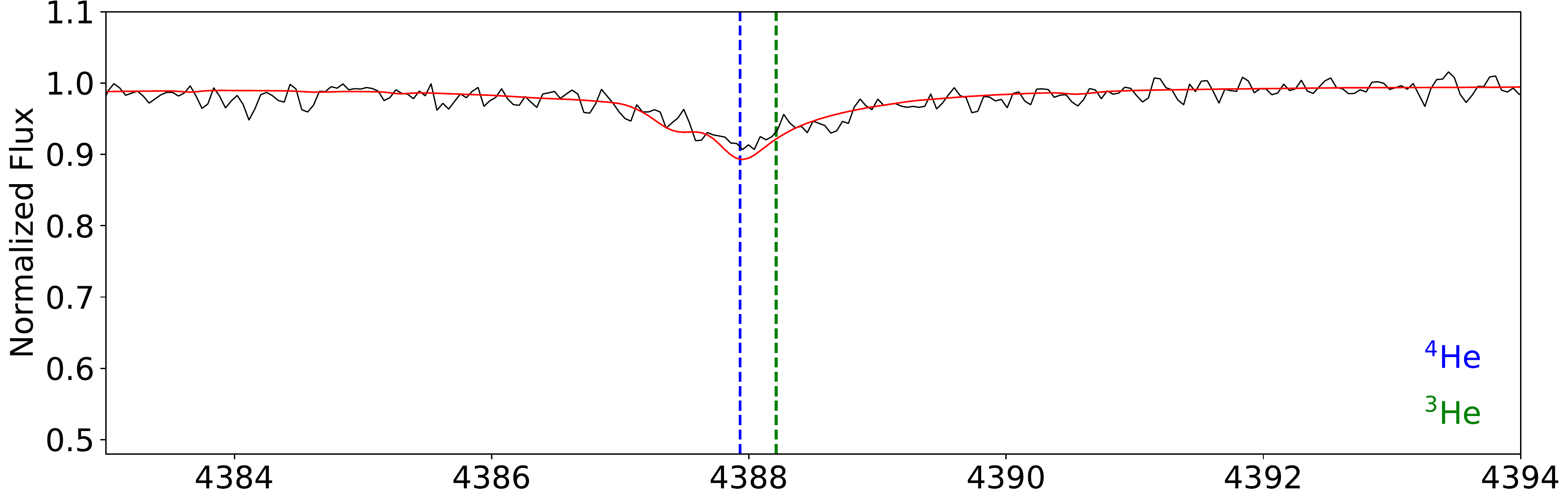}}
        \centering
        \end{minipage}\hfill
\begin{minipage}[b]{0.5\linewidth}      
\resizebox{\hsize}{!}{\includegraphics{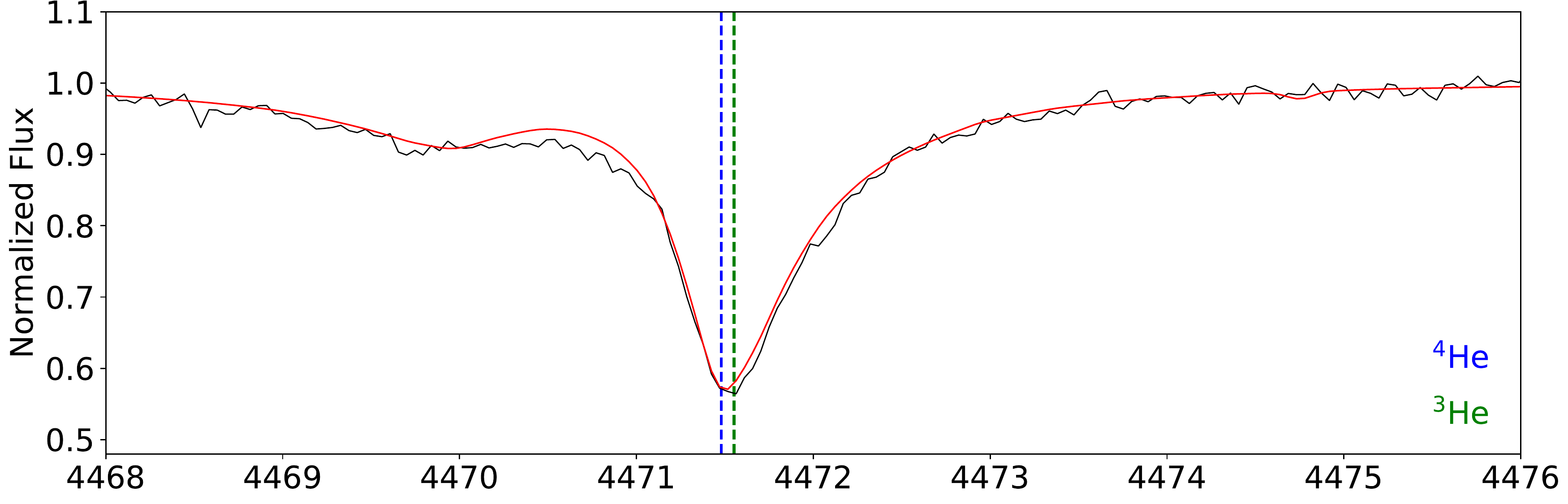}}
        \centering
        \end{minipage}\hfill
\begin{minipage}[b]{0.5\linewidth}      
\resizebox{\hsize}{!}{\includegraphics{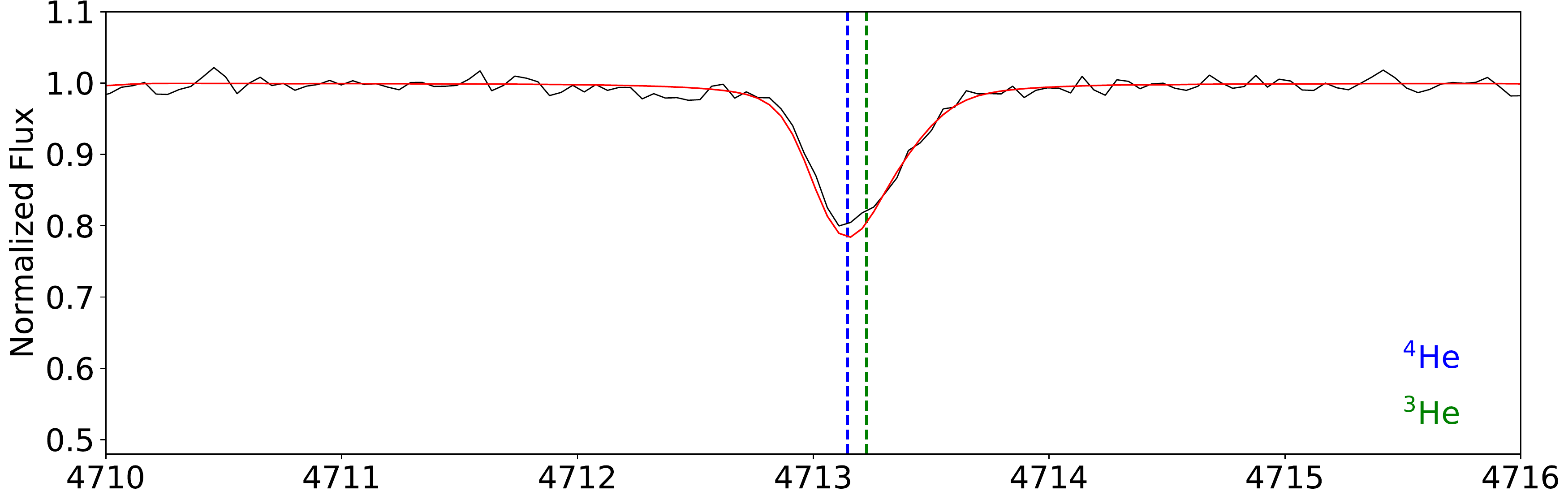}}
        \centering
        \end{minipage}\hfill
\begin{minipage}[b]{0.5\linewidth}      
\resizebox{\hsize}{!}{\includegraphics{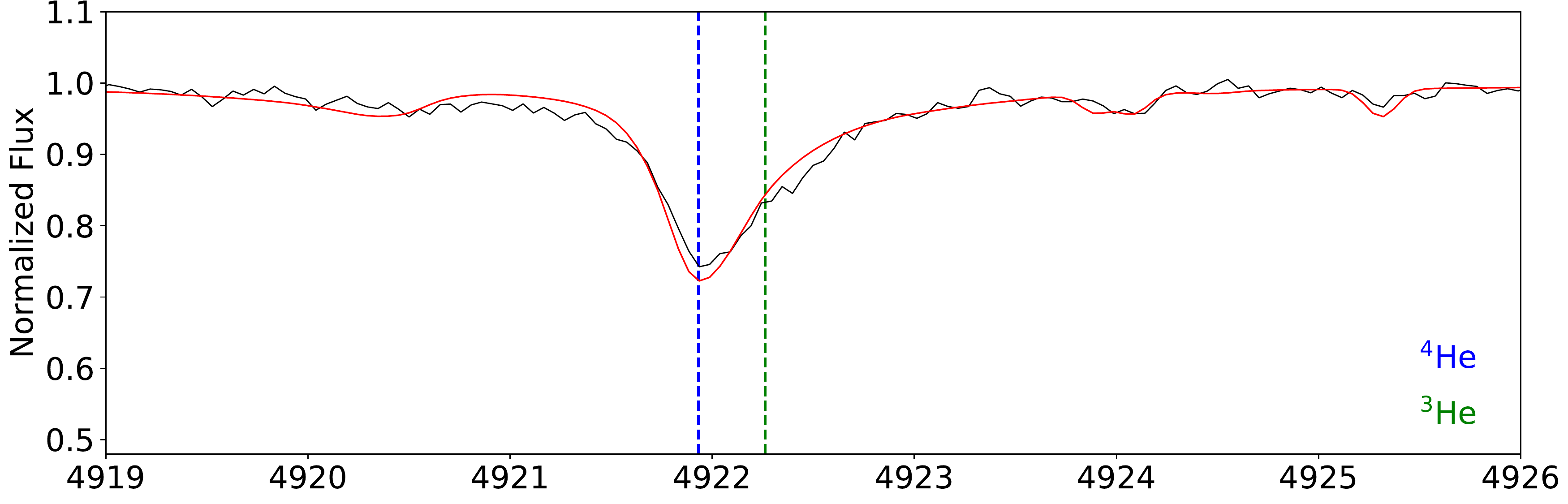}}
        \centering
        \end{minipage}\hfill
        \begin{minipage}[b]{0.5\linewidth}      
\resizebox{\hsize}{!}{\includegraphics{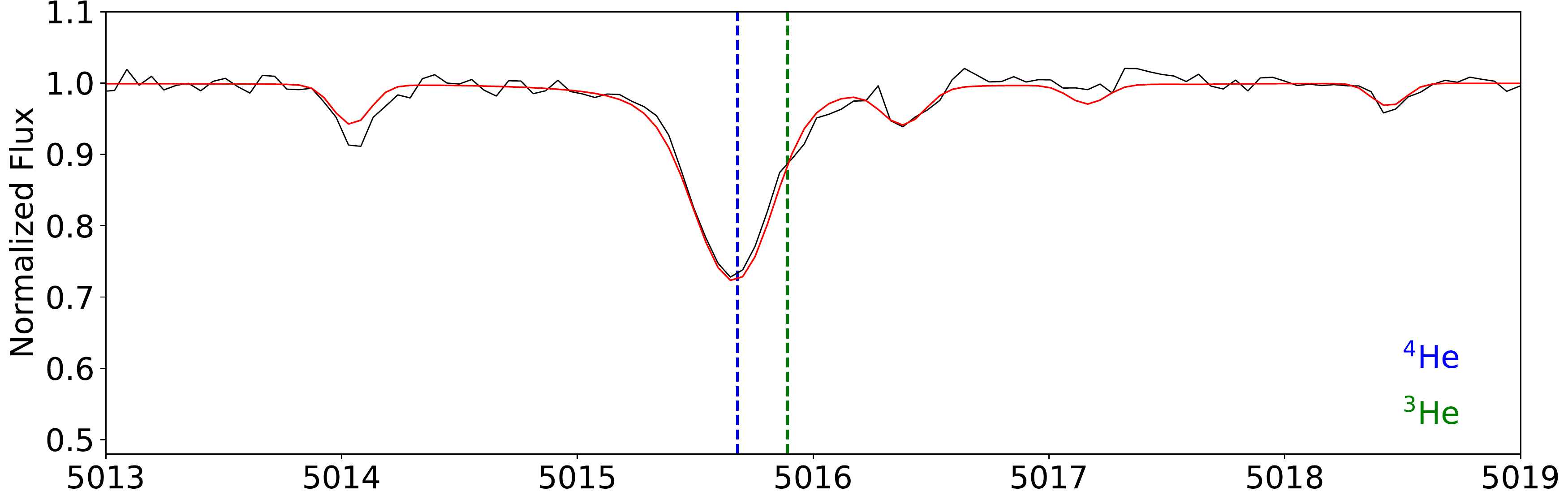}}
        \centering
        \end{minipage}\hfill
\begin{minipage}[b]{0.5\linewidth}      
\resizebox{\hsize}{!}{\includegraphics{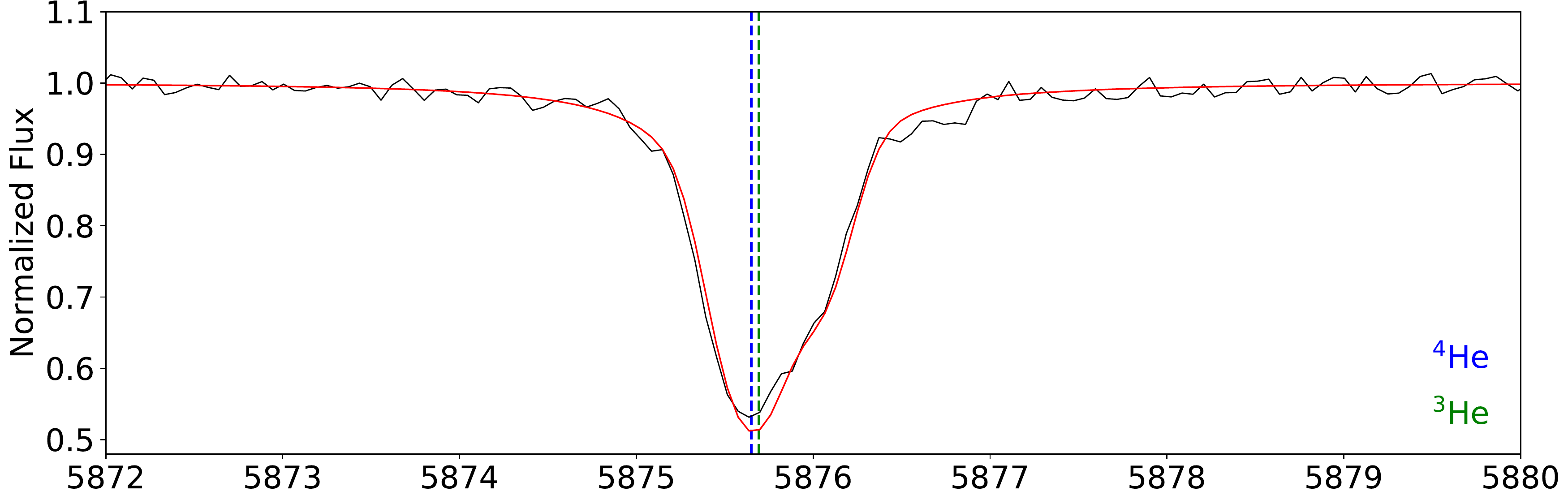}}
        \centering
        \end{minipage}\hfill
\begin{minipage}[b]{0.5\linewidth}      
\resizebox{\hsize}{!}{\includegraphics{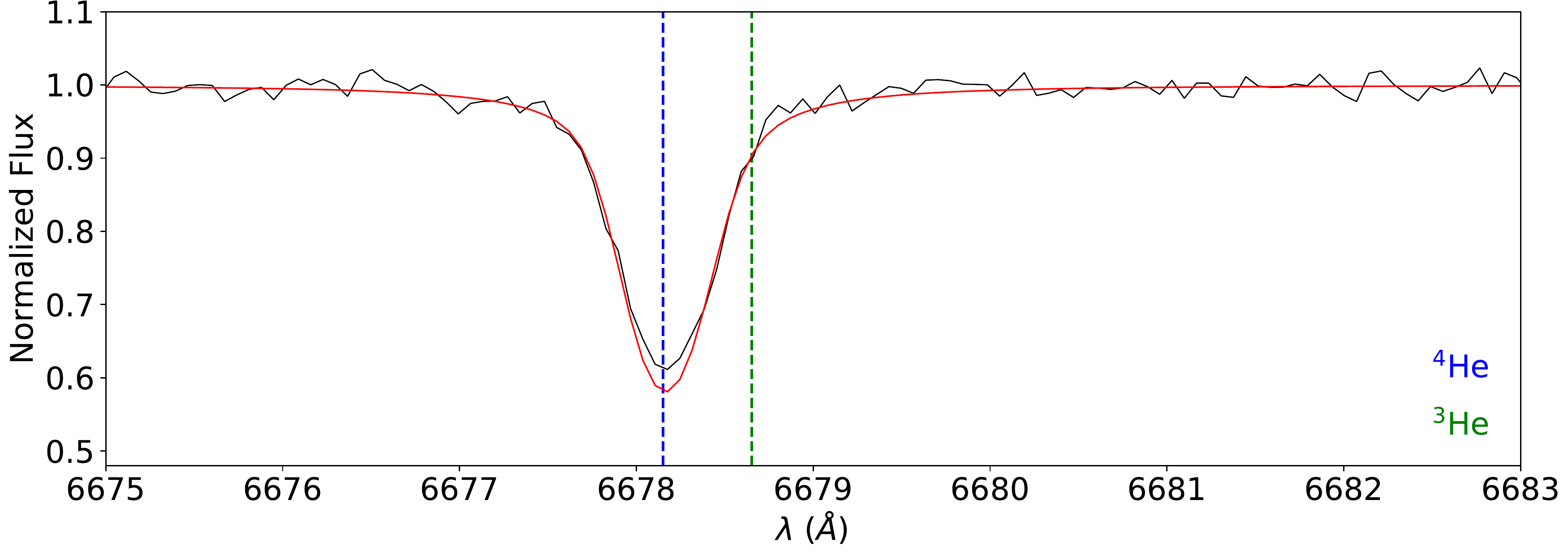}}
        \centering
        \end{minipage}\hfill
\begin{minipage}[b]{0.5\linewidth}      
\resizebox{\hsize}{!}{\includegraphics{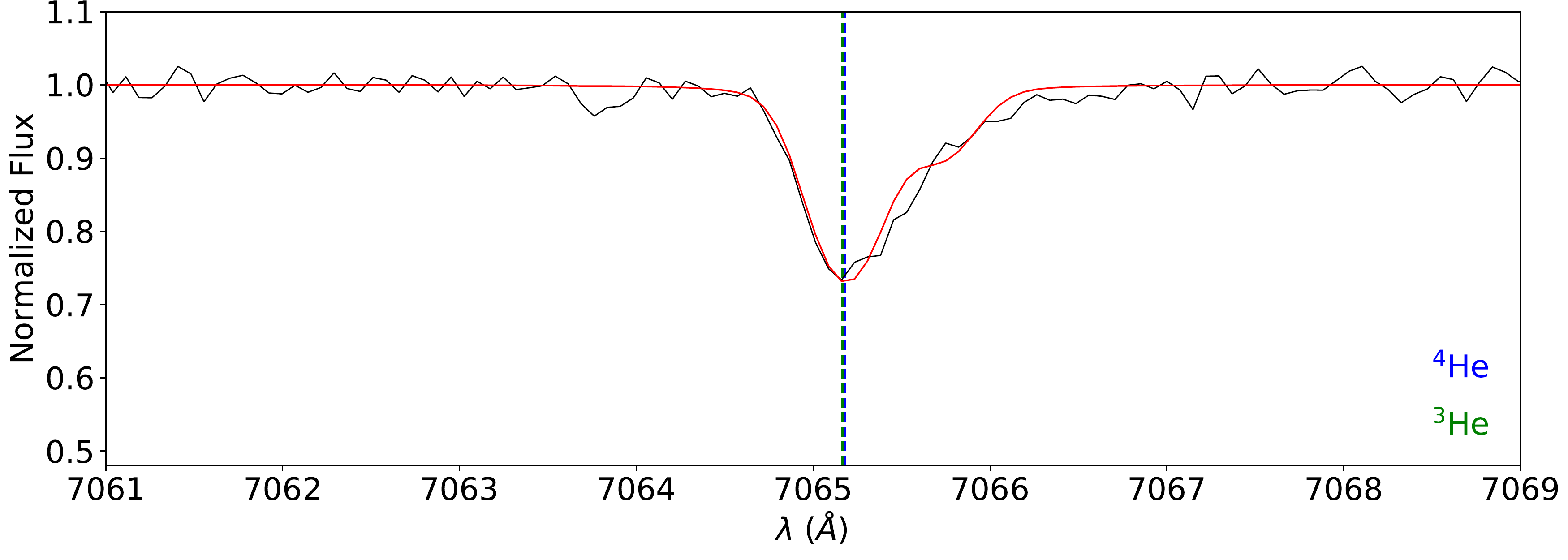}}
        \centering
        \end{minipage}\hfill    
\caption{Selected $\ion{He}{i}$ helium lines in the FEROS spectrum of the He-normal comparison star HD 4539. The observed spectrum (solid black line) and the best fit (solid red line) are shown. The dashed vertical lines display the $\isotope[3]{He}$ (green) and $\isotope[4]{He}$ (blue) component of the individual helium line according to the wavelengths listed in Table \ref{summary of isotopic shifts of neutral helium lines}.}\label{Feros HD4539 Hybrid LTE/NLTE Helium Line Fits}
\end{figure*}\noindent
\begin{figure*}
\begin{minipage}[b]{0.5\linewidth}
\resizebox{\hsize}{!}{\includegraphics{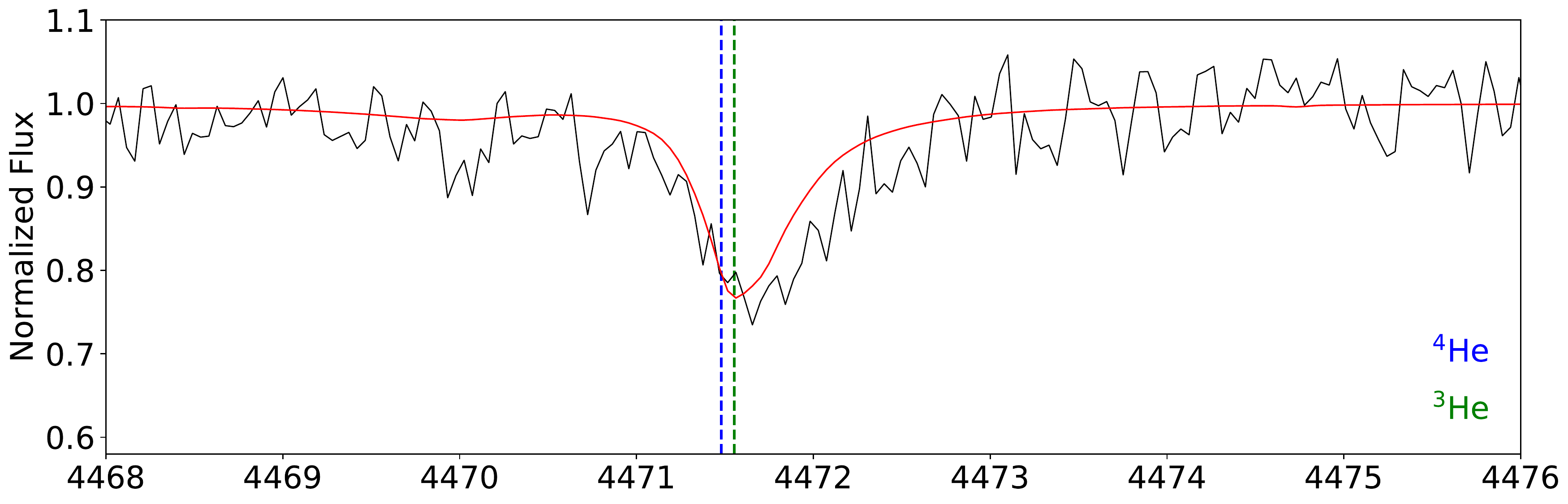}}
\centering
        \end{minipage}\hfill
\begin{minipage}[b]{0.5\linewidth}
\resizebox{\hsize}{!}{\includegraphics{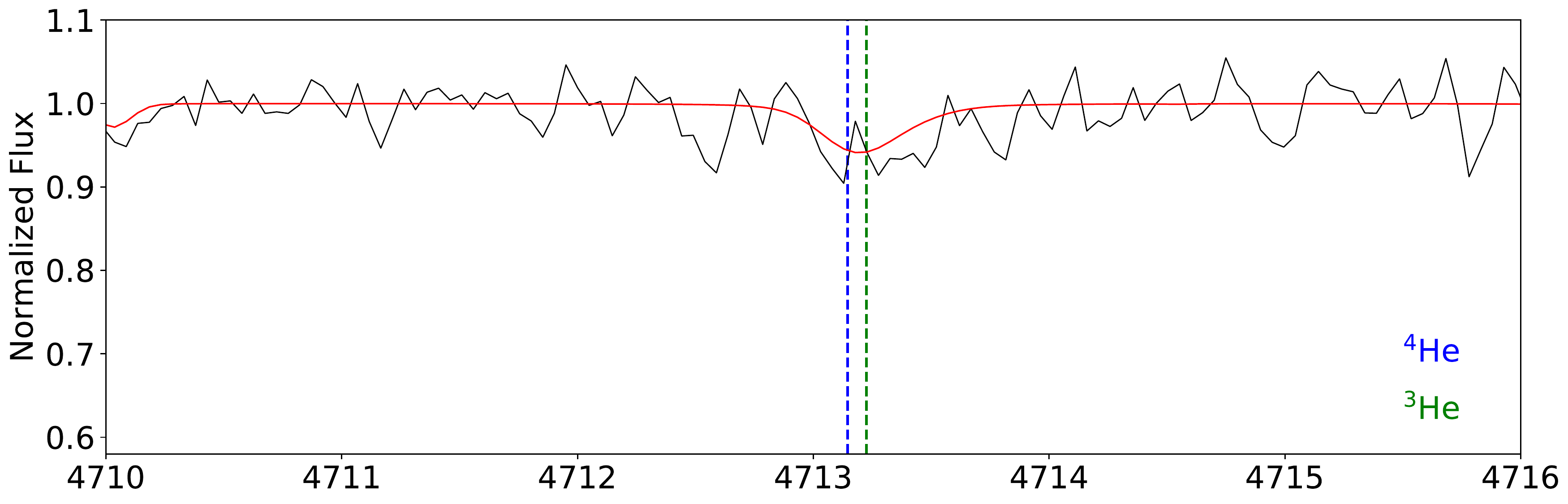}}
        \centering
        \end{minipage}\hfill
\begin{minipage}[b]{0.5\linewidth}      
\resizebox{\hsize}{!}{\includegraphics{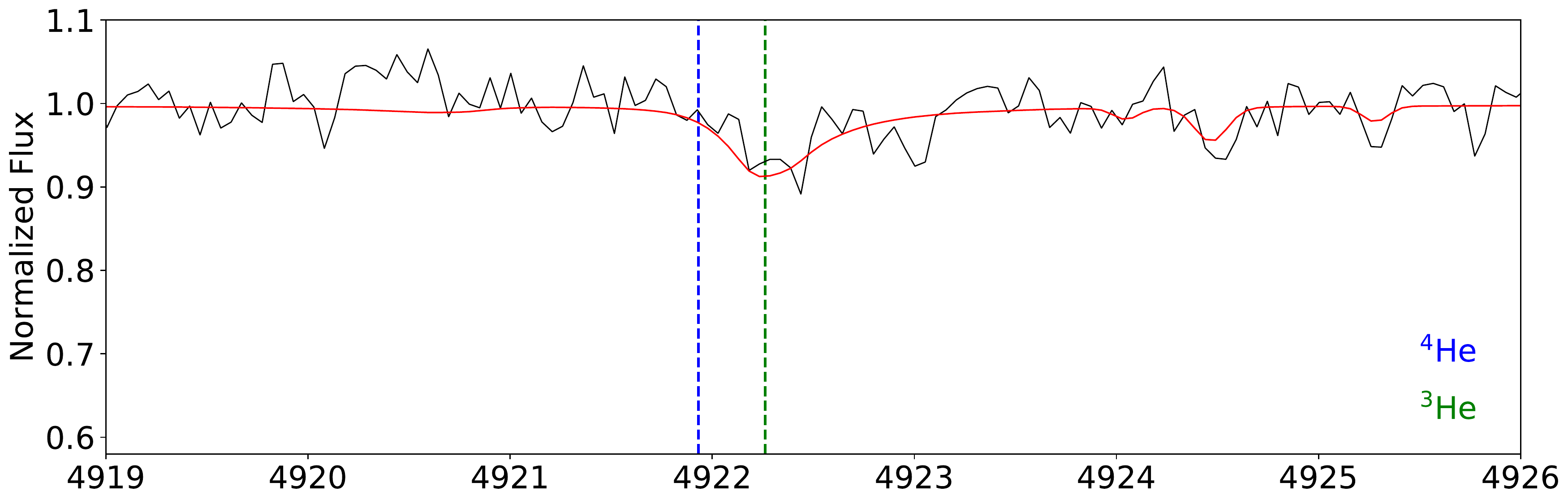}}
        \centering
        \end{minipage}\hfill
\begin{minipage}[b]{0.5\linewidth}      
\resizebox{\hsize}{!}{\includegraphics{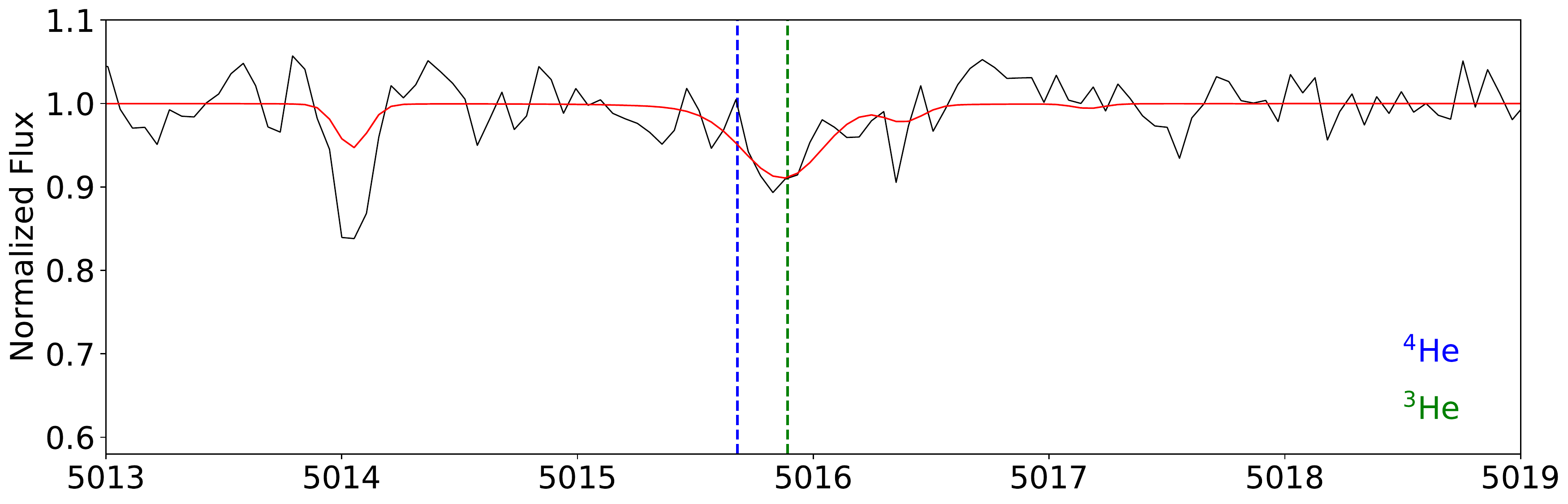}}
        \centering
        \end{minipage}\hfill
\begin{minipage}[b]{0.5\linewidth}      
\resizebox{\hsize}{!}{\includegraphics{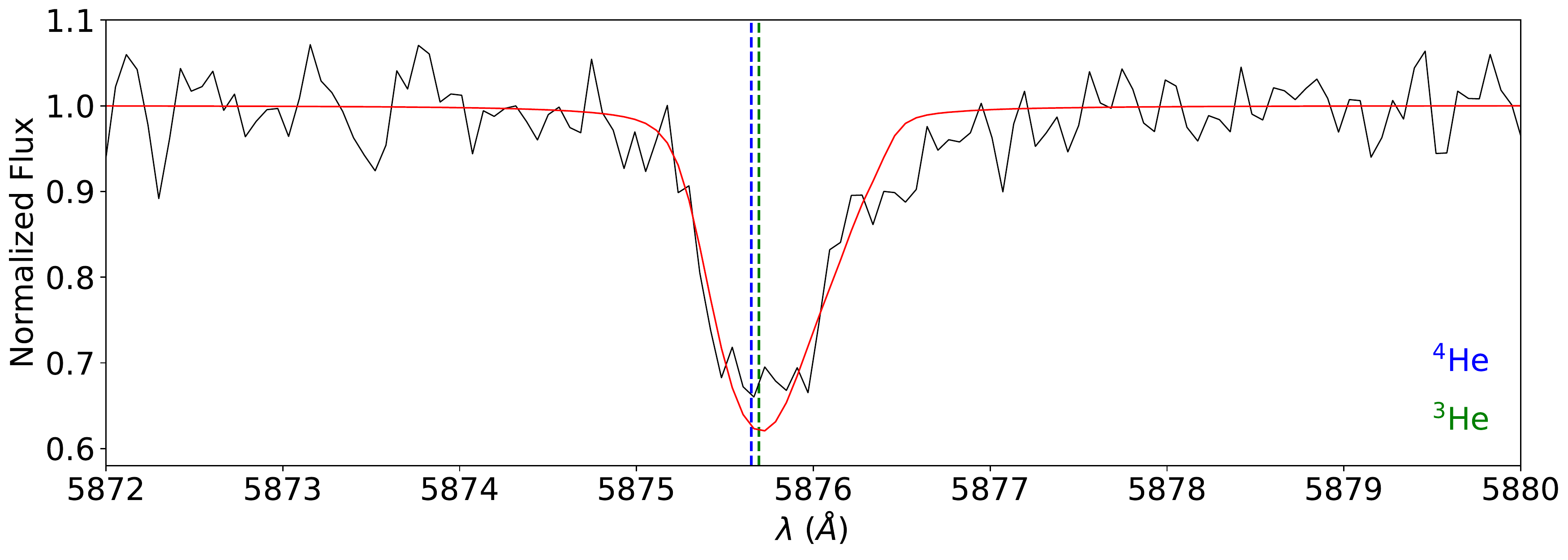}}
        \centering
        \end{minipage}\hfill
\begin{minipage}[b]{0.5\linewidth}      
\resizebox{\hsize}{!}{\includegraphics{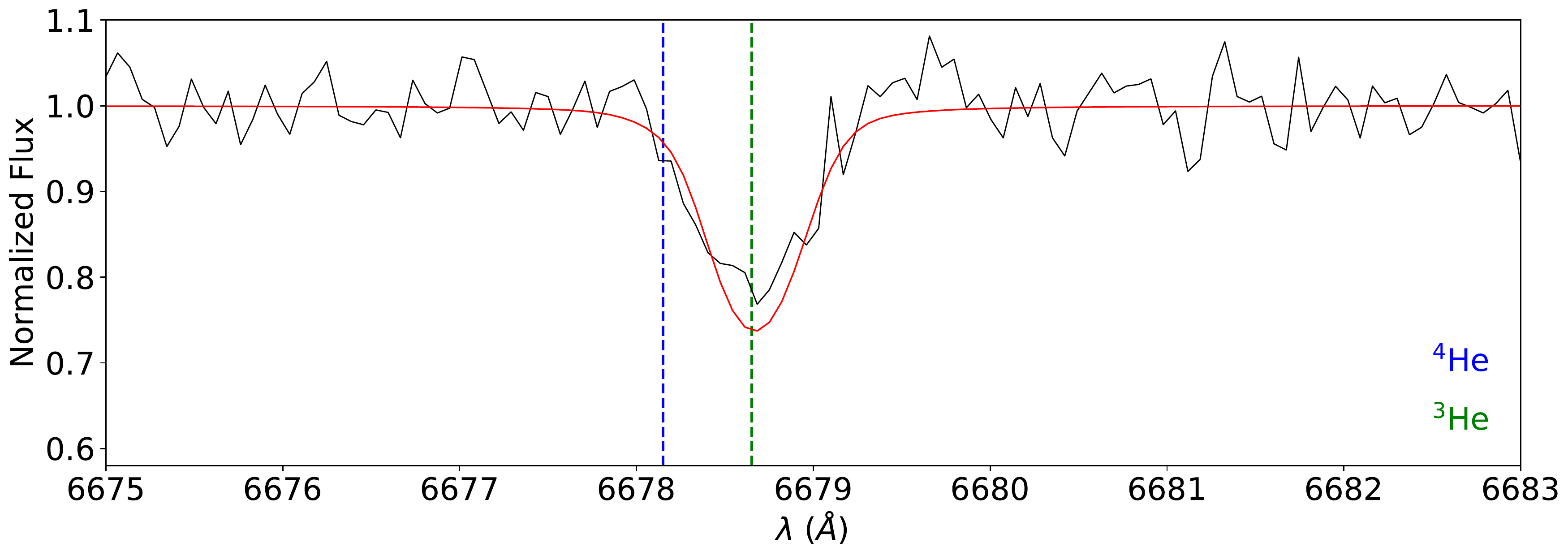}}
        \centering
        \end{minipage}\hfill
\caption{Same as Fig. \ref{Feros HD4539 Hybrid LTE/NLTE Helium Line Fits}, but for the FEROS spectrum of the $\isotope[3]{He}$ star EC 03263-6403.}\label{Feros EC03263M6403 Hybrid LTE/NLTE Helium Line Fits}
\end{figure*}\noindent

\begin{figure*}
\begin{minipage}[b]{0.5\linewidth}
\resizebox{\hsize}{!}{\includegraphics{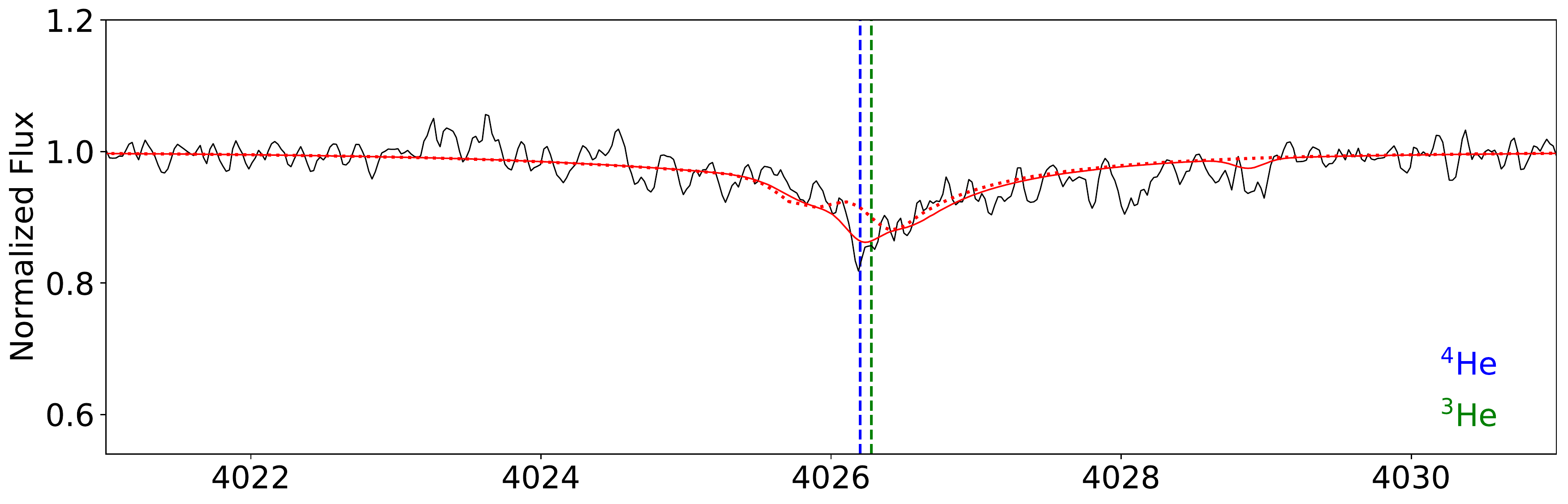}}
\centering
        \end{minipage}\hfill
\begin{minipage}[b]{0.5\linewidth}
\resizebox{\hsize}{!}{\includegraphics{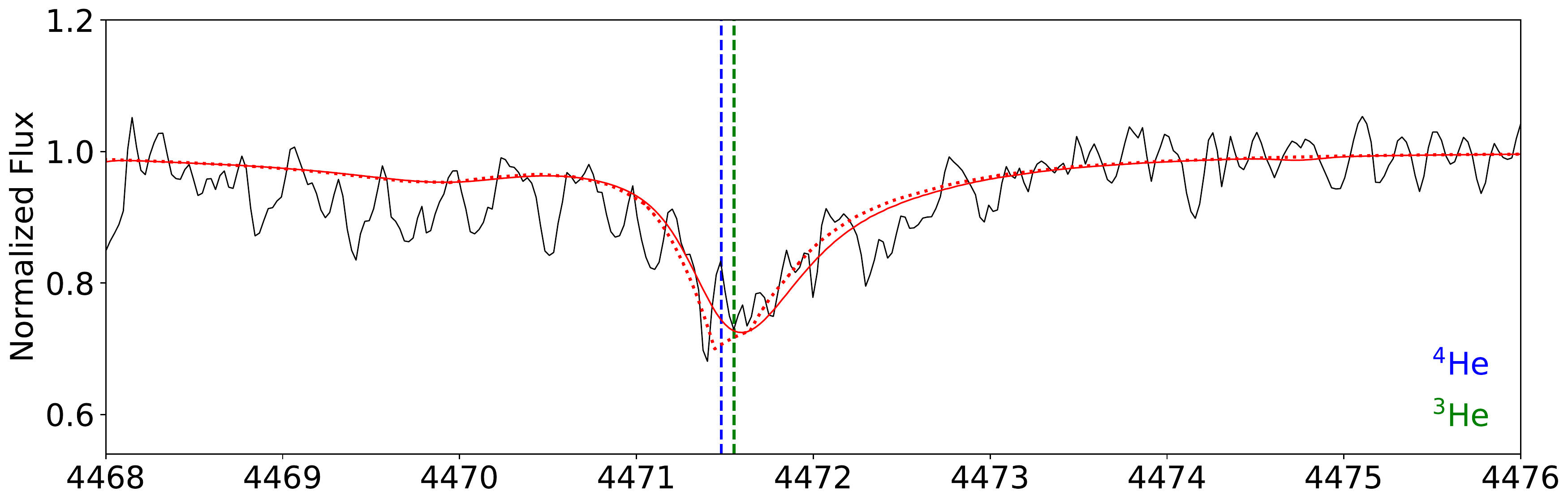}}
        \centering
        \end{minipage}\hfill
\begin{minipage}[b]{0.5\linewidth}      
\resizebox{\hsize}{!}{\includegraphics{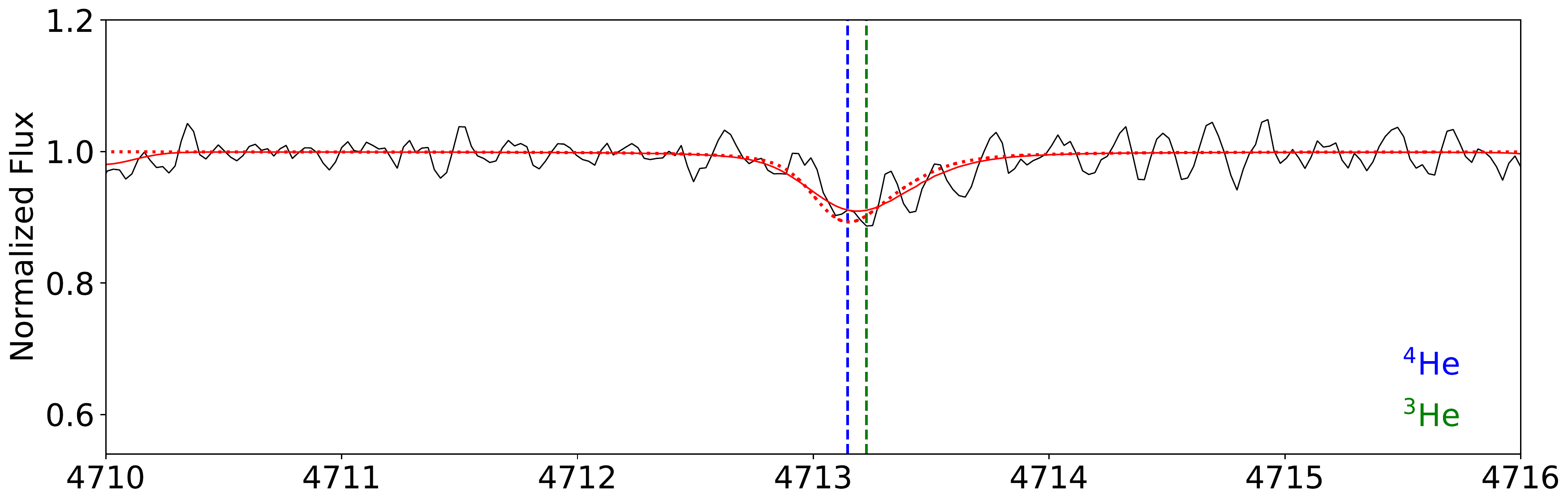}}
        \centering
        \end{minipage}\hfill
\begin{minipage}[b]{0.5\linewidth}      
\resizebox{\hsize}{!}{\includegraphics{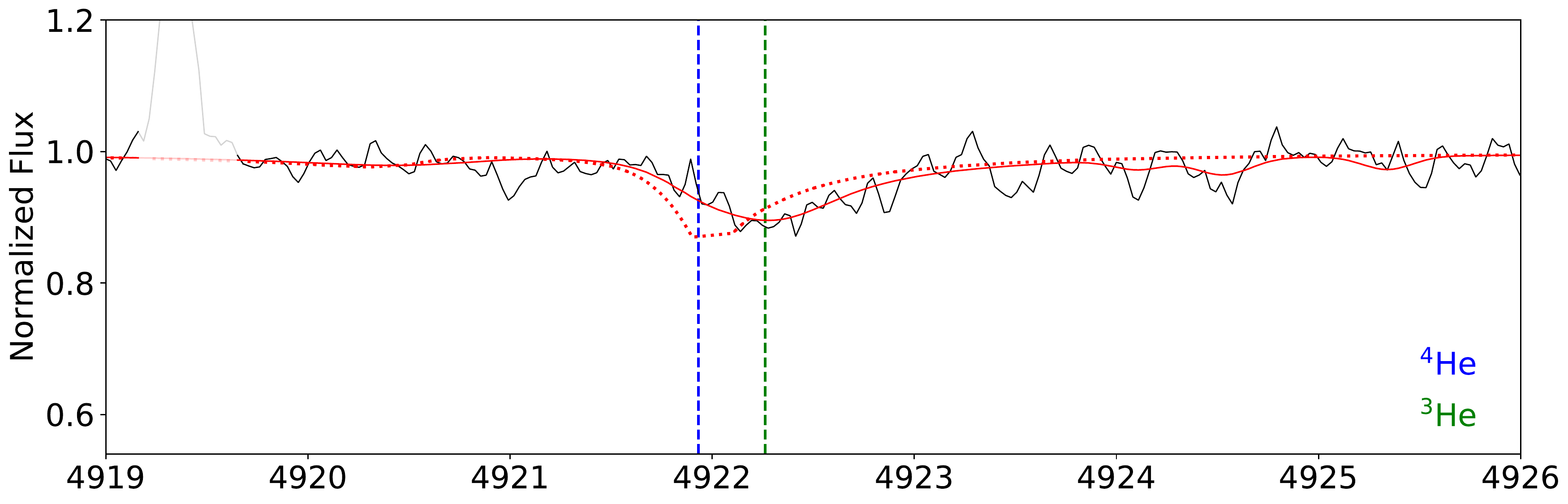}}
        \centering
        \end{minipage}\hfill
\begin{minipage}[b]{0.5\linewidth}      
\resizebox{\hsize}{!}{\includegraphics{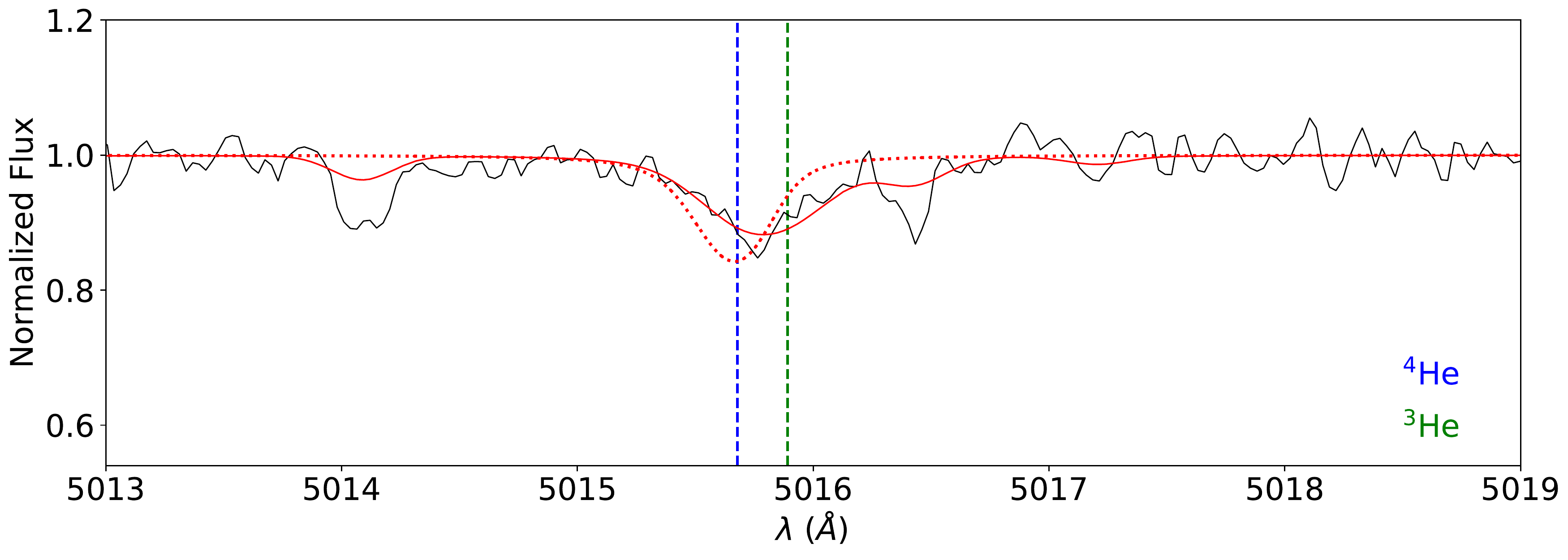}}
        \centering
        \end{minipage}\hfill    
\begin{minipage}[b]{0.5\linewidth}      
\resizebox{\hsize}{!}{\includegraphics{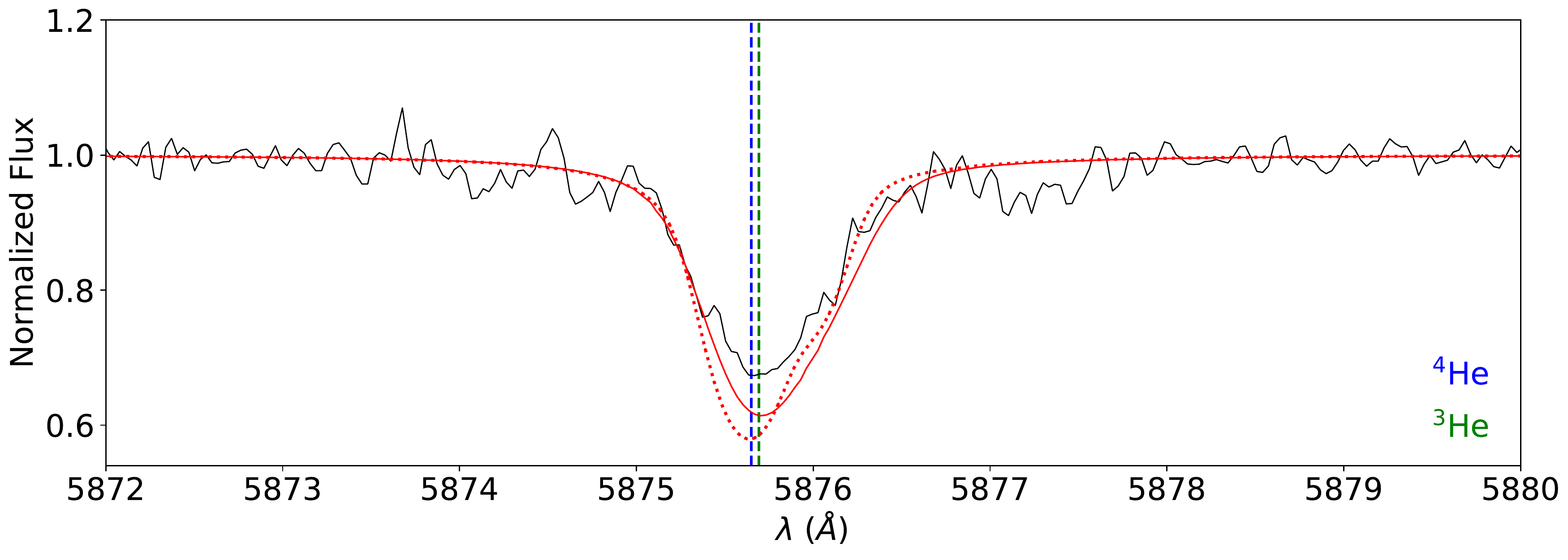}}
        \centering
        \end{minipage}\hfill    
\caption{Same as Fig. \ref{Feros HD4539 Hybrid LTE/NLTE Helium Line Fits}, but for the UVES spectrum of the formerly unclassified $\isotope[3]{He}$ star HE 1047-0436. The actual fit (solid red line) together with a model spectrum for the same atmospheric parameters but no $\isotope[3]{He}$ (dotted red line) is shown. No metals were synthesized for the latter.}\label{UVES HE1047M0436 Hybrid LTE/NLTE Helium Line Fits}
\end{figure*}\noindent
Using the calculated model spectra presented in Sect. \ref{Methods}, we investigated selected $\ion{He}{i}$ lines in the optical spectral range for all program stars. The selection criterion for each line under investigation was its respective strength, which obviously depends on the helium abundance. In consequence, we did not use the same $\ion{He}{i}$ lines for the abundance analysis of each individual star. The analysis focused on a detailed synthesis of the composite helium line profiles in order to derive both isotopic abundances ($\isotope[4]{He}$ and $\isotope[3]{He}$) and the abundance ratios ($\isotope[4]{He}$/$\isotope[3]{He}$). Unfortunately, $\ion{He}{i}$ \SI{7281}{\angstrom} either was not covered in the spectral range of the Echelle spectrographs used (FOCES, UVES) or was truncated due to the different diffraction orders (HRS). If covered (FEROS), $\ion{He}{i}$ \SI{7281}{\angstrom} was too weak to be useful. In order to study the $\isotope[3]{He}$ anomaly in all program stars, we therefore had to rely on the strong $\ion{He}{i}$ \SI{6678}{\angstrom} and $\ion{He}{i}$ \SI{4922}{\angstrom} lines as the most important signatures for $\isotope[4]{He}$/$\isotope[3]{He}$ in the optical spectral range.\\    
The derived $\isotope[4]{He}$ and $\isotope[3]{He}$ abundances and the isotopic abundance ratios are listed in Table \ref{briefly summarized 4He/3He table hybrid LTE/NLTE 1}. As expected, the changes due to the new broadening tables for $\ion{He}{i}$ are only minor (compare Table \ref{briefly summarized 4He/3He table hybrid LTE/NLTE 1} with Table 3 in \citealt{Schneider_2017}). There is also no systematic trend visible between the hybrid LTE/NLTE approach used here and LTE literature values regarding helium abundances.\\
Figure \ref{NLTE_abundances_plot.pdf} summarizes the isotopic helium abundances in a \teff$-\log{n(\isotope[3]{He})}$ and \teff$-\log{n(\isotope[4]{He})}$ diagram for all program stars. While all $\isotope[4]{He}$ abundances are clearly subsolar (solar $\isotope[4]{He}$ abundance: $\log{n(\text{\isotope[4]{He}})}=-1.11$;  \citealt{Asplund_2009}), the $\isotope[3]{He}$ abundances are strongly overabundant compared to the solar $\isotope[3]{He}$ value of $\log{n(\text{\isotope[3]{He}})}=-4.89$ \citep{Asplund_2009}.\\
In the following, the results on isotopic abundances and abundance ratios are presented for the He-normal comparison stars (Sect. \ref{The He-Normal sdB Stars HD 4539 and CD-35 15910}), five known $\isotope[3]{He}$ sdBs (Sect. \ref{3He sdB Stars with Known 3He Anomaly}), and two from the ESO SPY survey (Sect. \ref{Two 3He sdB Stars from the ESO SPY Project}). We find anomalous helium line profiles for all $\isotope[3]{He}$ BHB and three $\isotope[3]{He}$ sdB stars, which are discussed in Sect. \ref{Helium Line Profile Anomalies}. Last, a sensitivity study in order to verify the results is provided in Sect. \ref{Sensitivity study}.

\subsection{He-normal subdwarf B stars HD 4539 and CD-35$^\circ$ 15910}\label{The He-Normal sdB Stars HD 4539 and CD-35 15910}
\begin{figure*}
\begin{minipage}[b]{0.5\linewidth}
\resizebox{\hsize}{!}{\includegraphics{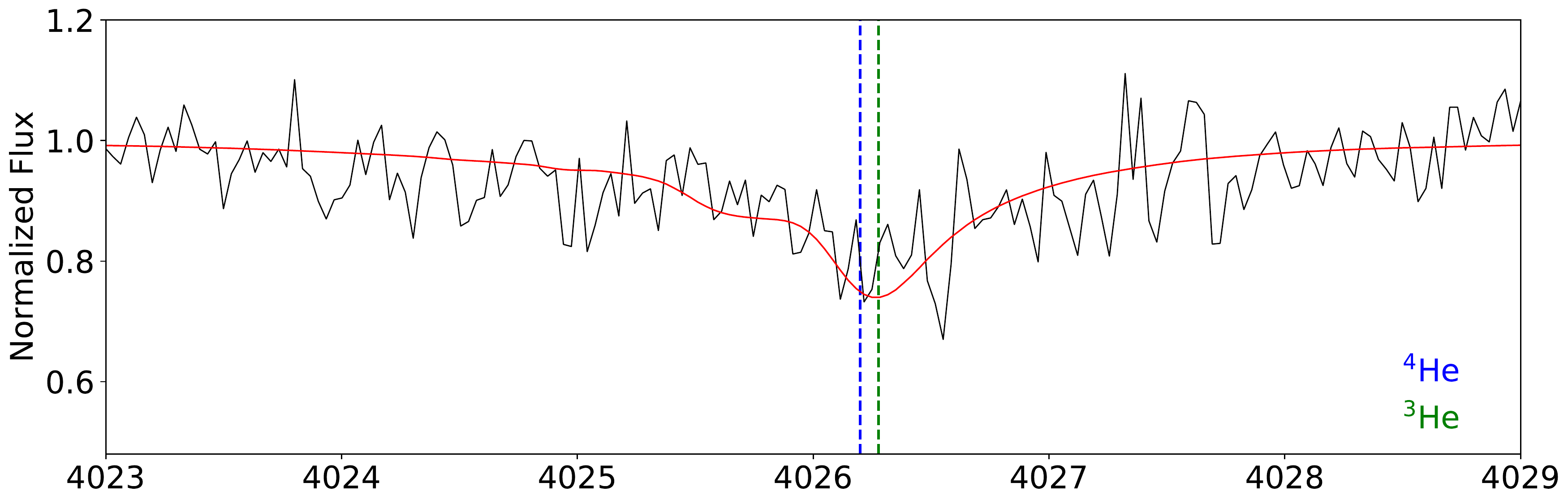}}
\centering
        \end{minipage}\hfill
\begin{minipage}[b]{0.5\linewidth}
\resizebox{\hsize}{!}{\includegraphics{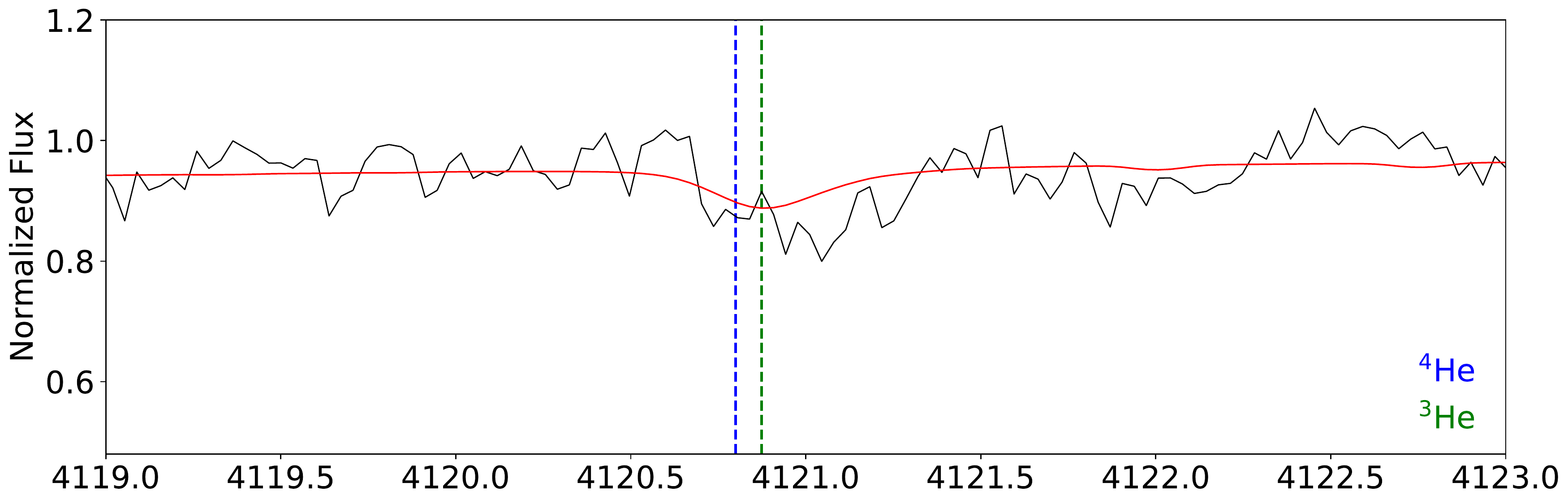}}
\centering
        \end{minipage}\hfill    
\begin{minipage}[b]{0.5\linewidth}
\resizebox{\hsize}{!}{\includegraphics{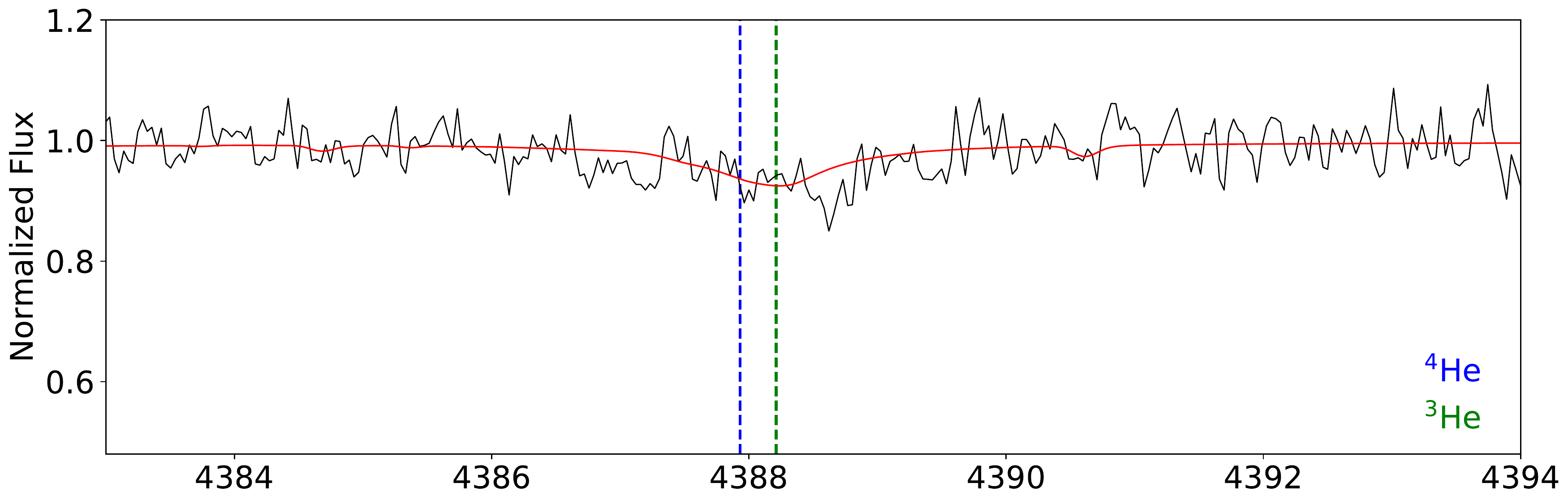}}
        \centering
        \end{minipage}\hfill
\begin{minipage}[b]{0.5\linewidth}      
\resizebox{\hsize}{!}{\includegraphics{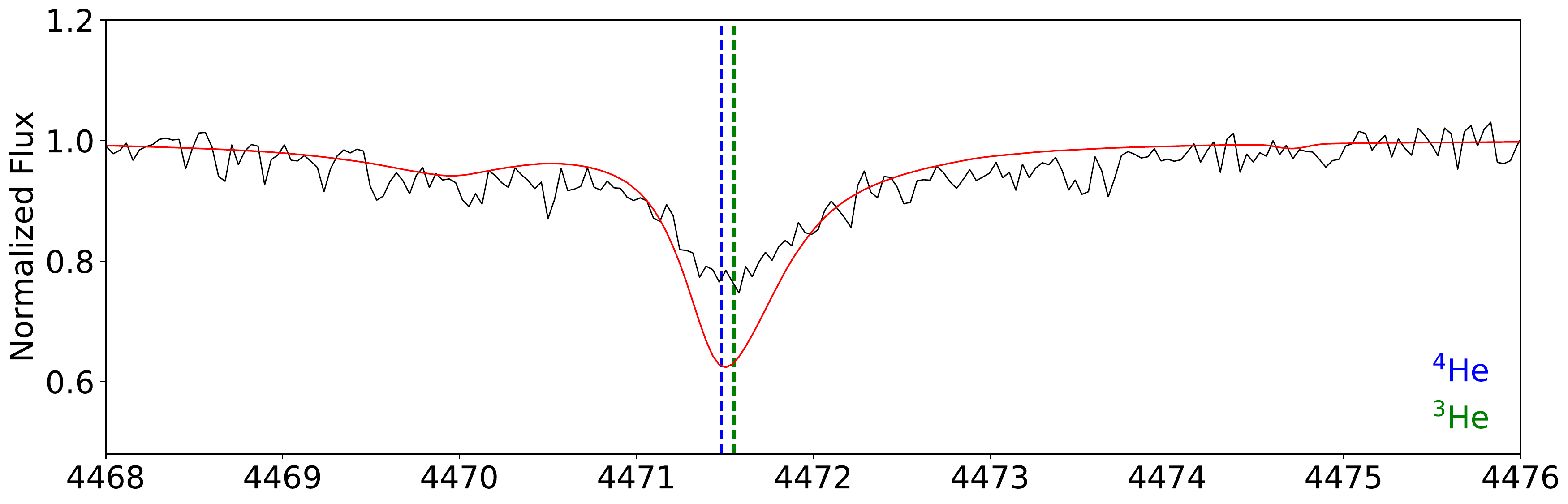}}
        \centering
        \end{minipage}\hfill
\begin{minipage}[b]{0.5\linewidth}      
\resizebox{\hsize}{!}{\includegraphics{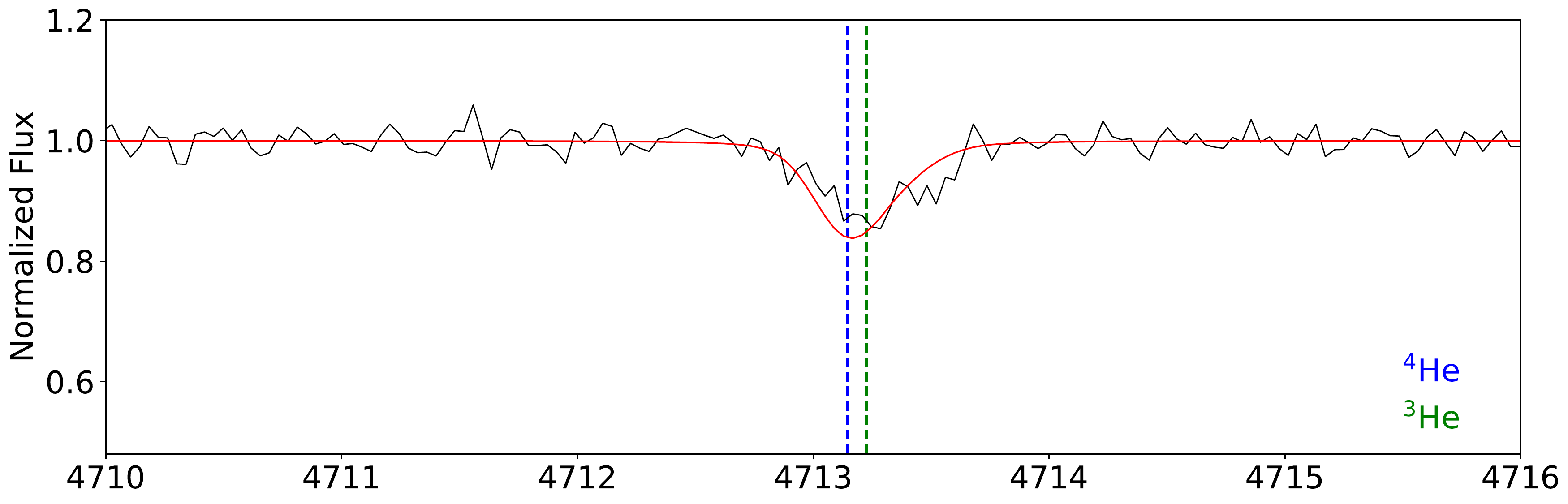}}
        \centering
        \end{minipage}\hfill
\begin{minipage}[b]{0.5\linewidth}      
\resizebox{\hsize}{!}{\includegraphics{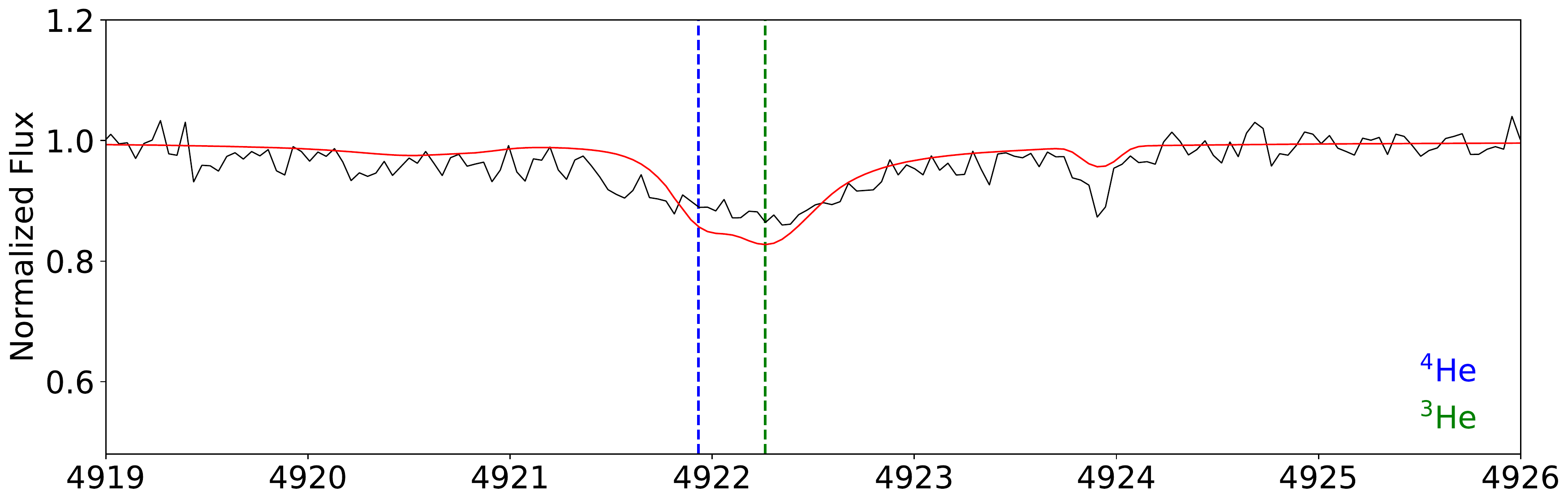}}
        \centering
        \end{minipage}\hfill
\begin{minipage}[b]{0.5\linewidth}      
\resizebox{\hsize}{!}{\includegraphics{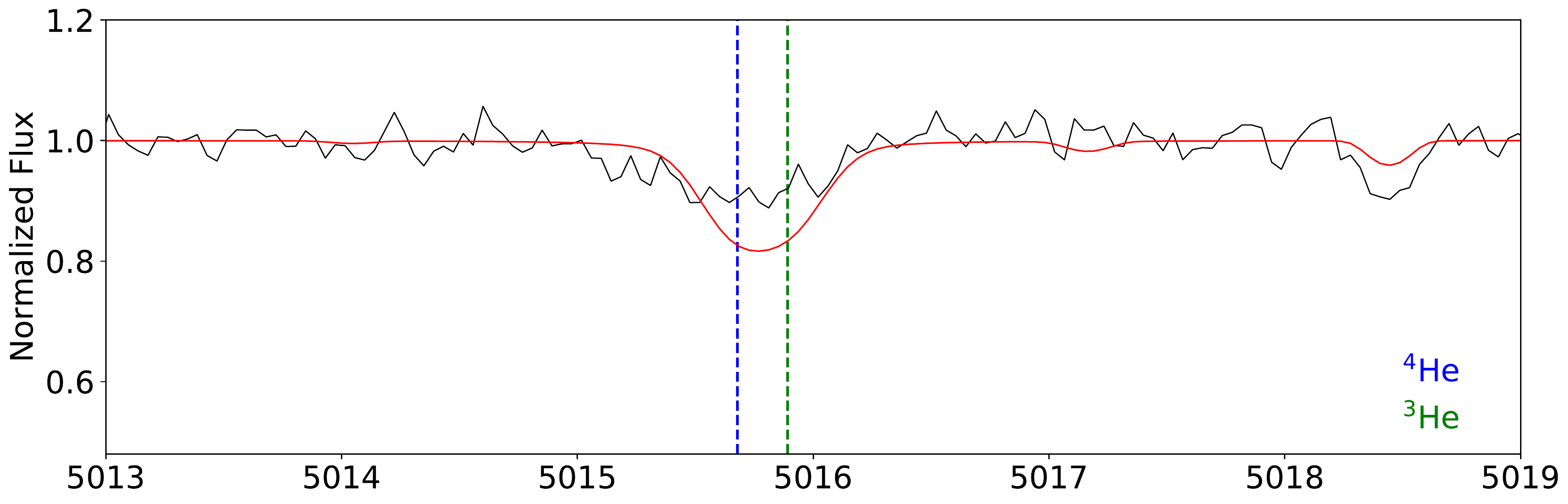}}
        \centering
        \end{minipage}\hfill
\begin{minipage}[b]{0.5\linewidth}      
\resizebox{\hsize}{!}{\includegraphics{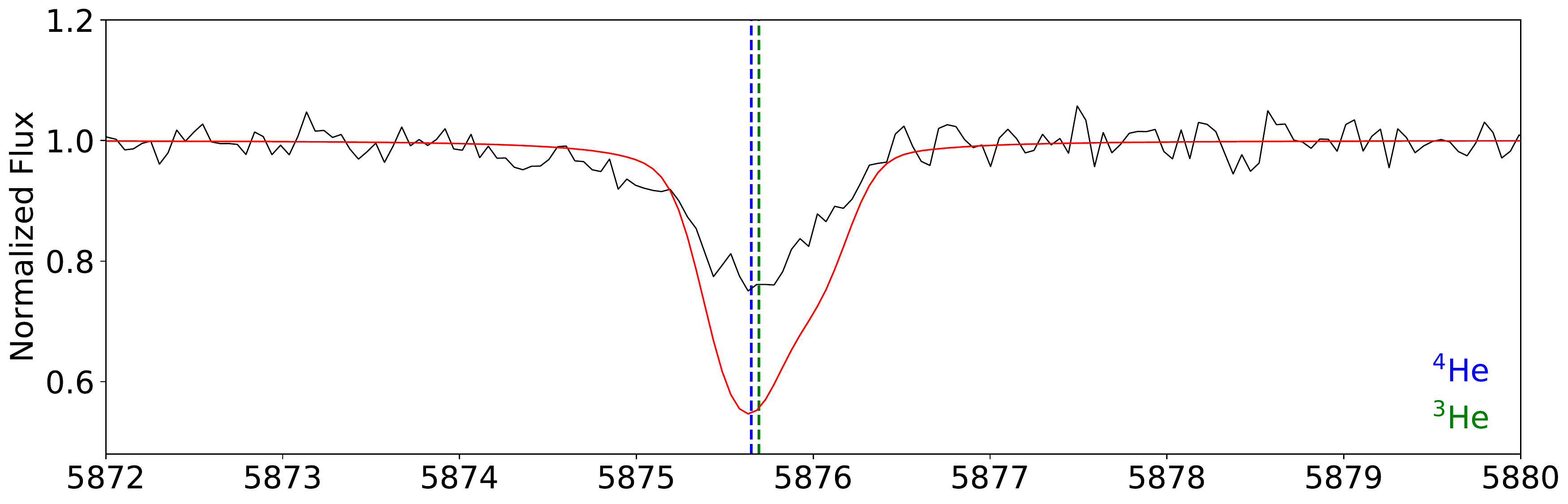}}
        \centering
        \end{minipage}\hfill
\begin{minipage}[b]{0.5\linewidth}      
\resizebox{\hsize}{!}{\includegraphics{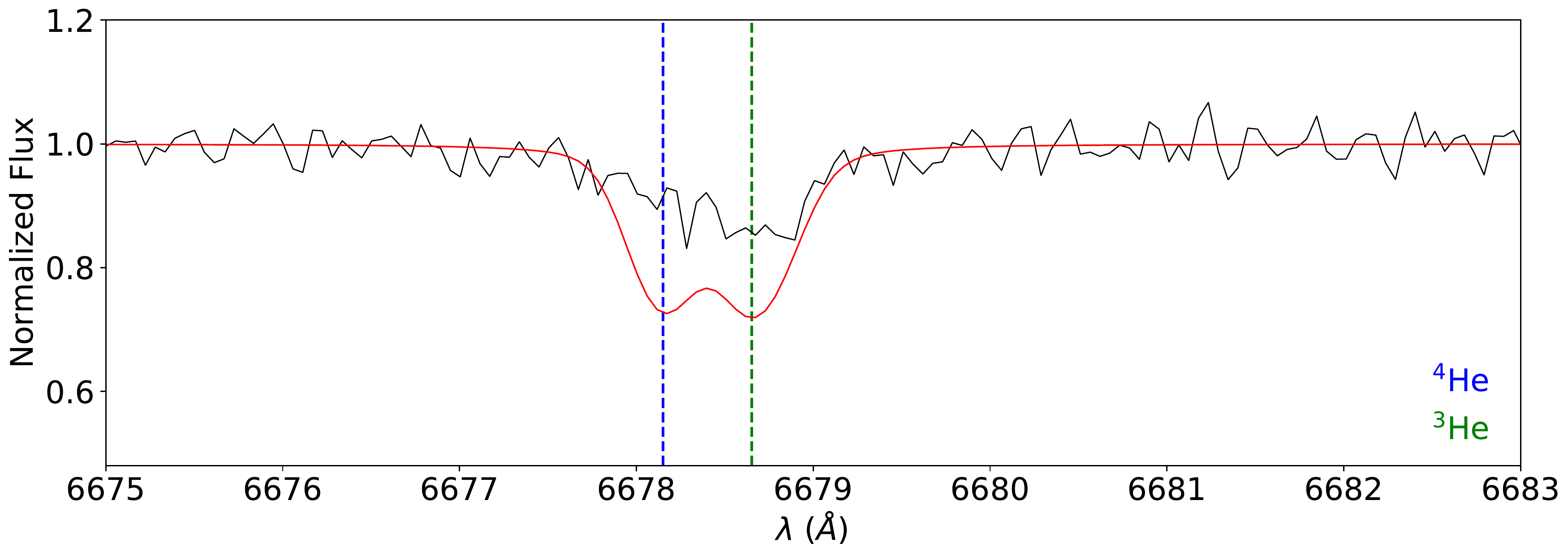}}
        \centering
        \end{minipage}\hfill
\begin{minipage}[b]{0.5\linewidth}      
\resizebox{\hsize}{!}{\includegraphics{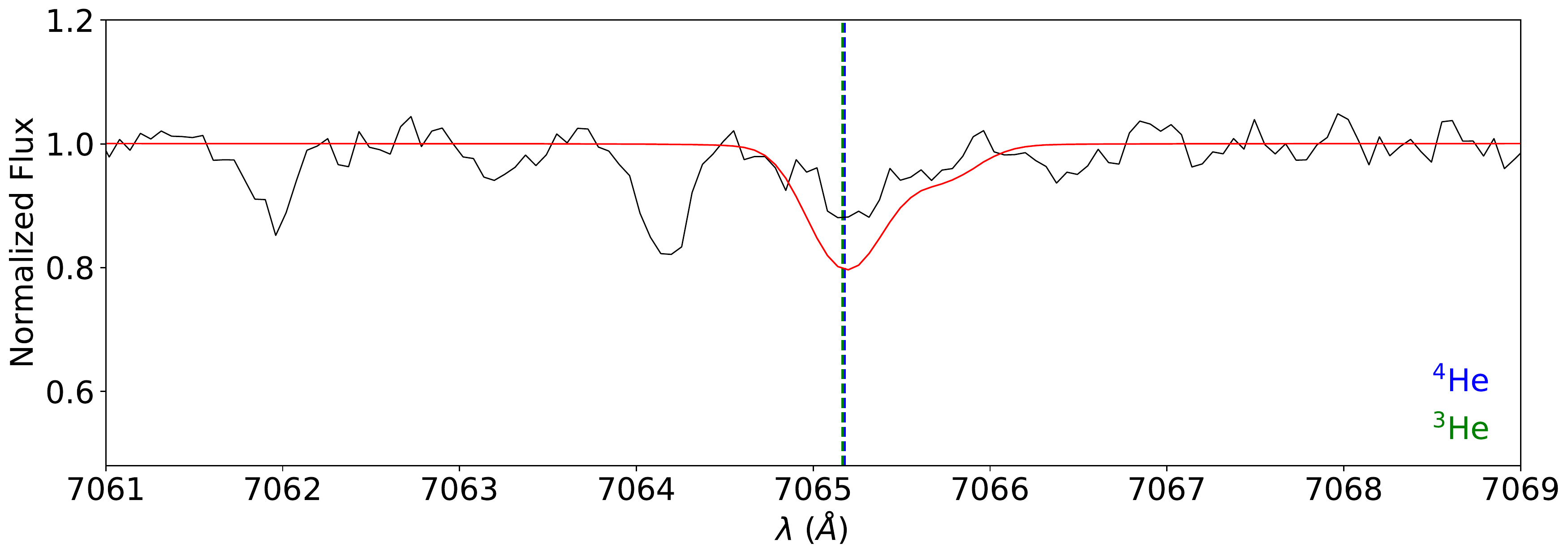}}
        \centering
        \end{minipage}\hfill                    
\caption{Same as Fig. \ref{Feros HD4539 Hybrid LTE/NLTE Helium Line Fits}, but for the HRS spectrum of the $\isotope[3]{He}$ star PHL 25. The line cores of $\ion{He}{i}$ \SI{4472}{\angstrom}, $\ion{He}{i}$ \SI{5875}{\angstrom}, $\ion{He}{i}$ \SI{6678}{\angstrom}, and $\ion{He}{i}$ \SI{7065}{\angstrom} have been excluded from the fit because of strong stratification effects and fitting problems, as are obvious from the strong mismatch (see Sect. \ref{Helium Line Profile Anomalies} for details).}\label{McDonald PHL25 Hybrid LTE/NLTE Helium Line Fits}
\end{figure*}\noindent
\begin{figure*}
\begin{minipage}[b]{0.5\linewidth}
\resizebox{\hsize}{!}{\includegraphics{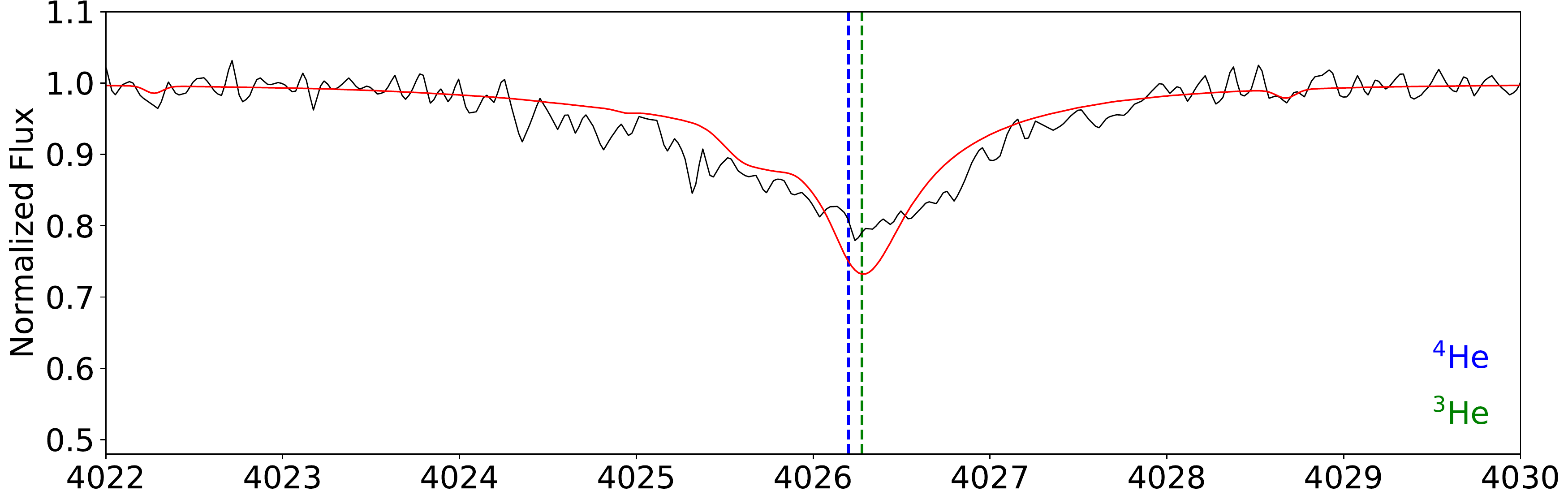}}
\centering
        \end{minipage}\hfill
\begin{minipage}[b]{0.5\linewidth}
\resizebox{\hsize}{!}{\includegraphics{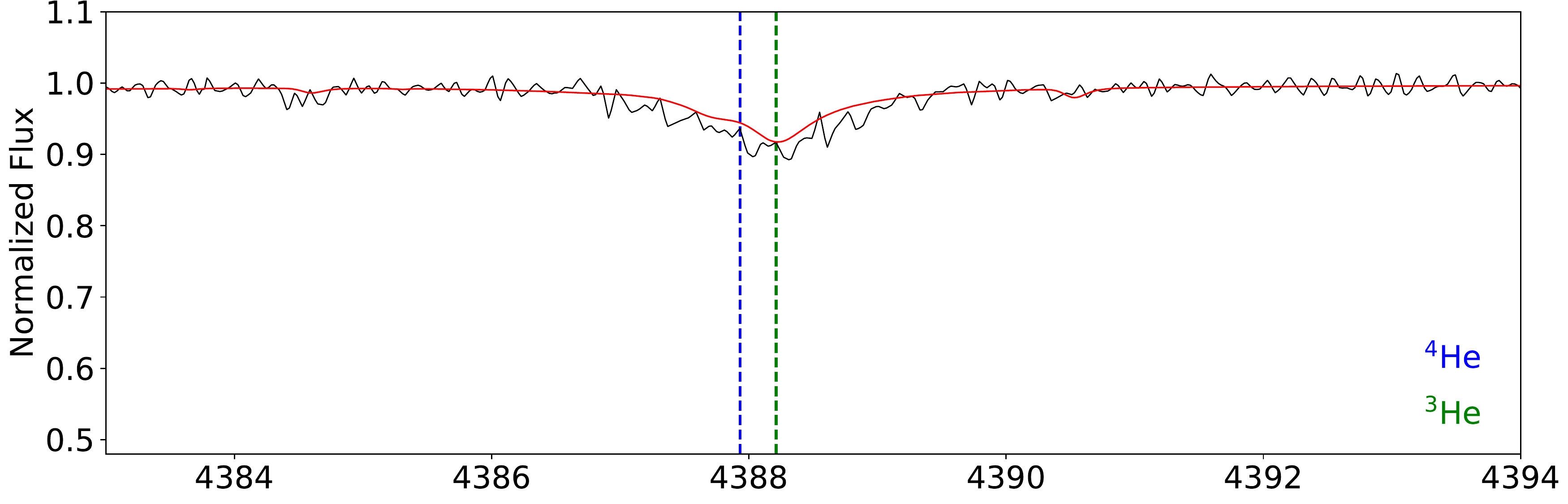}}
        \centering
        \end{minipage}\hfill
\begin{minipage}[b]{0.5\linewidth}      
\resizebox{\hsize}{!}{\includegraphics{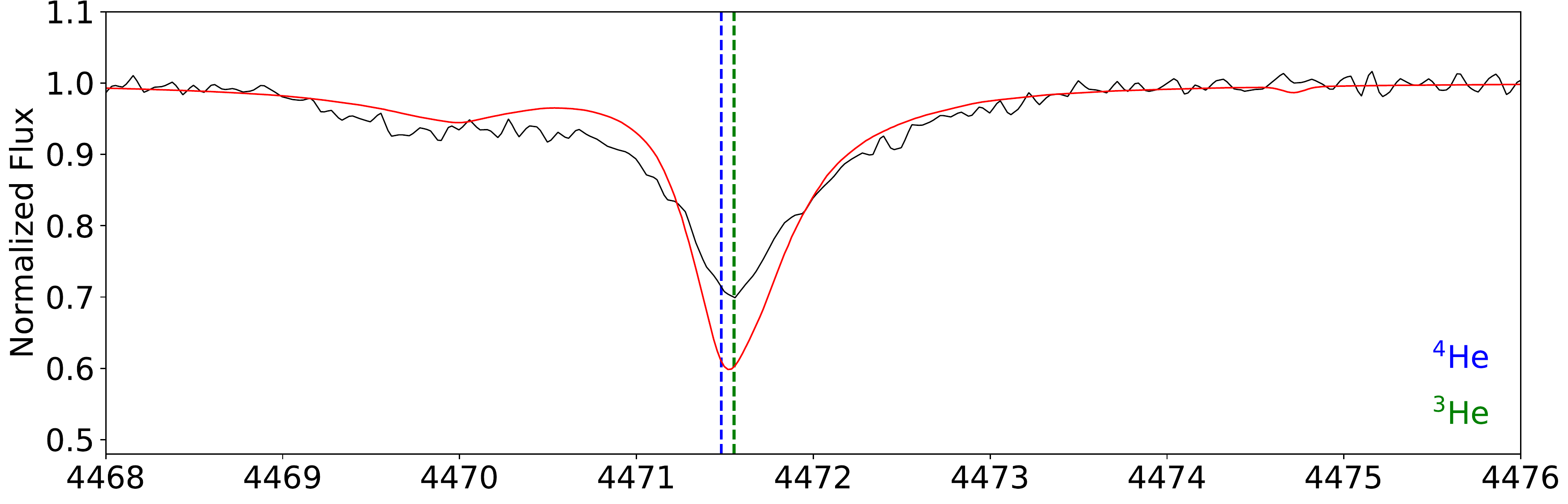}}
        \centering
        \end{minipage}\hfill
\begin{minipage}[b]{0.5\linewidth}      
\resizebox{\hsize}{!}{\includegraphics{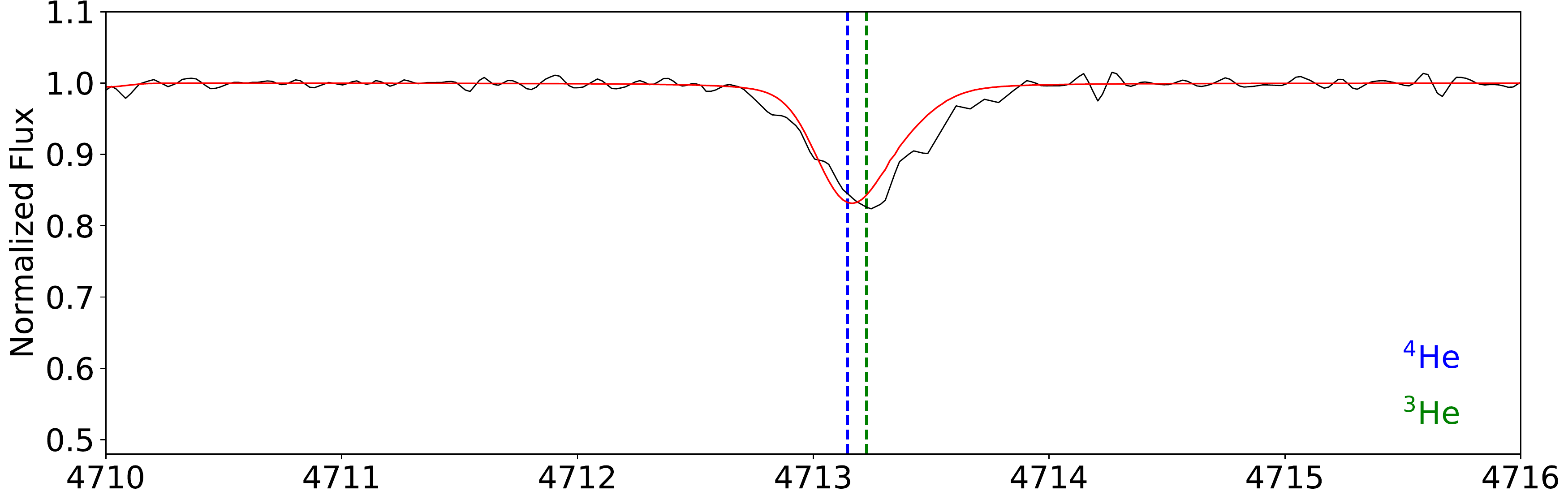}}
        \centering
        \end{minipage}\hfill
\begin{minipage}[b]{0.5\linewidth}      
\resizebox{\hsize}{!}{\includegraphics{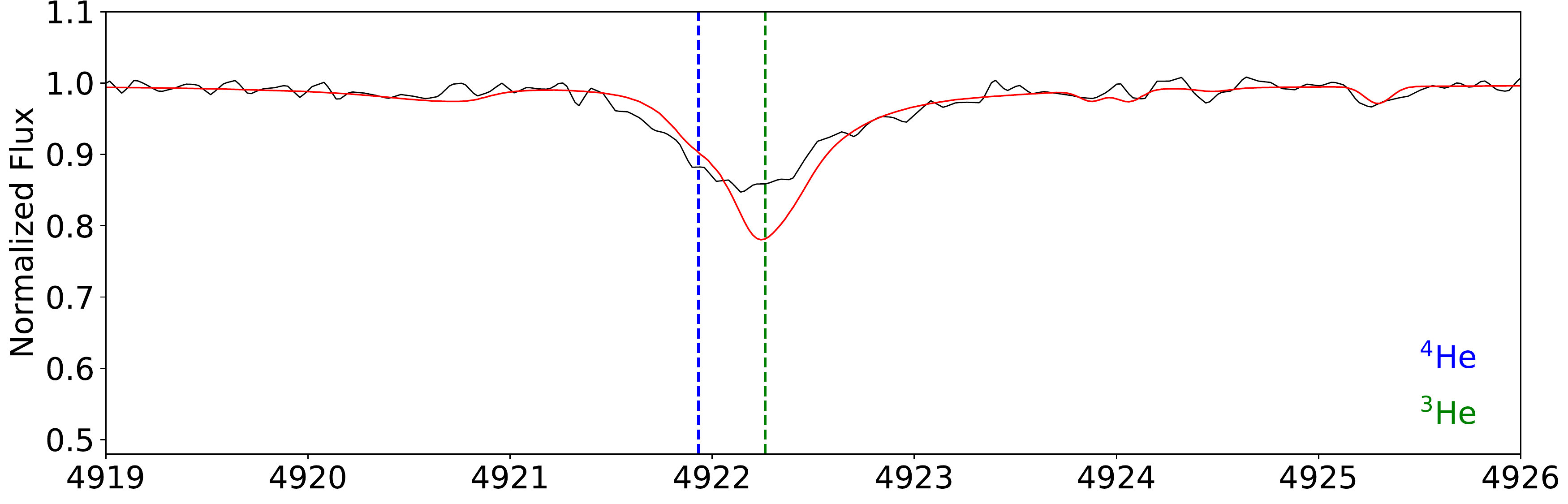}}
        \centering
        \end{minipage}\hfill
\begin{minipage}[b]{0.5\linewidth}      
\resizebox{\hsize}{!}{\includegraphics{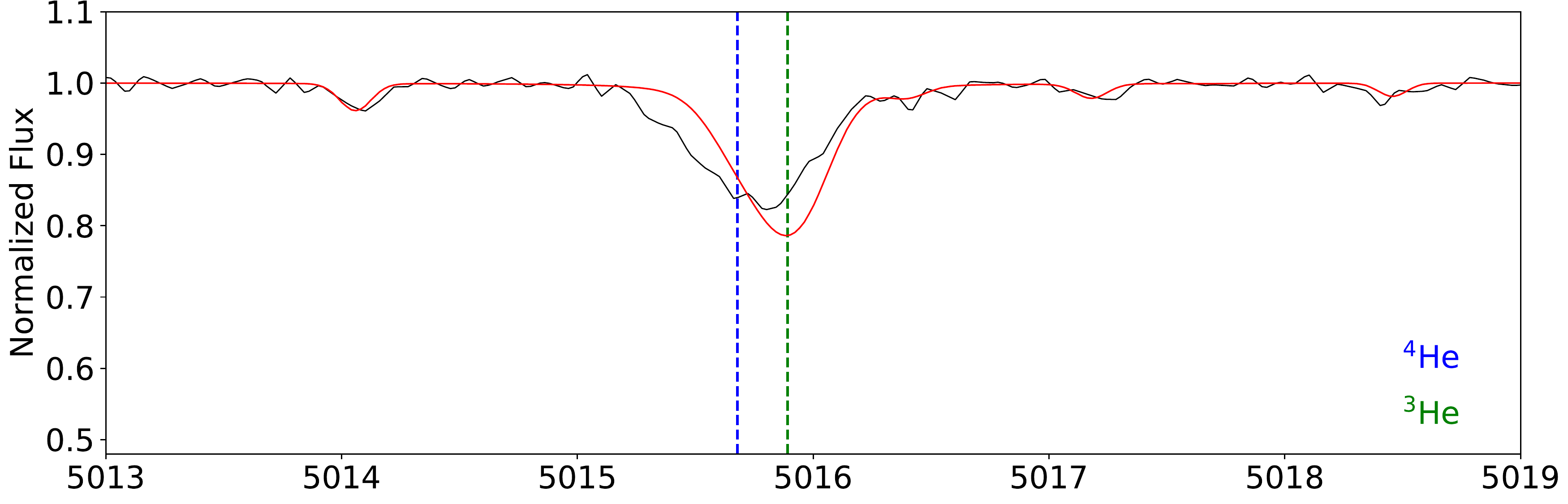}}
        \centering
        \end{minipage}\hfill
\begin{minipage}[b]{0.5\linewidth}      
\resizebox{\hsize}{!}{\includegraphics{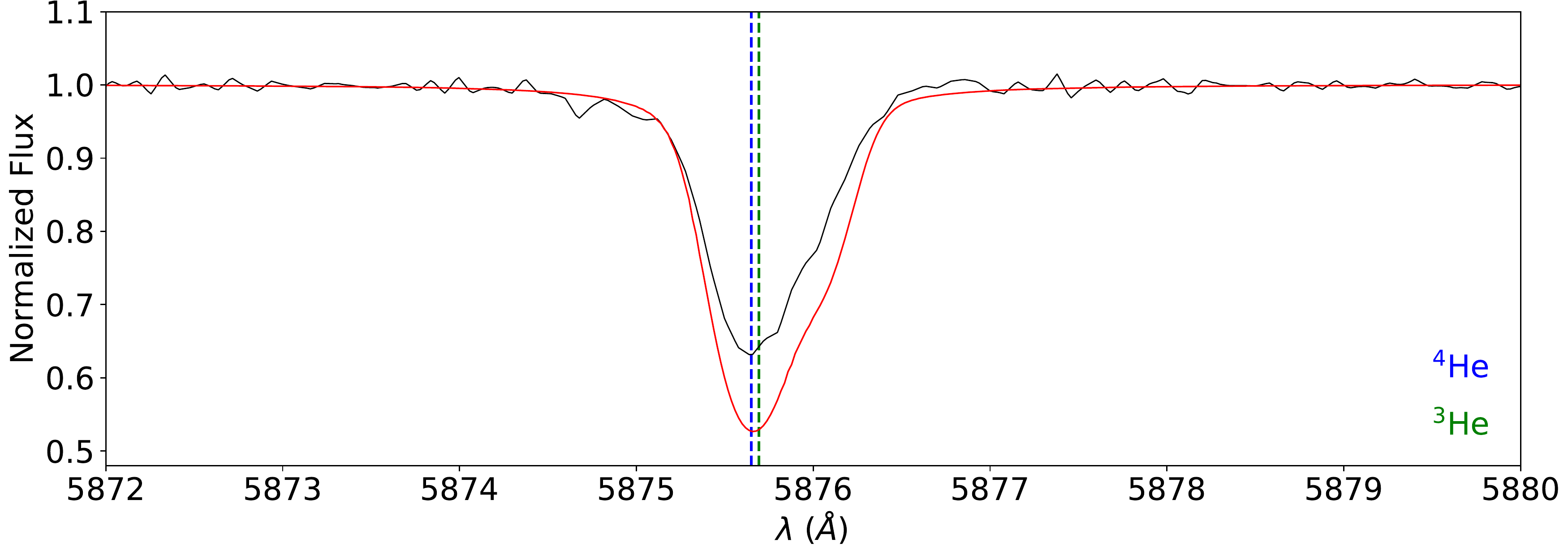}}
        \centering
        \end{minipage}\hfill
\begin{minipage}[b]{0.5\linewidth}      
\resizebox{\hsize}{!}{\includegraphics{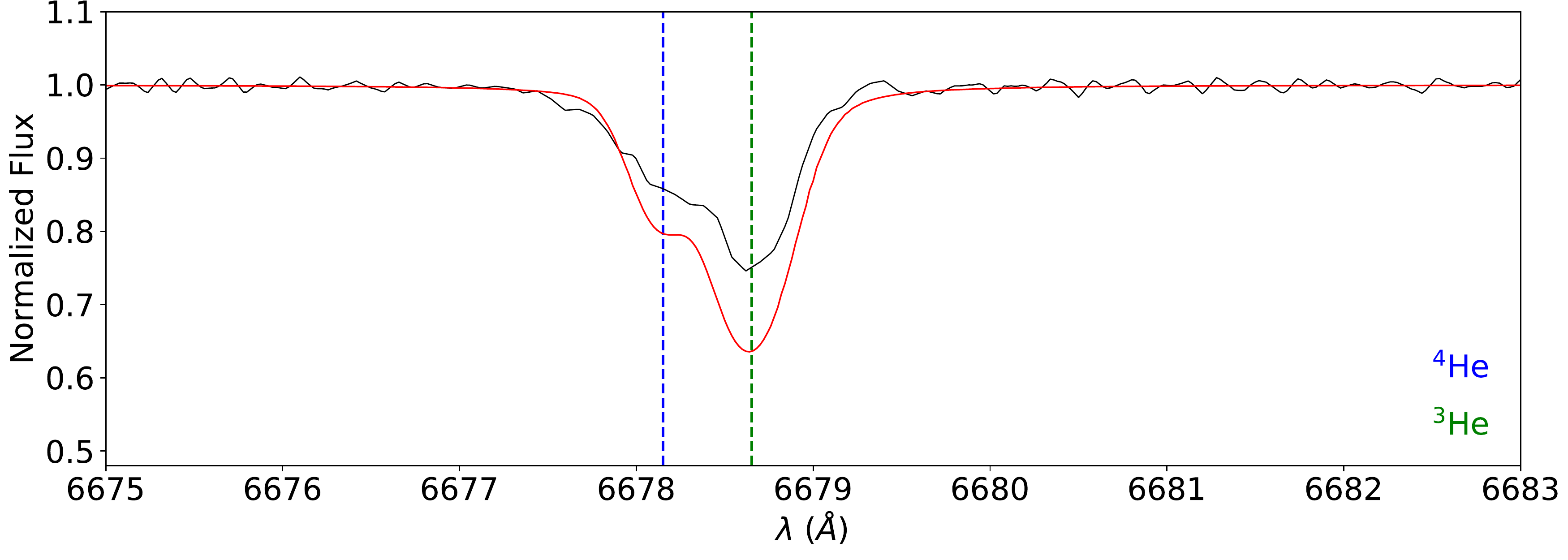}}
        \centering
        \end{minipage}\hfill            
\caption{Same as Fig. \ref{Feros HD4539 Hybrid LTE/NLTE Helium Line Fits}, but for the FOCES spectrum of the $\isotope[3]{He}$ star BD+48$^\circ$ 2721. The star shows strong helium stratification, as is obvious from the mismatch of the cores of many $\ion{He}{i}$ lines (see Sect. \ref{Helium Line Profile Anomalies} for details).}\label{Foces BDP482721 Hybrid LTE/NLTE Helium Line Fits}
\end{figure*}\noindent
The fit of all suitable $\isotope[4]{He}$ absorption lines of HD 4539 and CD-35$^\circ$ 15910 is precise. As an example, Fig. \ref{Feros HD4539 Hybrid LTE/NLTE Helium Line Fits} displays the line fits for the FEROS spectrum of HD 4539. Only small deviations between fit and observed spectrum are visible in the line wings of $\ion{He}{i}$ \SI{4472}{\angstrom} and $\ion{He}{i}$ \SI{4922}{\angstrom}, which are related to the respective forbidden component, and the predicted line cores of $\ion{He}{i}$ \SI{5875}{\angstrom} and $\ion{He}{i}$ \SI{6678}{\angstrom}, for instance, are slightly too strong. Neither comparison sdB shows any traces of $\isotope[3]{He}$ at all (see also Table \ref{briefly summarized 4He/3He table hybrid LTE/NLTE 1}). Synthetic spectra for HD 4539 and CD-35$^\circ$ 15910 were calculated using the solar value of $\log{n(\text{\isotope[3]{He}})=-4.89}$ \citep{Asplund_2009}.\\

\subsection{$\isotope[3]{He}$ subdwarf B stars}\label{3He sdB Stars with Known 3He Anomaly}
Most of the known $\isotope[3]{He}$ sdB stars (EC 03263-6403, EC 14338-1445, Feige 38, PG 1710+490, and Feige 36) show fits of similar quality as the He-normal stars. Fig. \ref{Feros EC03263M6403 Hybrid LTE/NLTE Helium Line Fits} shows the helium line fits for the FEROS spectrum of EC 03263-6403. Given the higher noise level of the spectra, these fits are satisfactory. Photospheric $\isotope[3]{He}$ is clearly detectable in these stars, as can be seen from the $\isotope[3]{He}$ and $\isotope[4]{He}$ abundances listed in Table \ref{briefly summarized 4He/3He table hybrid LTE/NLTE 1}. Feige 36 shows a balanced abundance ratio of $\isotope[4]{He}$/$\isotope[3]{He}\sim$ 0.98, whereas for EC 03263-6403, EC 14338-1445, Feige 38, and PG 1710+490, $\isotope[4]{He}$ is almost absent ($^4$He/$^3$He < 0.25).  

\subsection{Two new $\isotope[3]{He}$ subdwarf B stars from the ESO SPY project}\label{Two 3He sdB Stars from the ESO SPY Project}
The most comprehensive and homogeneous sample of sdB stars for which high-resolution spectra are available emerged from the ESO SPY project \citep{Napiwotzki_2001}. Overall, the sample included 76 sdBs for which UVES spectra at the ESO VLT were obtained. These spectra were analyzed by \citet{Lisker_2005}, but no search for the $\isotope[3]{He}$ anomaly has been carried out because the UVES spectra did not cover $\ion{He}{i}$ \SI{7281}{\angstrom} or $\ion{He}{i}$ \SI{6678}{\angstrom}. We revisited the list of classified sdBs in the framework of ESO SPY in order to spectroscopically study the $\isotope[3]{He}$ anomaly. Our focus had to be on $\ion{He}{i}$ \SI{4922}{\angstrom}, the strongest and most sensitive line to $\isotope[3]{He}$ in the spectral range of UVES (3290-6640\,\si{\angstrom}). In a first step, we preselected 26 candidates with effective temperatures between $\sim$ \SI{27000}{\kelvin} and $\sim$ \SI{31000}{\kelvin} typical for $\isotope[3]{He}$-enriched sdBs \citep{Geier_2013a}. Some of the candidates were too helium-deficient to show $\ion{He}{i}$ \SI{4922}{\angstrom} so that we could not investigate the $\isotope[3]{He}$ anomaly in those stars. We spectroscopically analyzed the remaining candidates and identified two of them as $\isotope[3]{He}$-enriched sdBs by means of their isotopic abundance ratios. These stars (HE 0929-0424, HE 1047-0436; see Table \ref{briefly summarized 4He/3He table hybrid LTE/NLTE 1}) were classified as close binaries by \citet{Karl_2006} and \citet{Napiwotzki_2001}, respectively. HE 0929-0424 has a semi-amplitude of $K=114.3\pm1.4$\,\si{\kilo\metre\per\second} and a period of $P=0.4400\pm0.0002$\,d, whereas HE 1047-0436 has $K=94.0\pm3.0$\,\si{\kilo\metre\per\second} and $P=1.21325\pm0.00001$\,d. With HE 0929-0424 and HE 1047-0436 being short-period sdB binaries, the total number of known close sdB binaries showing $\isotope[3]{He}$ increases to five (PG 1519+640, Feige 36, PG 0133+114, HE 0929-0424, and HE 1047-0436).\\
We also classified HE 2156-3927 and HE 2322-0617 as $\isotope[3]{He}$-enriched sdBs in the framework of the preliminary results presented in \citet{Schneider_2017}. However, \citet{Lisker_2005} found that both stars show features of cool companions, such as the $\ion{Mg}{i}$ triplet between \SI{5167}{\angstrom} and \SI{5184}{\angstrom}, which we can confirm here. In the case of HE 2156-3927, they determined the companion type to be K3, whereas the companion of HE 2322-0617 is of somewhat earlier spectral type (G9). Because the respective cool companion therefore already significantly contributes to the total flux at the spectral range of $\ion{He}{i}$ \SI{4922}{\angstrom}, we cannot be sure that our detection of $\isotope[3]{He}$ for HE 2156-3927 and HE 2322-0617 is real. Further investigations have to be conducted in order to consolidate the $\isotope[3]{He}$ hypothesis.\\
HE 0929-0424 and HE 1047-0436 have significantly higher isotopic abundance ratios than the stars discussed in Sect. \ref{3He sdB Stars with Known 3He Anomaly} except for Feige 36 (see Table \ref{briefly summarized 4He/3He table hybrid LTE/NLTE 1}). This is likely a selection effect, however, because the detection limit for $^4$He particularly increases when $\ion{He}{i}$ \SI{6678}{\angstrom} is not observed, as we discuss in Sect. \ref{Sensitivity study}. In addition, the analyzed UVES spectra had lower S/N ratios than those of most other stars (see Table \ref{summary of analyzed spectra}). We managed to fit most of the helium lines for HE 0929-0424 and HE 1047-0436. As an example, the line fits for the UVES spectrum of HE 1047-0436 are displayed in Fig. \ref{UVES HE1047M0436 Hybrid LTE/NLTE Helium Line Fits}. Based on the isotopic line shifts observable in the spectrum of the star, in particular for $\ion{He}{i}$ \SI{4922}{\angstrom} and $\ion{He}{i}$ \SI{5015}{\angstrom}, $\isotope[3]{He}$ enrichment is obvious.
\subsection{Helium line profile anomalies and vertical stratification}\label{Helium Line Profile Anomalies}
\begin{figure}
\begin{center}
\resizebox{\hsize}{!}{\includegraphics{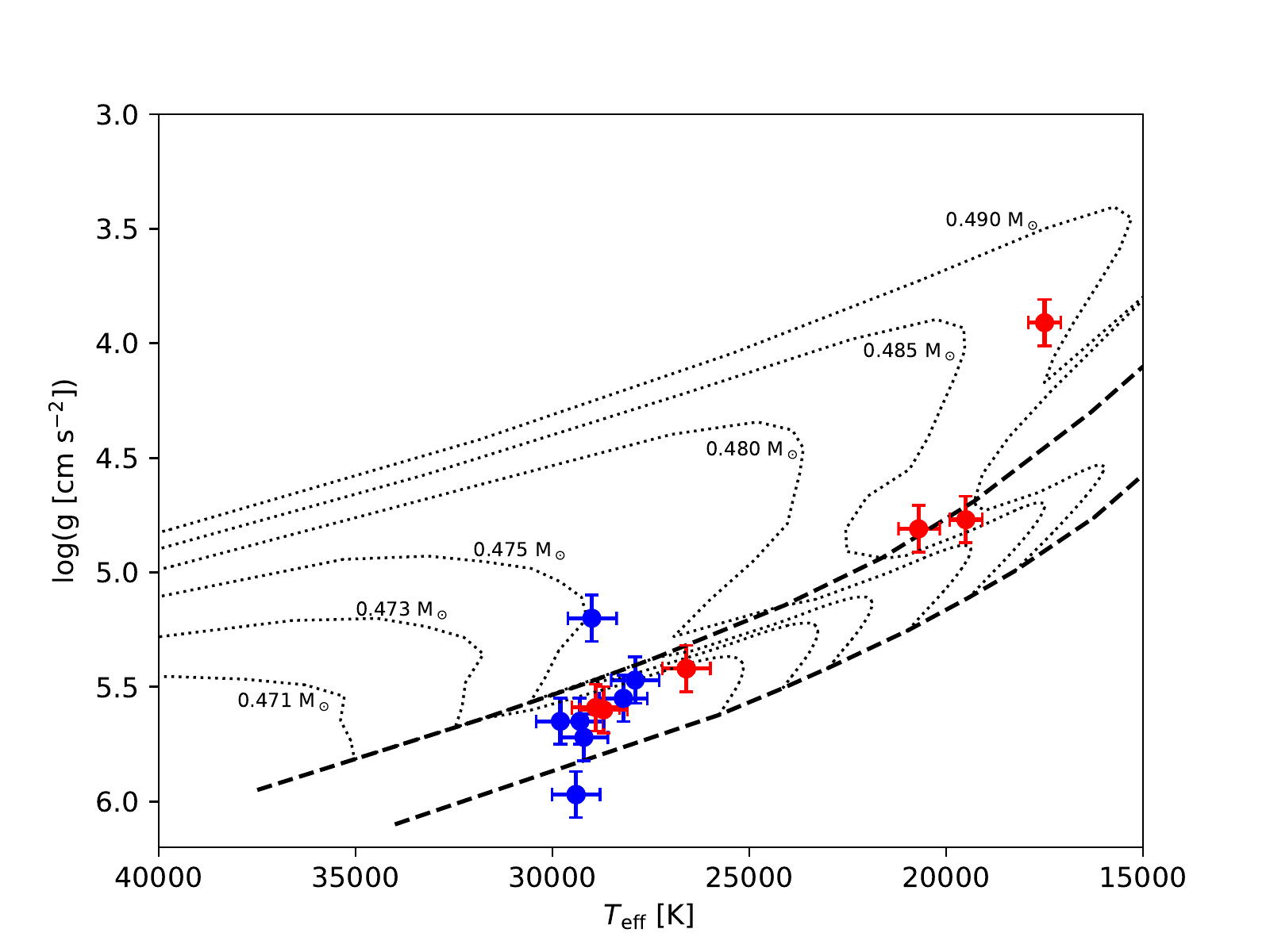}}
\caption{\teff-$\log{(g)}$ diagram of the program stars as in Fig. \ref{NLTE_Teff_logg_diagram_1.pdf}. $\isotope[3]{He}$ stars from our sample showing no evidence for helium stratification are marked in blue, and the stratified $\isotope[3]{He}$ ones are marked in red. The zero-age (ZAHB) and terminal-age horizontal branch (TAHB) as well as evolutionary tracks for different stellar masses but solar metallicity according to \citet{Dorman_1993} are also shown with dashed and dotted lines, respectively.}\label{NLTE_Teff_logg_diagram_2.pdf}
\end{center}
\end{figure}

We were able to fit the observed helium line profiles very precisely for the two He-normal comparison stars and five $^3$He sdBs. Most remarkably, however, we were unable to satisfactorily reproduce the helium line profiles in the case of all analyzed $\isotope[3]{He}$-enriched BHB stars (PHL 25, BD+48$^\circ$ 2721, and PHL 382; see Figs. \ref{McDonald PHL25 Hybrid LTE/NLTE Helium Line Fits}, \ref{Foces BDP482721 Hybrid LTE/NLTE Helium Line Fits}, and \ref{Feros PHL382 Hybrid LTE/NLTE Helium Line Fits}) and the $\isotope[3]{He}$ sdBs EC 03591-3232, EC 12234-2607, and SB 290 (see Figs. \ref{Feros EC03591M3232 Hybrid LTE/NLTE Helium Line Fits} and \ref{Feros SB290 Hybrid LTE/NLTE Helium Line Fits}). A significant mismatch of the cores of many strong $\ion{He}{i}$ lines is obvious. Only some of the weakest $\ion{He}{i}$ lines could be matched satisfactorily. The most prominent cases are PHL 25 and BD+48$^\circ$ 2721 (see Figs. \ref{McDonald PHL25 Hybrid LTE/NLTE Helium Line Fits} and \ref{Foces BDP482721 Hybrid LTE/NLTE Helium Line Fits}). To a lesser extent, discrepancies are also visible for PHL 382, EC 03591-3232, EC 12234-2607, and SB 290 (see Figs. \ref{Feros PHL382 Hybrid LTE/NLTE Helium Line Fits}, \ref{Feros EC03591M3232 Hybrid LTE/NLTE Helium Line Fits}, and \ref{Feros SB290 Hybrid LTE/NLTE Helium Line Fits}). Because of the high rotation velocity of $\sim$ \SI{50}{\kilo\metre\per\second} of SB 290 (see Sect. \ref{Rotational Broadening}) and the line broadening it causes, a more obvious mismatch as seen for the other relevant stars might be hidden to some extent. To test this, we convolved both the observed and the synthetic spectrum of the non-rotating $\isotope[3]{He}$ sdB EC 03591-3232 with a rotational profile for $v\sin{i}\sim$\,\SI{50}{\kilo\metre\per\second}. In this way, we reproduced similarly strong mismatches in the broadened helium line profiles of the star as seen for SB 290. Therefore, we conclude that the strong line broadening in the case of SB 290 indeed hides greater shortcomings in fitting.\\
Because of the insufficient line matches, the resulting abundance ratios for the relevant stars in Table \ref{briefly summarized 4He/3He table hybrid LTE/NLTE 1} are uncertain. Except for PHL 25 and EC 12234-2607, however, we conclude that $\isotope[3]{He}$ has to be the dominant isotope. Generally, we found no evidence that the stars with anomalous helium line profiles have more atmospheric helium than other program stars, although EC 03591-3232 and EC 12234-2607 belong to the most He-rich stars (see Table \ref{briefly summarized 4He/3He table hybrid LTE/NLTE 1}).\\
We calculated the entire helium line spectrum for a large variety of helium abundances, but none of them could simultaneously match both the wings and the cores of the analyzed helium absorption lines of the relevant stars. Specifically, $\ion{He}{i}$ \SI{4026}{\angstrom} and $\ion{He}{i}$ \SI{4472}{\angstrom} exhibit shallow cores in combination with unusually broad wings (see, e.g., Figs. \ref{Foces BDP482721 Hybrid LTE/NLTE Helium Line Fits} and \ref{Feros PHL382 Hybrid LTE/NLTE Helium Line Fits}), indicating that helium is not homogeneously distributed throughout the stellar atmosphere, but instead shows a vertical abundance stratification. The shallow line cores indicate a lower-than-average helium abundance in outer atmospheric layers, where the cores are formed. The strong line wings require a higher-than-average helium abundance in deeper atmospheric layers, where the wings are formed. The farther out in the stellar atmosphere the particular helium absorption line core is formed, that is, the stronger the individual line, the poorer the reproduction of the line core (see Figs. \ref{McDonald PHL25 Hybrid LTE/NLTE Helium Line Fits}, \ref{Foces BDP482721 Hybrid LTE/NLTE Helium Line Fits}, \ref{Feros PHL382 Hybrid LTE/NLTE Helium Line Fits}, \ref{Feros EC03591M3232 Hybrid LTE/NLTE Helium Line Fits}, and \ref{Feros SB290 Hybrid LTE/NLTE Helium Line Fits}). In addition to $\lambda$\SI{4026}{\angstrom} and $\lambda$\SI{4472}{\angstrom}, this particularly applies to $\lambda$\SI{4922}{\angstrom}, $\lambda$\SI{5016}{\angstrom}, $\lambda$\SI{5875}{\angstrom}, $\lambda$\SI{6678}{\angstrom}, and $\lambda$\SI{7065}{\angstrom} and indicates that the helium abundance indeed has to be higher in deeper layers of the atmospheres than in the outer ones. The apparent discrepancy in projected rotational velocities determined from helium lines being larger than from metals in the case of SB 290 (see Sect. \ref{Rotational Broadening}) can thus be explained as an effect of helium abundance stratification. Furthermore, Geier et al. (2013a) were not aware of the helium-stratified atmosphere of BD+48$^\circ$ 2721. Together with a different choice of helium lines for their spectral analysis, this might be the crucial factor for the discrepancy of the stellar atmospheric parameters (see Sect. \ref{Results from NLTE versus LTE analyses}).\\
Helium stratification has been reported for a few $\isotope[3]{He}$ B-type stars so far (see, e.g., \citealt{Bohlender_2005}). This includes B-type MS stars such as the helium-variable star aCen \citep{Leone_1997, Bohlender_2010, Maza_2014b}, the prototype HgMn star $\kappa$ Cancri \citep{Maza_2014}, and the chemically peculiar $\isotope[3]{He}$ star HD 185330 \citep{Niemczura_2018}. Helium stratification has also been found in Feige 86, a well-studied BHB star \citep{Bonifacio_1995, Cowley_2005, Cowley_2009, Nemeth_2017}, and in other chemically peculiar stars by \citet{Dworetsky_2004} and \citet{Castelli_2007}, for example. However, this is the first time that it has been detected in subdwarf B stars.\\
\begin{figure*}
\begin{minipage}[b]{0.5\linewidth}
\resizebox{\hsize}{!}{\includegraphics{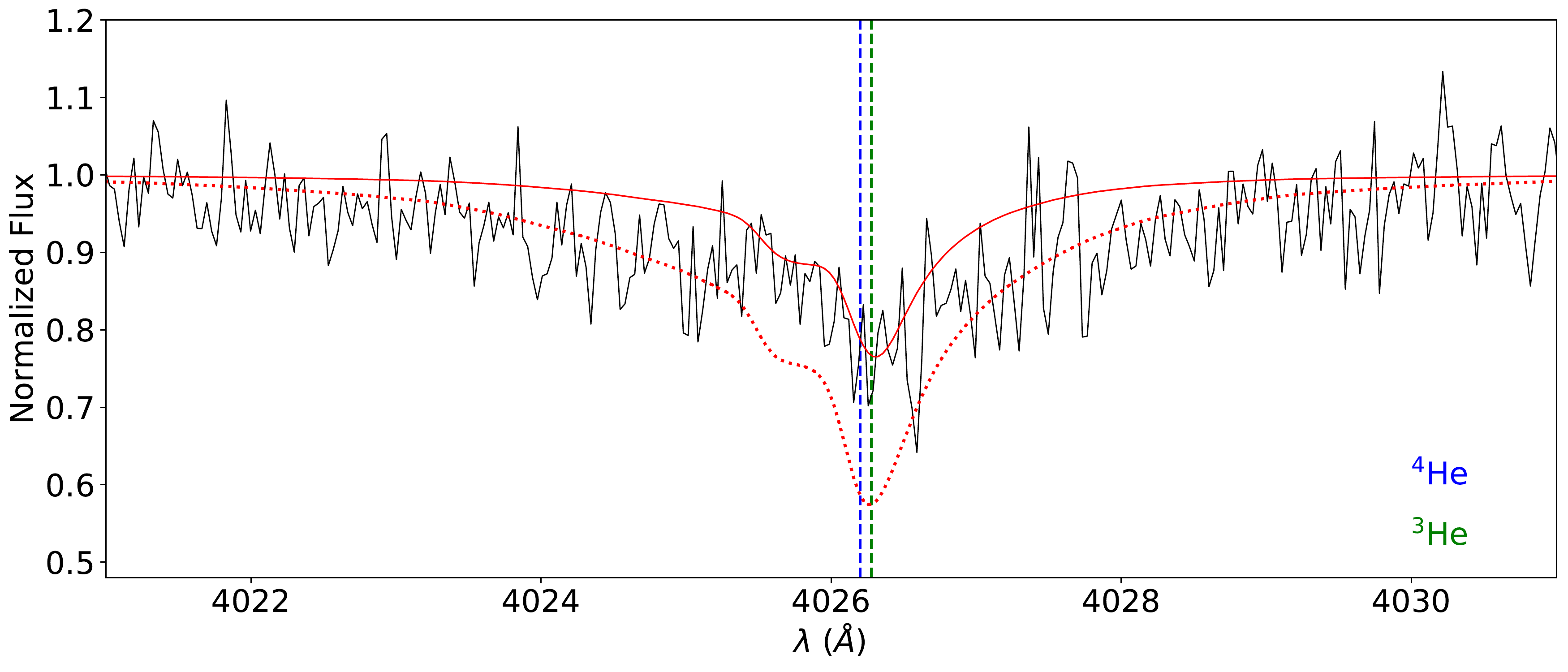}}
\centering
\end{minipage}\hfill
\begin{minipage}[b]{0.5\linewidth}
\resizebox{\hsize}{!}{\includegraphics{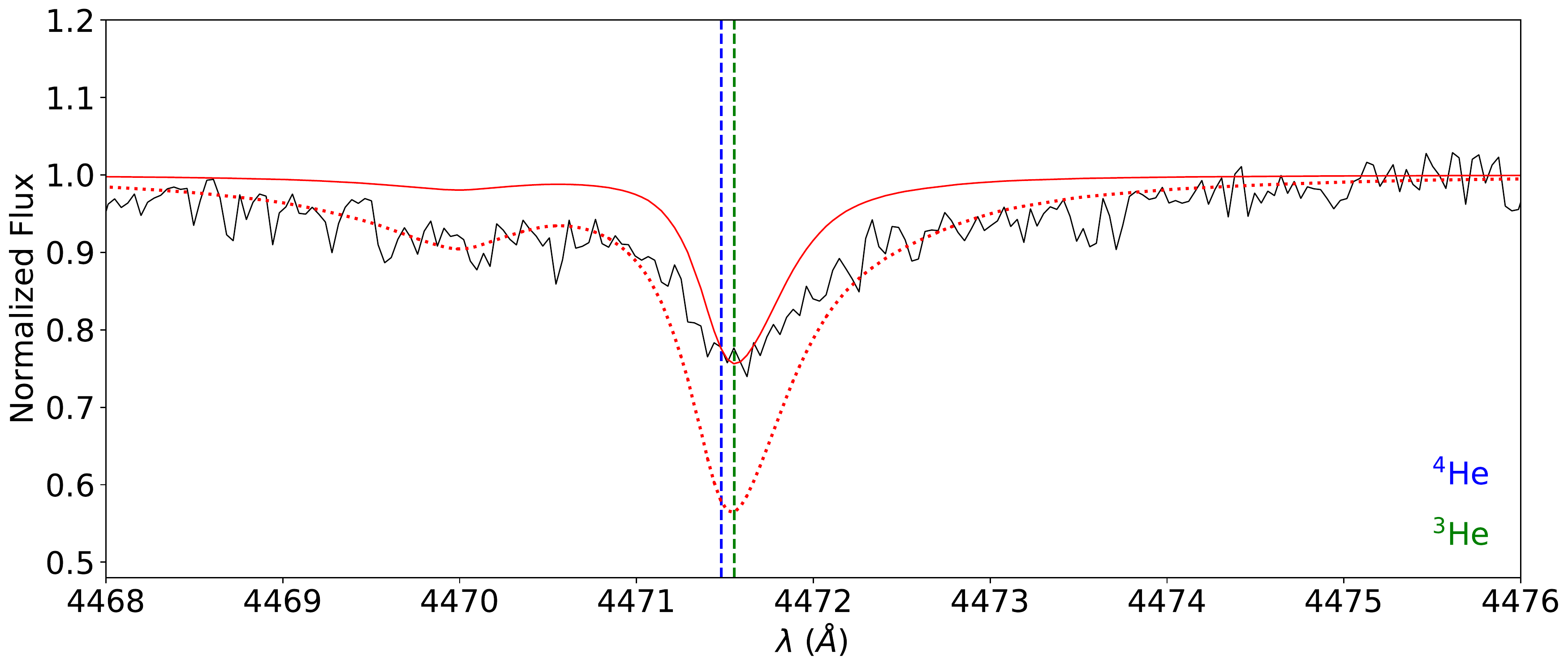}}
\centering
\end{minipage}\hfill
\captionof{figure}{\textit{Left-hand panel}: Best fit by eye for the line core (solid red line, $\log{n(\isotope[4]{He}+\isotope[3]{He})}=-2.50$) and the line wings (dotted red line, $\log{n(\isotope[4]{He}+\isotope[3]{He})}=-1.79$) of $\ion{He}{i}$ \SI{4026}{\angstrom} in the HRS spectrum of PHL 25 (solid black line). \textit{Right-hand panel}: Same as the left-hand panel, but for $\ion{He}{i}$ \SI{4472}{\angstrom}. The total helium abundances are $\log{n(\isotope[4]{He}+\isotope[3]{He})}=-3.00$ (solid red line) and $\log{n(\isotope[4]{He}+\isotope[3]{He})}=-2.10$ (dotted red line).}\label{fit by eye of HeI 4026 and HeI 4472 for stratification star PHL 25 HRS spectrum}
\end{figure*}\noindent
In order to reproduce the observed helium line profiles and to estimate the total helium abundance, $\log{n(\isotope[4]{He}+\isotope[3]{He})}$, in the outer and inner stellar atmospheres, we applied a two-component fit (see also \citealt{Maza_2014}). To this end, we chose the strong $\ion{He}{i}$ lines at $\lambda$\SI{4026}{\angstrom} and $\lambda$\SI{4472}{\angstrom}. Their line cores are formed farther out than their line wings. Trying to match the line cores and wings of both lines individually, we performed four fits by eye for the relevant stars, fixing the values for effective temperature, surface gravity, and projected rotation velocity determined in Sect. \ref{Effective Temperatures and Surface Gravities}. We also estimated errors for $\log{n(\isotope[4]{He}+\isotope[3]{He})}$ by varying the total helium abundance until clear mismatches in cores and wings, respectively, become obvious. Table \ref{fits by eye results for stratification stars} summarizes the results. As an example, Fig. \ref{fit by eye of HeI 4026 and HeI 4472 for stratification star PHL 25 HRS spectrum} compares the best-fit model spectrum for the investigated helium lines to the HRS spectrum of PHL 25.\\
The helium abundance overall increases from outer to inner atmospheric layers of the analyzed stars (see Table \ref{fits by eye results for stratification stars}). For $\ion{He}{i}$ \SI{4026}{\angstrom}, the total helium abundance increases by $\sim$\,0.15\,dex for SB 290 up to $\sim$\,0.83\,dex for PHL 382. Derived from $\ion{He}{i}$ \SI{4472}{\angstrom}, the helium abundance increases with depth even more significantly, by $\sim$\,0.46\,dex for SB 290 up to $\sim$\,0.90\,dex for PHL 25, EC 12234-2607, and BD+48$^\circ$ 2721, respectively. Depending on the individual stratified star, we hence estimate that the helium abundance increases by a factor of $\sim$\,1.4-8.0 from the outer to the inner atmosphere. This is a clear indication for an inhomogeneous distribution of helium in the atmosphere, or in other words, for vertical abundance stratification.\\
Last but not least, we highlight the position of the helium-stratified program stars in the \teff-$\log{(g)}$ plane (see Fig. \ref{NLTE_Teff_logg_diagram_2.pdf}). They populate the whole effective temperature sequence, and in contrast to the $\isotope[3]{He}$ sdBs, they do not cluster in a certain temperature regime.\\
\noindent
\begin{table*}
\caption{Total helium abundance, $\log{n(\isotope[4]{He}+\isotope[3]{He})}$, of the individually performed fits by eye for the helium-stratified program stars (see Sect. \ref{Helium Line Profile Anomalies} for details).}\label{fits by eye results for stratification stars}
\centering
\begin{tabular}{ccccccc}
\hline\hline
Fitted Line & PHL 25 & PHL 382 & BD+48$^\circ$ 2721 & SB 290 & EC 03591-3232 & EC 12234-2607\\
\hline
$\ion{He}{i}$ \SI{4026}{\angstrom} core & $-2.50^{+0.30}_{-0.20}$ & $-2.65^{+0.13}_{-0.10}$ & $-2.50\pm0.15$ & $-2.45^{+0.15}_{-0.10}$ & $-2.15\pm0.15$ & $-1.80^{+0.40}_{-0.30}$\\
$\ion{He}{i}$ \SI{4026}{\angstrom} wings & $-1.79^{+0.60}_{-0.40}$ & $-1.82^{+0.32}_{-0.38}$ & $-1.80\pm0.30$ & $-2.30^{+0.40}_{-0.20}$ & $-1.80^{+0.25}_{-0.30}$ & $-1.45^{+0.30}_{-0.45}$\\
$\ion{He}{i}$ \SI{4472}{\angstrom} core & $-3.00^{+0.20}_{-0.10}$ & $-2.68^{+0.18}_{-0.14}$ & $-2.80^{+0.20}_{-0.10}$ & $-2.75^{+0.20}_{-0.15}$ & $-2.40^{+0.20}_{-0.10}$ & $-2.35\pm0.20$\\
$\ion{He}{i}$ \SI{4472}{\angstrom} wings & $-2.10^{+0.30}_{-0.20}$ & $-1.95^{+0.15}_{-0.20}$ & $-1.90\pm0.20$ & $-2.29^{+0.24}_{-0.36}$ & $-1.85^{+0.25}_{-0.15}$ & $-1.45^{+0.30}_{-0.34}$\\
\hline
\end{tabular}
\tablefoot{
Given errors result from the fits by eye (see Sect. \ref{Helium Line Profile Anomalies} for details).
}
\end{table*}\noindent
\subsection{Sensitivity study}\label{Sensitivity study}
Most of the known $\isotope[3]{He}$ sdB stars from \citet{Geier_2013a} show $\isotope[4]{He}$/$\isotope[3]{He}$ $<0.25$ (see Table \ref{briefly summarized 4He/3He table hybrid LTE/NLTE 1}). Given the range in S/N of the observations, what are the detection limits for $\isotope[4]{He}$?\\
We added Gaussian noise to spectra computed with the parameters and S/N ratios given in Tables \ref{briefly summarized 4He/3He table hybrid LTE/NLTE 1} and \ref{summary of analyzed spectra} for three program stars: EC03263-6403 $=$ Model I, BD+48$^\circ$ 2721 $=$ Model II, and HE 1047-0436 $=$ Model III. BD+48$^\circ$ 2721 was chosen for its high S/N ratio ($\sim$\,84; see Table \ref{summary of analyzed spectra}), EC 03263-6403  for its low isotopic ratio ($\isotope[4]{He}$/$\isotope[3]{He}\sim0.01$, see Table \ref{briefly summarized 4He/3He table hybrid LTE/NLTE 1}), and HE 1047-0436 for having the lowest $\isotope[4]{He}$ abundance of the $\isotope[3]{He}$ ESO SPY sdBs (where the sensitive $\ion{He}{i}$ \SI{6678}{\angstrom} line has not been observed; see also Table \ref{briefly summarized 4He/3He table hybrid LTE/NLTE 1}).\\
The Gaussian noise was simulated using samples ($p$), drawn from a parameterized normal distribution centered around zero (mean value $\mu=0$) and with a standard deviation of one ($\sigma=1$). The fluxes of the individual mock spectra, $F_{\mathrm{mock}}$, then were calculated from model fluxes, $F_{\mathrm{model}}$, according to the following formula:
\begin{equation}  
F_{\mathrm{mock}}=\left(\frac{p}{S/N}\right)\cdot F_{\mathrm{model}} + F_{\mathrm{model}}\,. 
\end{equation}
Here, $S/N$ is the individual signal-to-noise ratio. We chose different values for S/N: {\sc i}) the original S/N of the individual observed spectra; {\sc ii}) S/N$=$100; {\sc iii}) S/N$=$200; and {\sc iv}) S/N$=$300. We simultaneously fitted effective temperature, surface gravity, and both isotopic helium abundances, making use of the same analysis technique as presented in Sect. \ref{Spectroscopic Analysis Technique}.
\subsubsection{Detectability and error estimation}\label{Detectability and Error Estimation}
The results of the sensitivity study for the three models together with those of the abundance analysis are shown in Table \ref{results of the sensitivity study}. In the following, we highlight three important results.\\
First, we are able to reproduce the results of the previous abundance analysis. This confirms that the abundances given in Table \ref{briefly summarized 4He/3He table hybrid LTE/NLTE 1} are reliable detections of the small traces of $\isotope[4]{He}$. Values and uncertainties for $\log{n(\isotope[4]{He})}$ and $\log{n(\isotope[3]{He})}$ derived from mock spectra are overall in good agreement with the observed ones if the same S/N and the same number of investigated helium lines are used. Increasing the S/N ratio, however, results in a significant improvement of accuracy.\\
Second, if the strong and sensitive $\ion{He}{i}$ \SI{6678}{\angstrom} line is ignored, we are not able to reproduce the observed $\isotope[4]{He}$ abundances for EC 03263-6403 and BD+48$^\circ$ 2721. In particular, this results in large statistical uncertainties on $\log{n(\isotope[4]{He})}$, even for better S/N. Hence, it is not possible to determine $\log{n(\isotope[4]{He})}$ of the $\isotope[4]{He}$-deficient sdB star EC 03263-6403 without $\ion{He}{i}$ \SI{6678}{\angstrom}.\\
Last, we also derived similar isotopic helium abundances and statistical uncertainties from mock spectra as from the previous abundance analysis in the case of HE 1047-0436, that is, for the most $\isotope[4]{He}$-deficient but $\isotope[3]{He}$-enriched sdB star from the ESO SPY project (see Table \ref{briefly summarized 4He/3He table hybrid LTE/NLTE 1}). We simulated the case of the analyzed UVES spectra in which $\ion{He}{i}$ \SI{6678}{\angstrom} as the most important signature for $\isotope[4]{He}$/$\isotope[3]{He}$ was also not available for investigation. This result substantiates the discovery of HE 1047-0436 as a $\isotope[3]{He}$-enriched sdB. Adding observations of $\ion{He}{i}$ \SI{6678}{\angstrom} (and $\ion{He}{i}$ \SI{10830}{\angstrom}, see Sect. \ref{The infrared HeI 10830 Line}) would still significantly improve the accuracy of the analysis and might reveal other unclassified $\isotope[3]{He}$ sdBs in the ESO SPY sample at lower $\isotope[4]{He}$/$\isotope[3]{He}$ ratios.\\
In addition, it is worthwhile to deduce a detection limit for $\isotope[3]{He}$. Although $\ion{He}{i}$ \SI{6678}{\angstrom} is not observed for HE 1047-0436 and the S/N is mediocre, Fig. \ref{UVES HE1047M0436 Hybrid LTE/NLTE Helium Line Fits} demonstrates that the $\isotope[3]{He}$ anomaly is detectable at an $\isotope[4]{He}$/$\isotope[3]{He}$ abundance ratio as high as $\sim0.91$ even for such poor data. In consequence, we conclude that all other cases where the abundance ratio is lower than $\sim0.91$ are indeed reliable detections of $\isotope[3]{He}$, even if the $\isotope[3]{He}$ abundance is as low as $\log{n(\text{\isotope[3]{He}})}\sim\,-3.10,$ as in the case of EC 14338-1445. Hence, the latter is a reliable upper limit for the detection limit for $\isotope[3]{He}$. HE 0929-0424 has an even higher abundance ratio ($\isotope[4]{He}$/$\isotope[3]{He}\sim2.51$) than HE 1047-0436, but at the same time shows significantly more helium (see Table \ref{briefly summarized 4He/3He table hybrid LTE/NLTE 1}). Thus, the discovery of HE 0929-0424 as a $\isotope[3]{He}$-enriched sdB is reliable as well. 
\begin{table}
\small
\caption{Results of the sensitivity study.}\label{results of the sensitivity study}
\centering
\begin{tabular}{ccccc}
\hline\hline
Mock spectrum & S/N & $\log{n(\isotope[4]{He})}$ & $\log{n(\isotope[3]{He})}$ & $\frac{n(\text{\isotope[4]{He}})}{n(\text{\isotope[3]{He}})}$\\
\hline
Model I+ & 23 & $-4.72^{+0.31}_{-0.49}$ & $-2.87^{+0.01}_{-0.02}$ & $0.01^{+0.02}_{-0.01}$\\
Model I+ & 100 & $-4.74^{+0.29}_{-0.46}$ & $-2.85^{+0.01}_{-0.02}$ & $0.01\pm0.01$\\
Model I+ & 200 & $-4.75^{+0.25}_{-0.41}$ & $-2.84\pm0.01$ & $0.01\pm0.01$\\
Model I+ & 300 & $-4.75^{+0.18}_{-0.37}$ & $-2.85\pm0.02$ & $0.01\pm0.01$\\
\hline
EC 03263-6403\tablefootmark{a} & 23 & $-4.75^{+0.29}_{-0.32}$ & $-2.85^{+0.03}_{-0.02}$ & $0.01\pm0.01$\\
\hline
\hline
Model II+ & 84 & $-3.40^{+0.08}_{-0.12}$ & $-2.55^{+0.09}_{-0.12}$ & $0.14^{+0.07}_{-0.08}$\\
Model II+ & 100 & $-3.40^{+0.05}_{-0.04}$ & $-2.56\pm0.06$ & $0.14\pm0.06$\\
Model II+ & 200 & $-3.38\pm0.02$ & $-2.57\pm0.02$ & $0.15\pm0.06$\\
Model II+ & 300 & $-3.35\pm0.02$ & $-2.57\pm0.01$ & $0.17\pm0.06$\\
\hline
BD+48$^\circ$ 2721\tablefootmark{a} & 84 & $-3.34^{+0.09}_{-0.11}$ & $-2.57^{+0.09}_{-0.11}$ & $0.17^{+0.08}_{-0.09}$\\
\hline
\hline
Model III- & 25 & $-2.77\pm0.04$ & $-2.70\pm0.03$ & $0.86\pm0.30$\\
\hline
HE 1047-0436\tablefootmark{a} & 25 & $-2.76\pm0.04$ & $-2.72\pm0.03$ & $0.91\pm0.32$\\
\hline
\end{tabular}
\tablefoot{
$1\sigma$ statistical single parameter errors on $\log{n(\isotope[4]{He})}$ and $\log{n(\isotope[3]{He})}$ are given. The systematic uncertainties are $\pm0.10$ in all cases. The given uncertainties on $n(\text{\isotope[4]{He}})/n(\text{\isotope[3]{He}})$ result from statistical and systematic errors, for which Gaussian error propagation was used.\\
From Table \ref{briefly summarized 4He/3He table hybrid LTE/NLTE 1}: Model I+ (\teff$=29\,000$\,\si{\kelvin}, $\log{(g)}=5.21$, $\log{n(\text{\isotope[4]{He}})}=-4.75$, $\log{n(\text{\isotope[3]{He}})}=-2.85$, $\ion{He}{i}$ \SI{6678}{\angstrom} included), Model II+ (20\,700\,\si{\kelvin}, 4.81, -3.34, -2.57, $\ion{He}{i}$ \SI{6678}{\angstrom} included), and Model III- (29\,800\,\si{\kelvin}, 5.65, -2.76, -2.72, $\ion{He}{i}$ \SI{6678}{\angstrom} excluded).\\ 
\tablefoottext{a}{Observed spectrum}   
}
\end{table}\noindent
\begin{table}
\small
\caption{Influence of $\ion{He}{i}$ \SI{10830}{\angstrom} on the sensitivity study for BD+48$^\circ$ 2721.}\label{results of the sensitivity study 2}
\centering
\begin{tabular}{ccccc}
\hline\hline
Mock spectrum & S/N & $\log{n(\isotope[4]{He})}$ & $\log{n(\isotope[3]{He})}$ & $\frac{n(\text{\isotope[4]{He}})}{n(\text{\isotope[3]{He}})}$\\
\hline
Model II++ & 84 & $-3.35^{+0.05}_{-0.04}$ & $-2.56\pm0.05$ & $0.16\pm0.06$\\
Model II++ & 100 & $-3.34^{+0.03}_{-0.04}$ & $-2.58\pm0.02$ & $0.17\pm0.06$\\
Model II++ & 200 & $-3.34\pm0.02$ & $-2.57\pm0.01$ & $0.17\pm0.06$\\
Model II++ & 300 & $-3.34\pm0.01$ & $-2.57\pm0.01$ & $0.17\pm0.06$\\
\hline
BD+48$^\circ$ 2721\tablefootmark{a} & 84 & $-3.34^{+0.09}_{-0.11}$ & $-2.57^{+0.09}_{-0.11}$ & $0.17^{+0.08}_{-0.09}$\\
\hline
\end{tabular}
\tablefoot{
Given uncertainties are the same as in Table \ref{results of the sensitivity study}.\\
From Table \ref{briefly summarized 4He/3He table hybrid LTE/NLTE 1}: Model II++ (\teff$=20\,700$\,\si{\kelvin}, $\log{(g)}=4.81$, $\log{n(\text{\isotope[4]{He}})}=-3.34$, $\log{n(\text{\isotope[3]{He}})}=-2.57$, $\ion{He}{i}$ \SI{6678}{\angstrom} and $\ion{He}{i}$ \SI{10830}{\angstrom} included).\\ 
\tablefoottext{a}{Observed spectrum}  
}
\end{table}\noindent
\subsubsection{Infrared $\ion{He}{i}$ \SI{10830}{\angstrom} line}\label{The infrared HeI 10830 Line}
$\ion{He}{i}$ \SI{10830}{\angstrom} in the near-infrared is known to show the largest isotopic line shift of $\sim$ \SI{1.32}{\angstrom}, that is, more than twice as large as for $\ion{He}{i}$ \SI{6678}{\angstrom} ($\sim$ \SI{0.50}{\angstrom}). Thus, the $\ion{He}{i}$ \SI{10830}{\angstrom} line should improve the detectability of $\isotope[3]{He}$ and in turn the sensitivity to $\isotope[4]{He}$/$\isotope[3]{He}$ significantly. As $\ion{He}{i}$ \SI{5875}{\angstrom} and $\ion{He}{i}$ \SI{6678}{\angstrom}, this line is strongly affected by departures from LTE. However, the helium model atom used here in combination with the hybrid LTE/NLTE approach have been shown to be appropriate to match the observed line profiles in early B-type MS stars well \citep{Przybilla_2005}. Hence, we included $\ion{He}{i}$ \SI{10830}{\angstrom} into the sensitivity study for BD+48$^\circ$ 2721 (see Table \ref{results of the sensitivity study 2}) in order to test its influence on the derived $\isotope[4]{He}$/$\isotope[3]{He}$ abundance ratio.\\
$\ion{He}{i}$ \SI{10830}{\angstrom} indeed particularly leads to a better accuracy in determining both isotopic abundances for a given S/N (compare the results in Table \ref{results of the sensitivity study 2} to those in Table \ref{results of the sensitivity study}). Both isotopic abundances are already well determined if $\ion{He}{i}$ \SI{6678}{\angstrom} as well as $\ion{He}{i}$ \SI{10830}{\angstrom} are available and the S/N ratio is on the order of $\sim$ 85. An even higher S/N decreases the derived statistical uncertainties on both isotopic abundances if $\ion{He}{i}$ \SI{10830}{\angstrom} is included.

\section{Conclusion and outlook}\label{Summary}
We carried out a quantitative spectral analysis of 13 subluminous B stars that show the $\isotope[3]{He}$ anomaly (three BHBs, eight known sdBs, and two newly discovered sdBs from the ESO Supernova Ia Progenitor Survey), as well as of two prototypical He-normal sdBs for reference, making use of high-quality optical spectra and a grid of synthetic spectra calculated in a hybrid LTE/NLTE approach based on modified versions of the model atmosphere code ATLAS12 and the NLTE line formation and synthesis codes DETAIL and SURFACE. Well-tested model atoms were supplemented by a $\isotope[3]{He}$ model atom from \citet{Maza_2014}.\\
We redetermined effective temperatures, surface gravities, and helium abundances of nine stars previously analyzed from the same spectra but using LTE model atmospheres, different metal contents, and a different minimization procedure to determine the best fit. In general, the results were found to be consistent for all but one program star (BD+48$^\circ$ 2721), indicating that systematic effects by the cumulative impact of  departures from LTE, different metal contents, and the different analysis strategies are small. However, BD+48$^\circ$ 2721 is much cooler than previously deduced, and was therefore reclassified as a BHB star.\\
Isotopic abundance ratios, $^4$He/$^3$He, were determined from the helium line spectra. Both $\ion{He}{i}$ \SI{6678}{\angstrom} and $\ion{He}{i}$ \SI{5875}{\angstrom} play a crucial role because they show the largest and smallest isotopic line shift, respectively. However, these lines are known to be strengthened by departures from LTE, in particular $\ion{He}{i}$ \SI{6678}{\angstrom}, which requires an appropriately detailed atomic model and sophisticated line broadening tables. We verified our synthetic spectra against the benchmark sdB stars HD 4539 and CD-35$^\circ$ 15910.\\
As expected, $\isotope[4]{He}$ is almost absent ($\isotope[4]{He}$/$\isotope[3]{He}<0.25$) in most of the known $\isotope[3]{He}$ sdB stars from \citet{Geier_2013a}. We identified the $\isotope[3]{He}$ anomaly in the stellar atmospheres of HE 0929-0424 and HE 1047-0436, resulting in $\isotope[4]{He}$/$\isotope[3]{He}$ $\sim2.51$ and $\isotope[4]{He}$/$\isotope[3]{He}$ $\sim0.91$, respectively. This substantiates the helium abundance results for the known $\isotope[3]{He}$ stars. The interesting question arises whether there is a continuum of $\isotope[3]{He}$-enriched sdBs/BHBs at higher abundance ratios ($\isotope[4]{He}$/$\isotope[3]{He}\sim$\,1.0-3.0), which could be answered from a homogenous sample with excellent data. Such a sample may be the one of \citet{Geier_2013a}, who examined 44 bright sdB stars that have been observed at high spectral resolution and good S/N, covering $\ion{He}{i}$ \SI{6678}{\angstrom} (e.g., with FEROS) with a sharp eye on the isotopic anomaly. The sample is somewhat biased, however, because the previously known $\isotope[3]{He}$ sdBs were included \textit{\textup{a priori}}. All new $\isotope[3]{He}$ sdBs found by \citet{Geier_2013a} have very low $\isotope[4]{He}$ abundances ($\log{n(\text{\isotope[4]{He}})}\la\,-3.30$), except for EC 12234-2607. A larger unbiased sample is needed in order to draw sound conclusions on a potential continuum.\\
Anomalous helium line profiles (broad wings, shallow cores) were found in all three BHB (PHL 25, PHL 382, and BD+48$^\circ$ 2721) and in three of the $\isotope[3]{He}$ sdB stars (EC 03591-3232, EC 12234-2607, and SB 290). This phenomenon can be explained by vertical stratification of the atmospheres. We estimate that the helium abundance increases from the outer atmospheric layers, where the cores of strong helium lines form, to the deeper ones, where the line wings form, by factors ranging from about 1.4 in the case of the sdB SB 290 to a factor of 8.0 in the case of the BHB star BD+48$^\circ$ 2721. Such a layered distribution of helium has been found previously in late B-type stars of peculiar chemical composition as well as in other BHB stars, but this is the first time that helium stratification is reported for sdB stars. A particularly interesting case is SB 290, because it is the only rapidly rotating sdB in the sample. \citet{Geier_SB290_2013} derived $v\sin{i}=48.0\pm2.0$\,\si{\kilo\metre\per\second} from metal, but $v\sin{i}=58.0\pm1.0$\,\si{\kilo\metre\per\second} from helium lines. We confirmed this discrepancy and traced it back to the anomalous helium line profiles. The projected rotational velocity of $v\sin{i}=58.0\pm1.0$\,\si{\kilo\metre\per\second} derived from helium lines by \citet{Geier_SB290_2013} is therefore overestimated because of vertical helium abundance stratification.\\
The simplest way to carry out a quantitative spectral analysis of the stratification profile is to make use of a smoothed step function (see \citealt{Farthmann_1994}), which sets the helium abundance in the outer atmospheric layers to a lower level than farther in. In this way, the optical depth at which the change in helium abundance occurs can also be identified. As an example, stratification analyses have been carried out successfully for carbon and nitrogen in the post-HB star HD 76431 by \citet{Khalack_2014} and for nitrogen, sulfur, titanium, and in particular for iron, in BHB stars by \citet{Khalack_2007, Khalack_2008, Khalack_2010} and \citet{LeBlanc_2010}. The strategy to derive the stratified helium abundance profile throughout the photosphere with the hybrid LTE/NLTE approach was reported in \citet{Maza_2014}, who tested it successfully on $\kappa$ Cancri.\\
The $\isotope[3]{He}$ anomaly for MS stars is restricted to a narrow temperature strip of \SI{14000}{\kelvin} $\la$ \teff\,$\la$ \SI{21000}{\kelvin} \citep{Sargent_1961, Hartoog_1979b}. The same is true for sdB stars, but at hotter temperatures (\SI{26000}{\kelvin} $\la$ \teff\,$\la$ \SI{30000}{\kelvin}). Otherwise, no correlation can be found. The anomaly occurs for single as well as for close binary sdBs (PG 1519+640, Feige 36, PG 0133+114, HE 0929-0424, and HE 1047-0436). We also found evidence that it occurs for objects that have evolved off the canonical horizontal branch, as in the case of EC 03263-6403 and PHL 382.\\
The $\isotope[3]{He}$ isotopic anomaly was previously considered to be a rare phenomenon. Although many sdB stars have been analyzed by means of high-resolution spectra, observations of the crucial $\ion{He}{i}$ \SI{6678}{\angstrom} line are lacking in many cases, in particular for the largest homogeneous sdB sample from the ESO SPY project \citep{Lisker_2005}. The fraction of $\isotope[3]{He}$ stars among sdBs could be best constrained from high-resolution spectroscopy of the $\ion{He}{i}$ \SI{6678}{\angstrom} line. However, even more promising for investigating the isotopic anomaly would be observations of the near-infrared helium line, $\ion{He}{i}$ \SI{10830}{\angstrom}, which now becomes possible with modern spectrographs.

\begin{acknowledgements}
D. S. was supported by the Deutsche Forschungsgemeinschaft (DFG) under grant HE 1356/70-1. M. F. N. acknowledges support by the Austrian Science Fund (FWF) in the form of a Lise-Meitner Fellowship under project number N-1868-NBL. We thank S. Geier and H. Edelmann for sharing their data with us. We wish to thank the anonymous referee for providing us with a number of detailed comments that greatly improved the clarity of this manuscript. We made use of ISIS functions provided by ECAP/Remeis observatory and MIT (\url{http://www.sternwarte.uni-erlangen.de/isis/}). We thank J. E. Davis for the development of the \texttt{slxfig} module, which has been used to prepare part of the figures in this work. For the other part \texttt{matplotlib} \citep{Hunter_2007} and \texttt{NumPy} \citep{van_der_Walt_2011} were used. Based on observations at the La Silla Observatory of the European Southern Observatory (ESO) and on observations made at the Centro Astron\'{o}mico Hispano Alem\'{a}n (CAHA) at Calar Alto, operated jointly by the Max-Planck-Institut f\"ur Astronomie and the Instituto de Astrof\'{i}sica de Andaluc\'{i}a (CSIC). Based on observations made at the McDonald Observatory in Austin operated by the University of Texas and on observations conducted at the W. M. Keck Observatory on Hawaii.      
\end{acknowledgements}

%
%

\bibliographystyle{aa}
\bibliography{spec_analysis_3he_anomaly_B_type_stars_ref}

\begin{appendix}
\section{Helium line fits}\label{helium line fits}

\begin{minipage}[b]{2.0\linewidth}
\begin{minipage}[b]{0.5\linewidth}
\resizebox{\hsize}{!}{\includegraphics{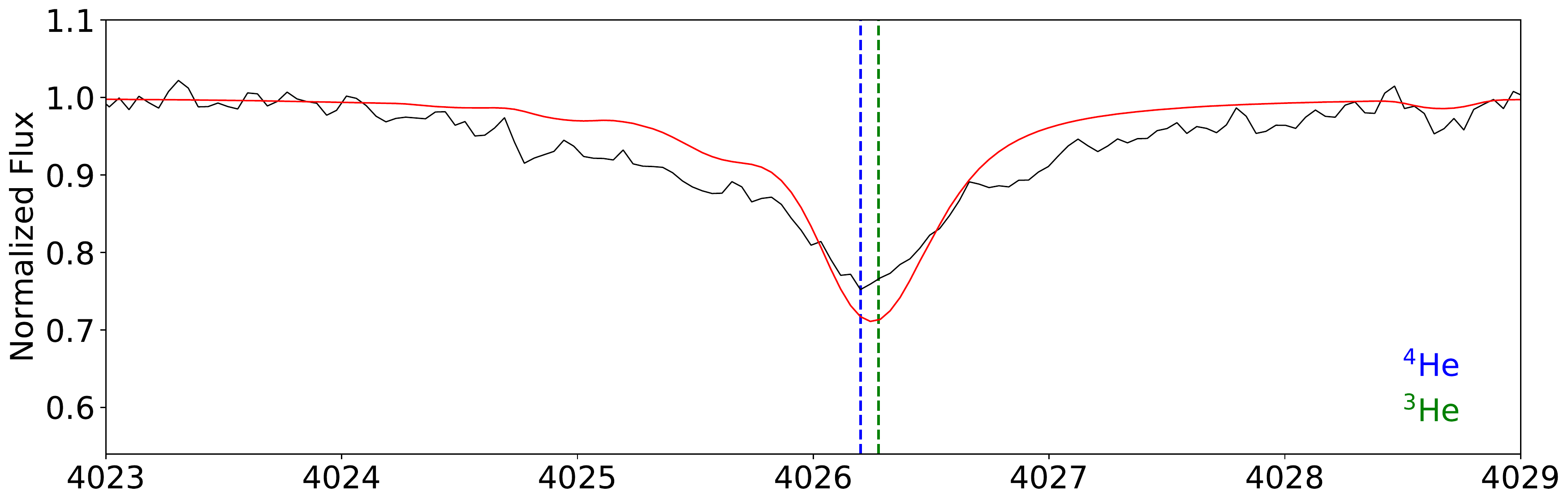}}
\centering
        \end{minipage}\hfill
\begin{minipage}[b]{0.5\linewidth}
\resizebox{\hsize}{!}{\includegraphics{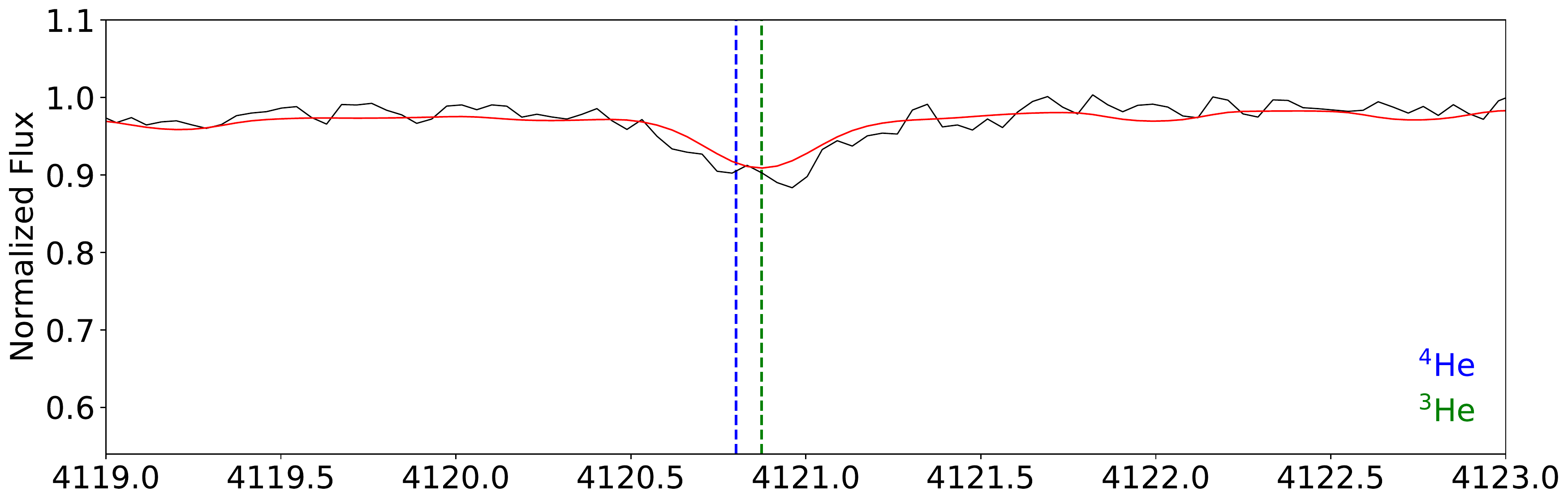}}
        \centering
        \end{minipage}\hfill    
\begin{minipage}[b]{0.5\linewidth}
\resizebox{\hsize}{!}{\includegraphics{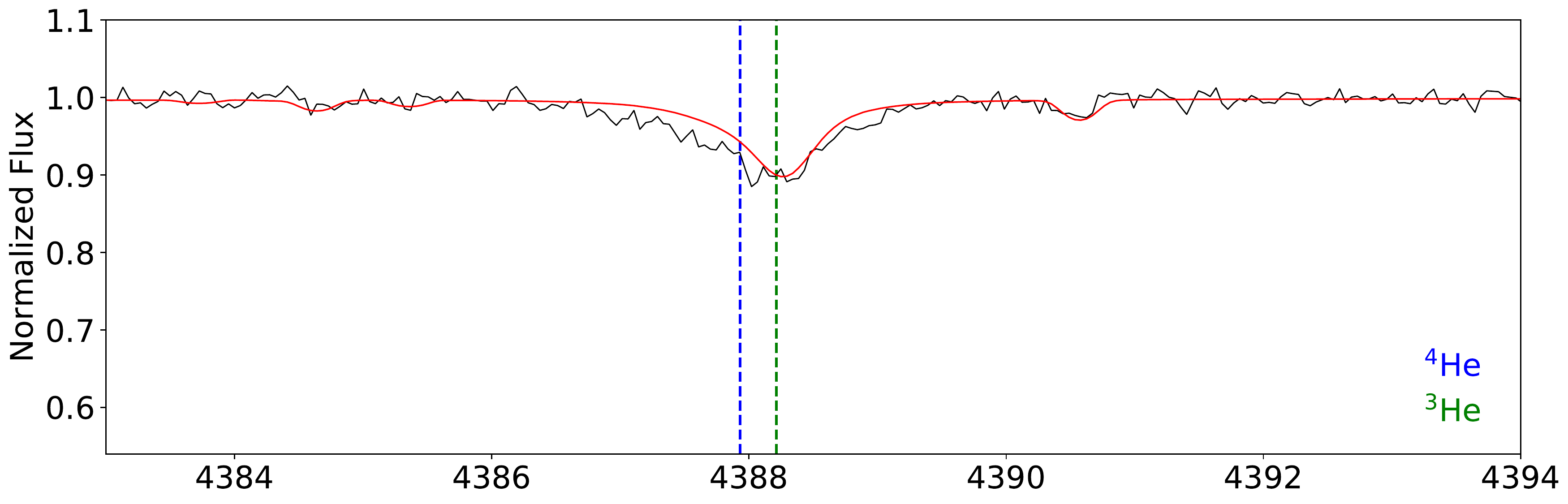}}
        \centering
        \end{minipage}\hfill
\begin{minipage}[b]{0.5\linewidth}      
\resizebox{\hsize}{!}{\includegraphics{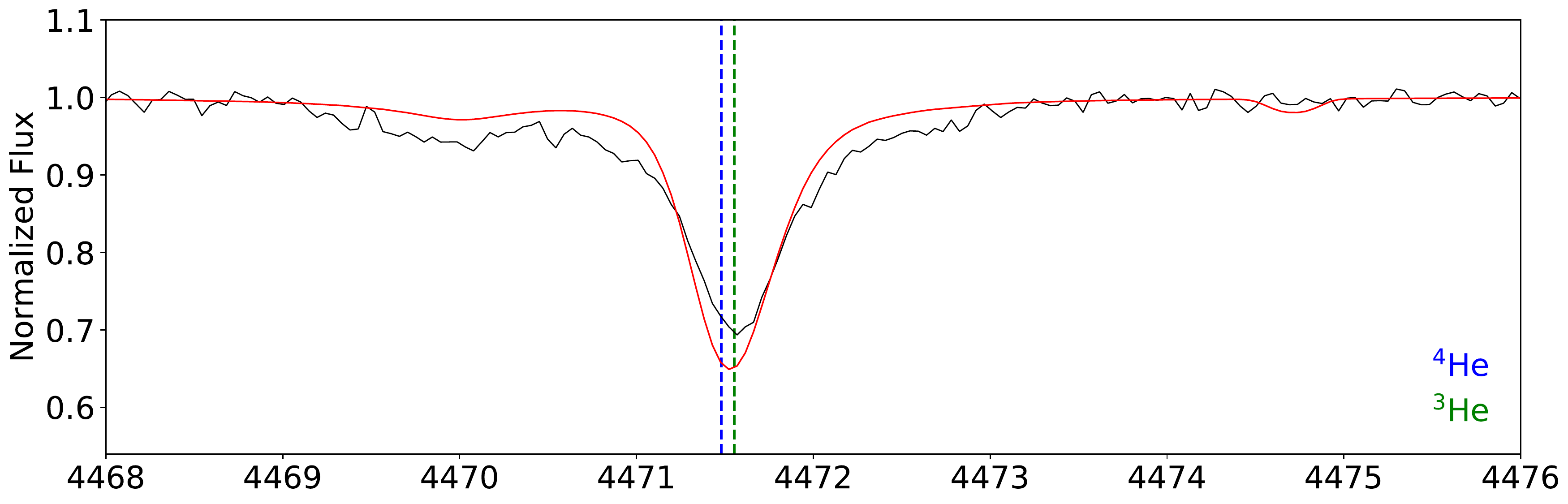}}
        \centering
        \end{minipage}\hfill
\begin{minipage}[b]{0.5\linewidth}      
\resizebox{\hsize}{!}{\includegraphics{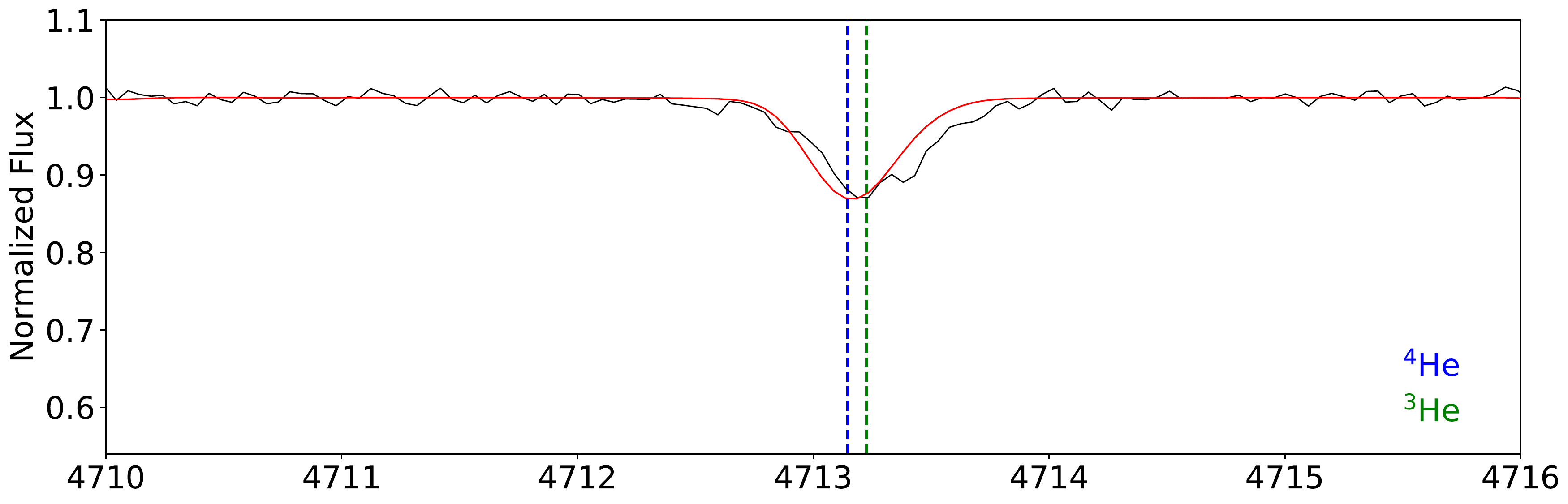}}
        \centering
        \end{minipage}\hfill    
\begin{minipage}[b]{0.5\linewidth}      
\resizebox{\hsize}{!}{\includegraphics{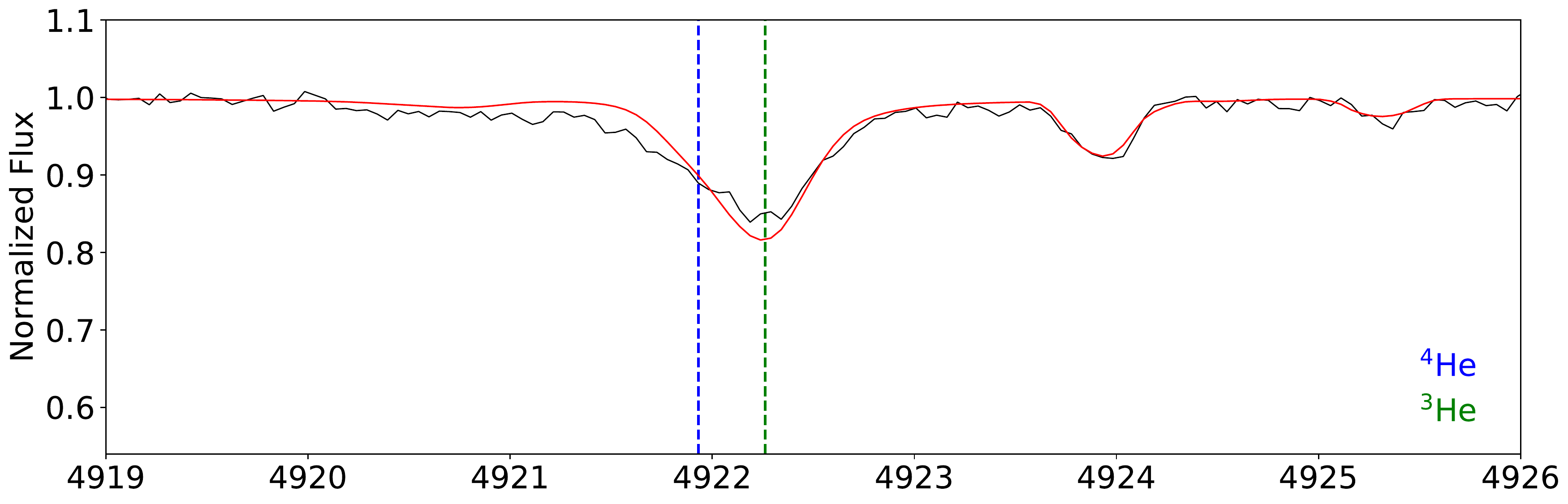}}
        \centering
        \end{minipage}\hfill
\begin{minipage}[b]{0.5\linewidth}      
\resizebox{\hsize}{!}{\includegraphics{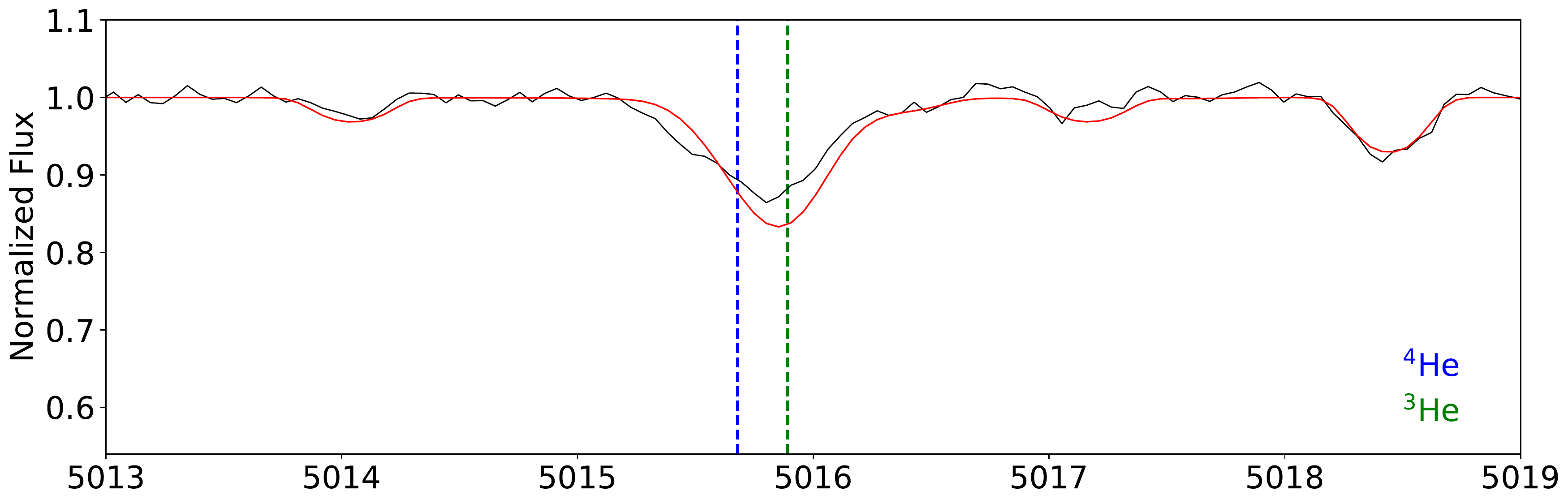}}
        \centering
        \end{minipage}\hfill
\begin{minipage}[b]{0.5\linewidth}      
\resizebox{\hsize}{!}{\includegraphics{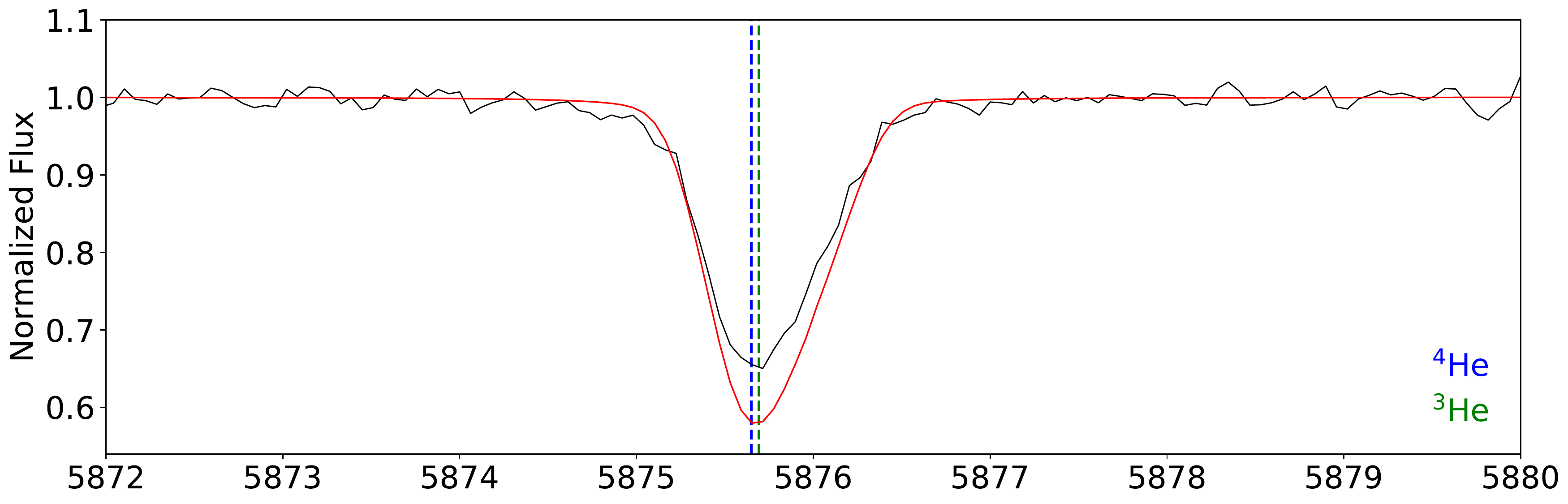}}
        \centering
        \end{minipage}\hfill
\begin{minipage}[b]{0.5\linewidth}      
\resizebox{\hsize}{!}{\includegraphics{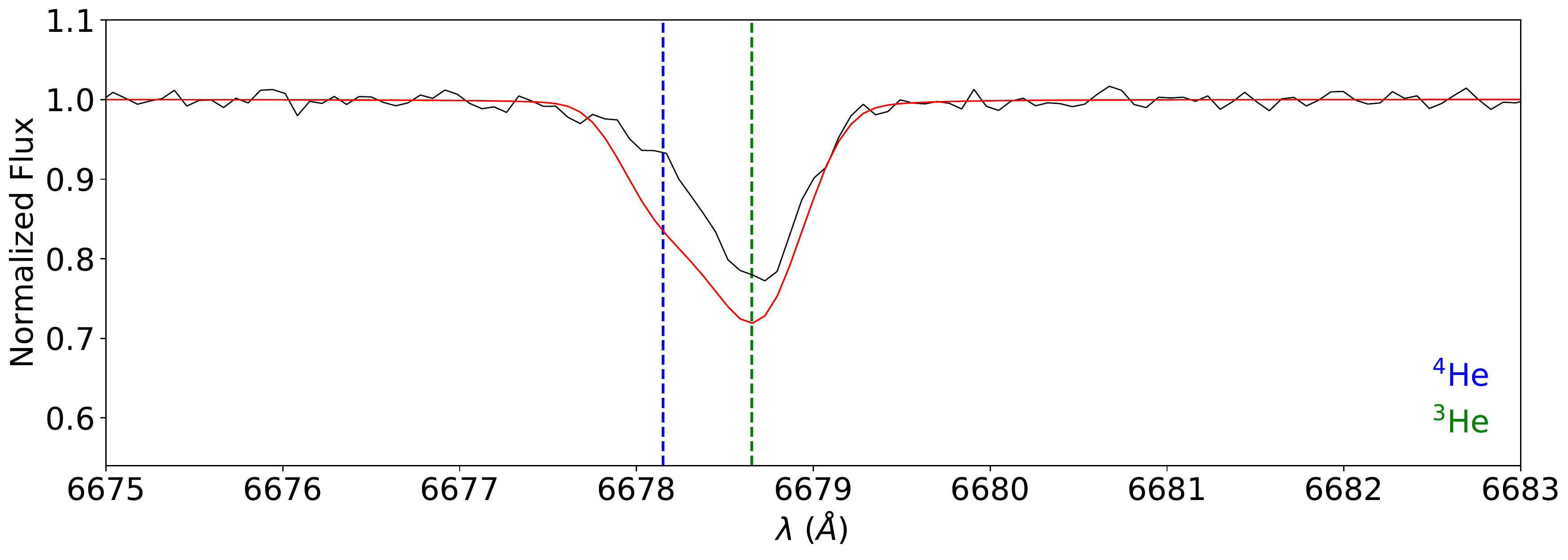}}
        \centering
        \end{minipage}\hfill
\begin{minipage}[b]{0.5\linewidth}      
\resizebox{\hsize}{!}{\includegraphics{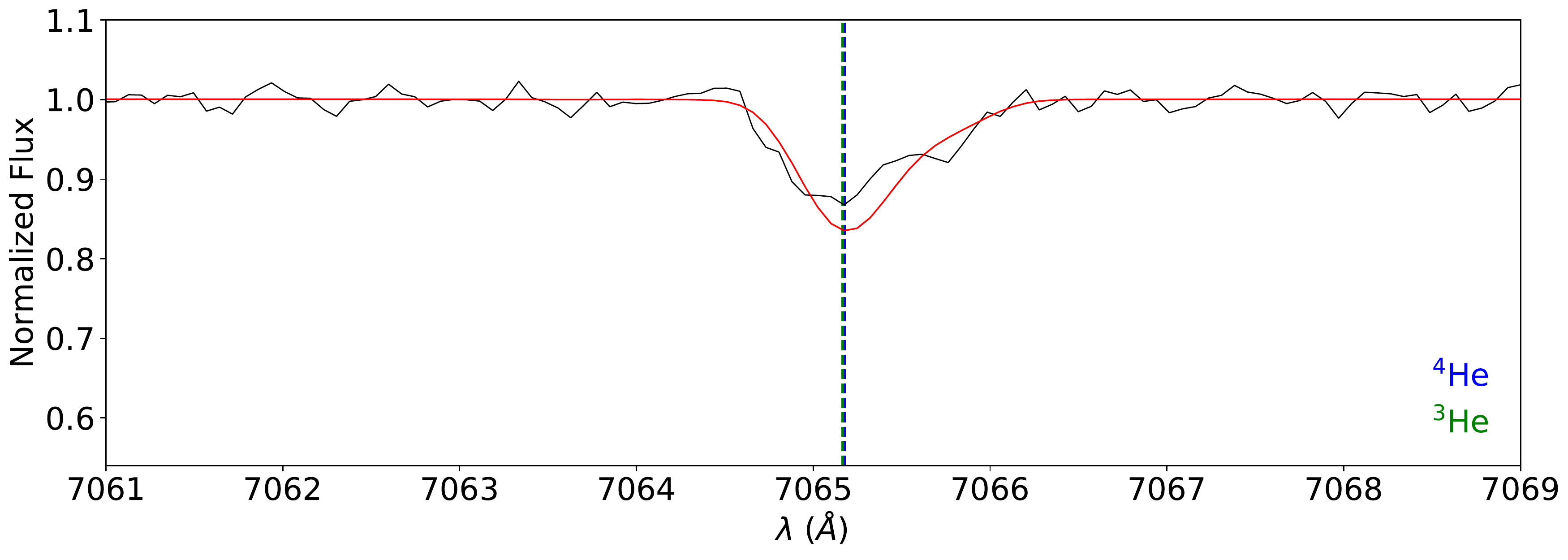}}
        \centering
        \end{minipage}\hfill            
\captionof{figure}{Same as Fig. \ref{Feros HD4539 Hybrid LTE/NLTE Helium Line Fits}, but for the FEROS spectrum of the $\isotope[3]{He}$ star PHL 382. The star shows strong helium stratification, as is obvious from the mismatch of the cores of many $\ion{He}{i}$ lines (see Sect. \ref{Helium Line Profile Anomalies} for details).}\label{Feros PHL382 Hybrid LTE/NLTE Helium Line Fits}
\end{minipage}\hfill

\begin{figure*}

\begin{minipage}[b]{0.5\linewidth}
\resizebox{\hsize}{!}{\includegraphics{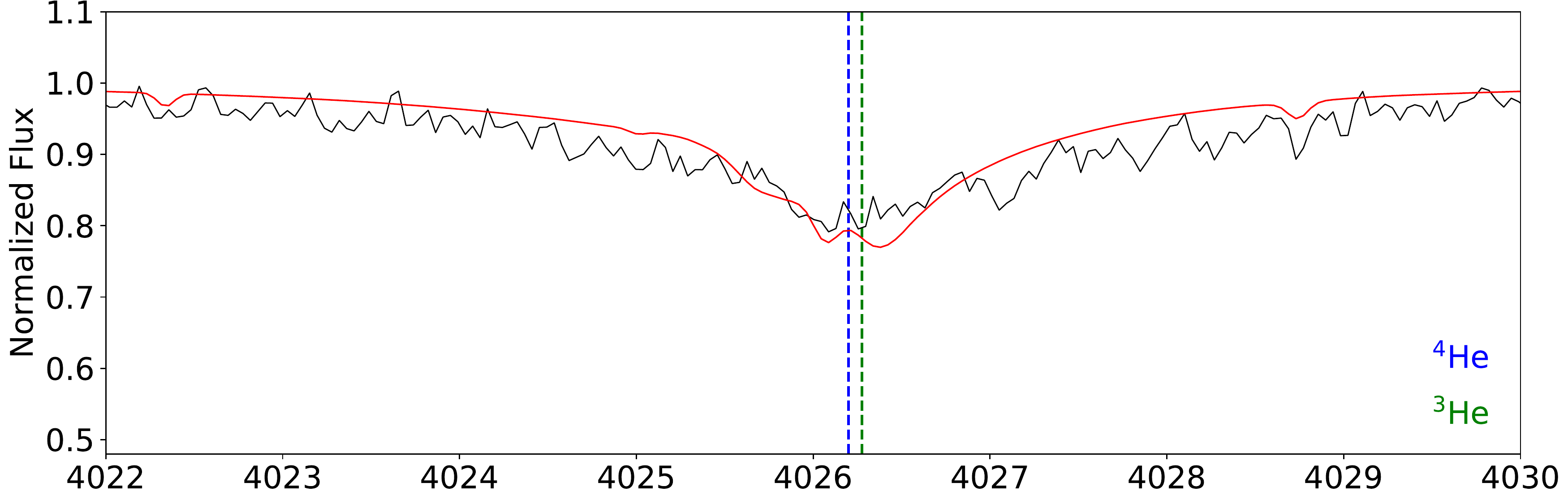}}
        \centering
        \end{minipage}\hfill
        \begin{minipage}[b]{0.5\linewidth}
\resizebox{\hsize}{!}{\includegraphics{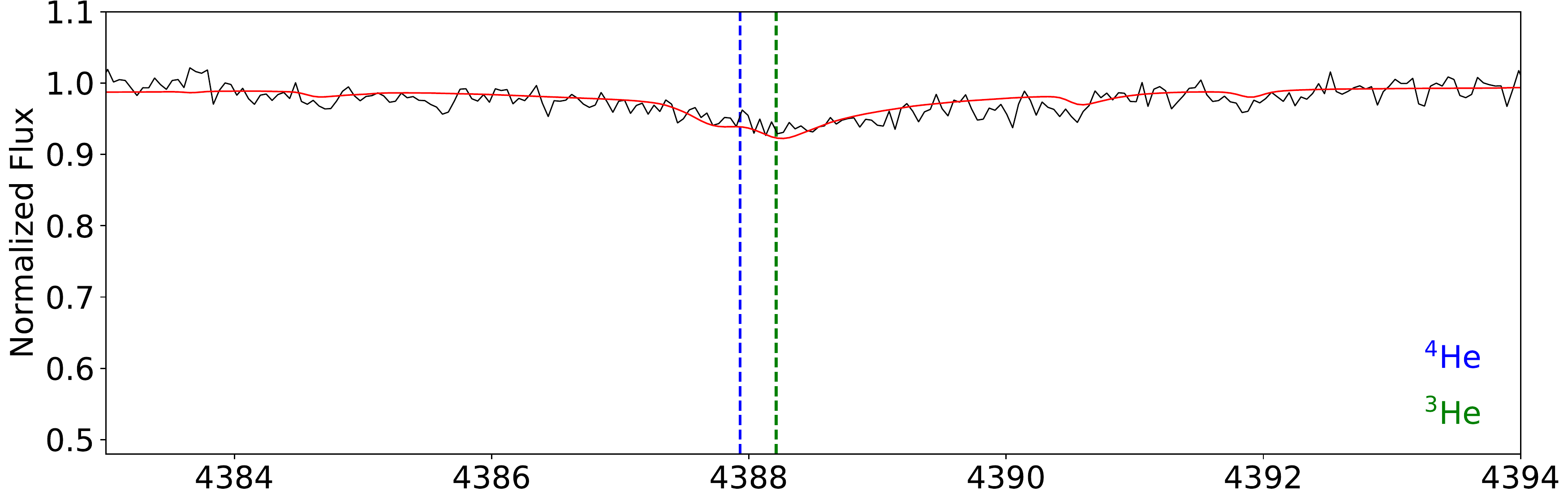}}
        \centering
        \end{minipage}\hfill
        \begin{minipage}[b]{0.5\linewidth}
\resizebox{\hsize}{!}{\includegraphics{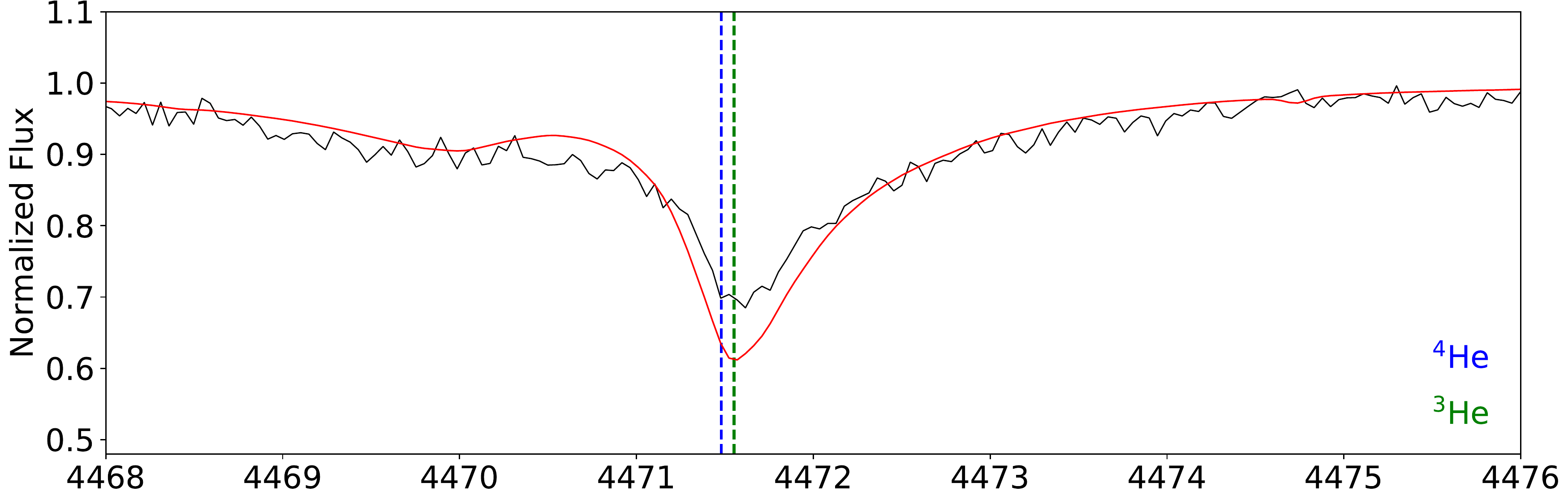}}
        \centering
        \end{minipage}\hfill
\begin{minipage}[b]{0.5\linewidth}
\resizebox{\hsize}{!}{\includegraphics{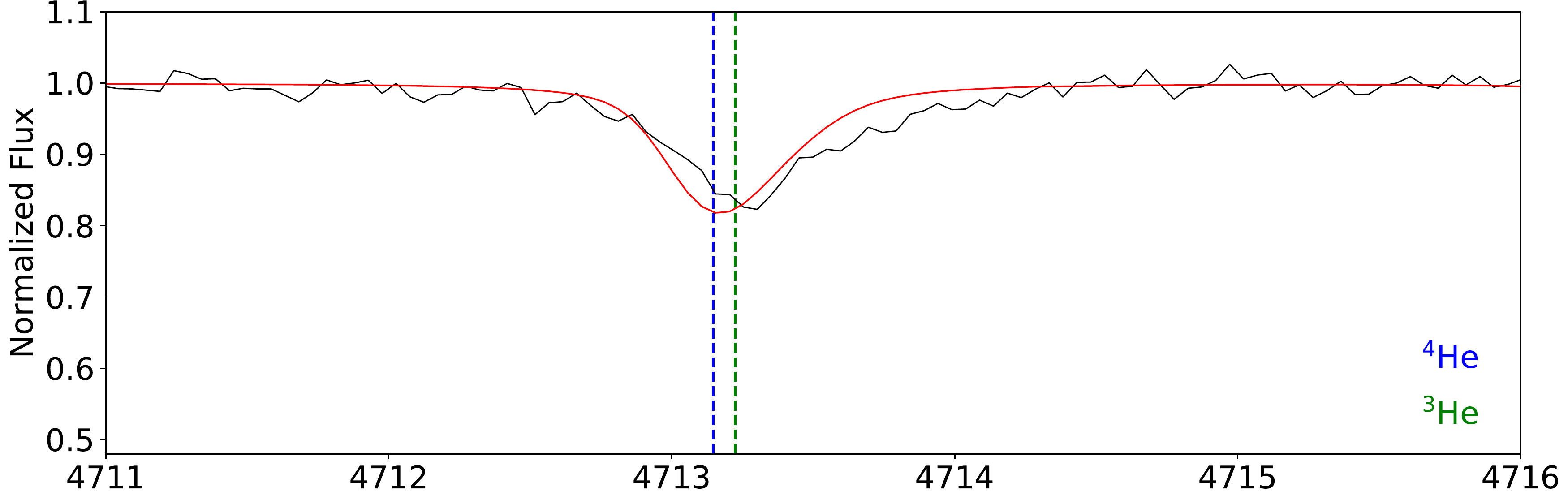}}
        \centering
        \end{minipage}\hfill    
        \begin{minipage}[b]{0.5\linewidth}
\resizebox{\hsize}{!}{\includegraphics{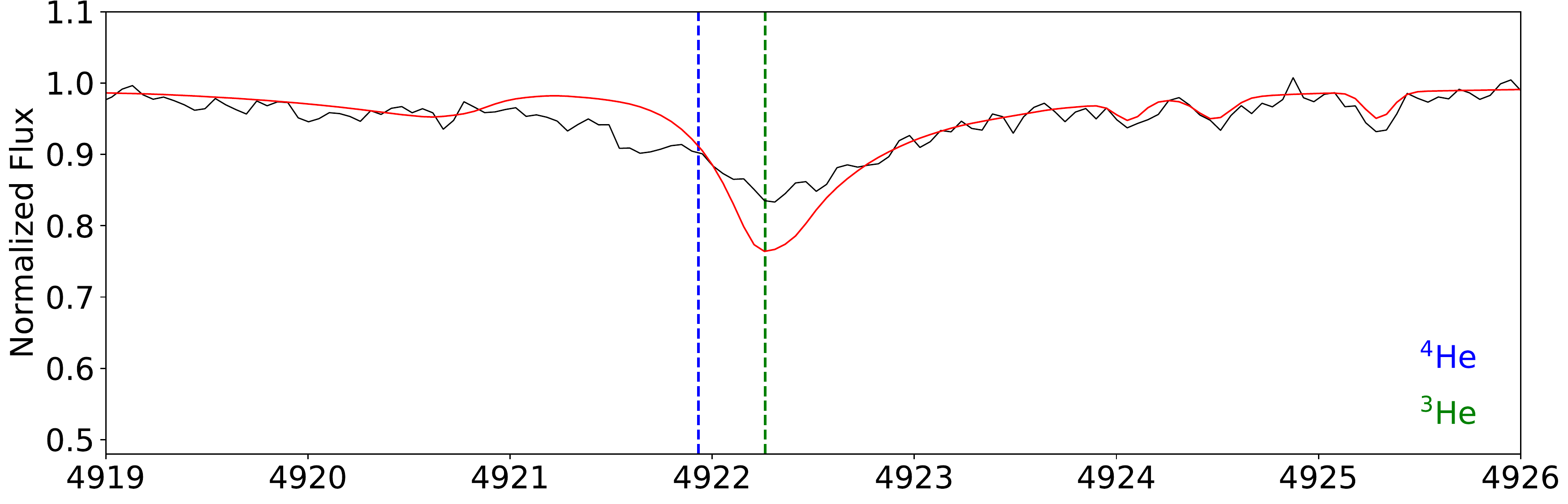}}
        \centering
        \end{minipage}\hfill
        \begin{minipage}[b]{0.5\linewidth}
\resizebox{\hsize}{!}{\includegraphics{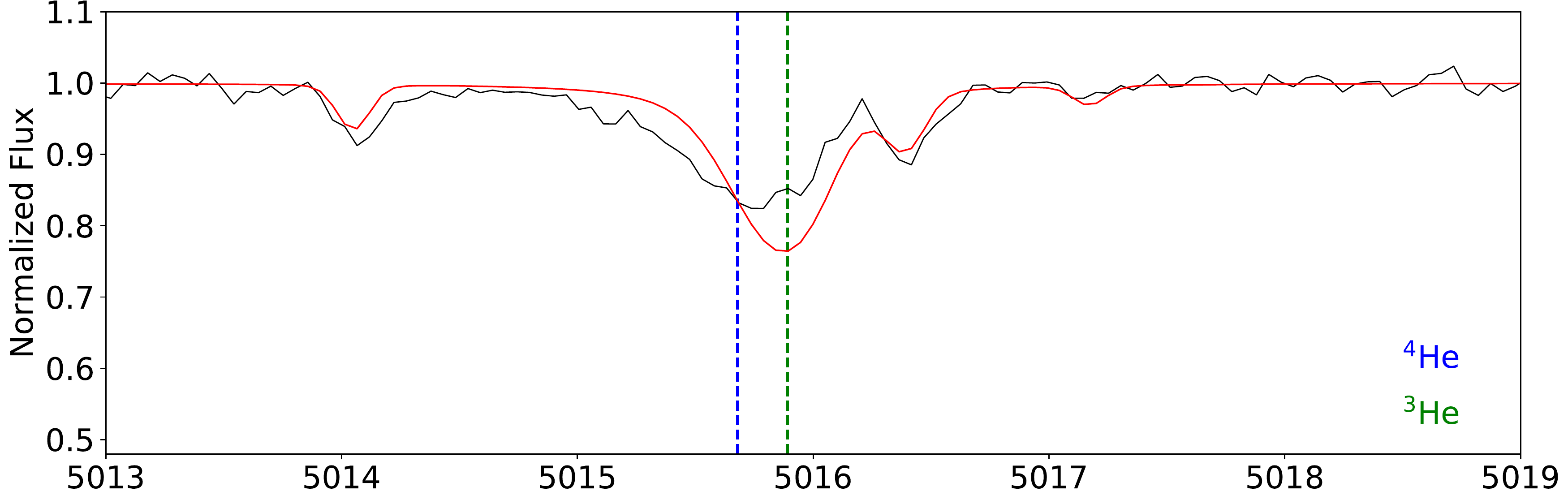}}
        \centering
        \end{minipage}\hfill    
        \begin{minipage}[b]{0.5\linewidth}
\resizebox{\hsize}{!}{\includegraphics{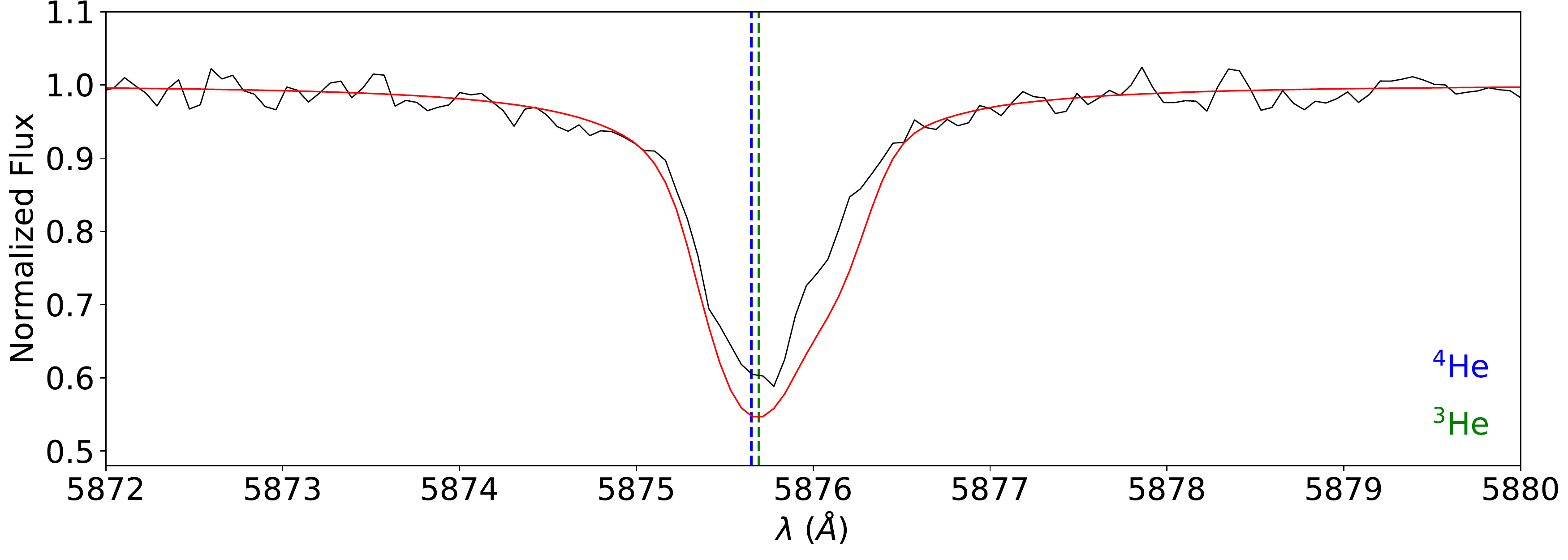}}
        \centering
        \end{minipage}\hfill
        \begin{minipage}[b]{0.5\linewidth}
\resizebox{\hsize}{!}{\includegraphics{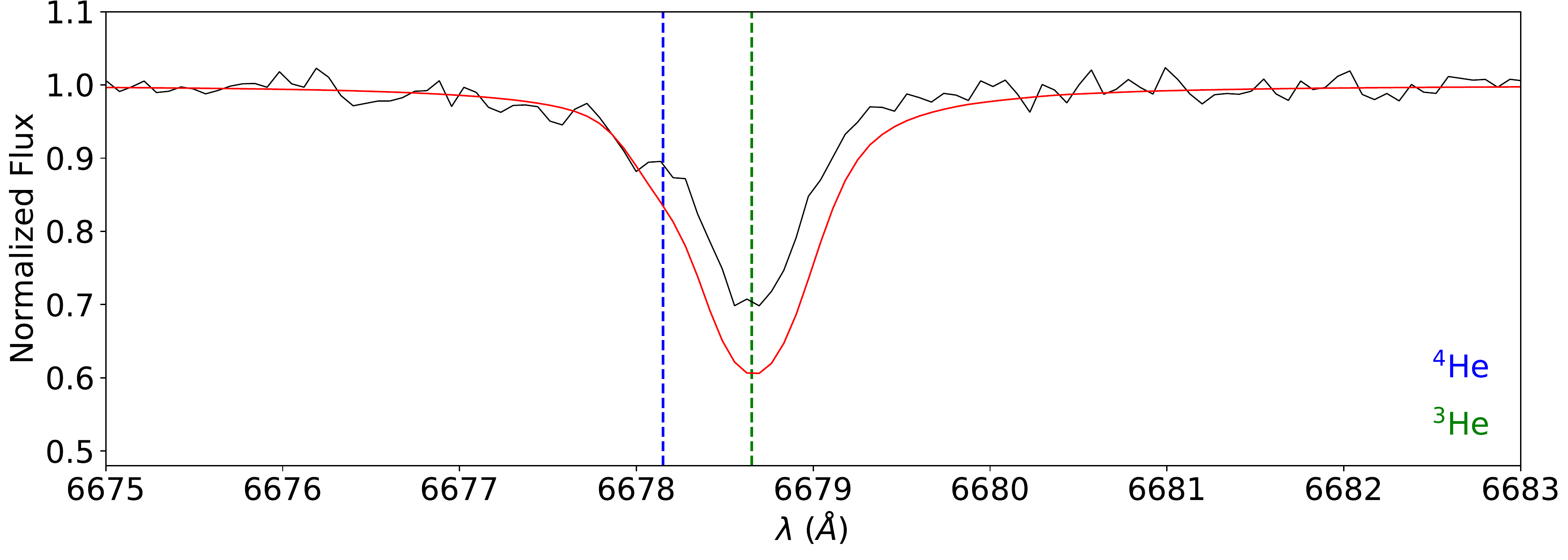}}
        \centering
        \end{minipage}\hfill
        \caption{Same as Fig. \ref{Feros HD4539 Hybrid LTE/NLTE Helium Line Fits}, but for the FEROS spectrum of the $\isotope[3]{He}$ star EC 03591-3232. The star shows helium stratification, as is obvious from the mismatch of the cores of several strong $\ion{He}{i}$ lines (see Sect. \ref{Helium Line Profile Anomalies} for details).}\label{Feros EC03591M3232 Hybrid LTE/NLTE Helium Line Fits}
\end{figure*}\noindent

\begin{figure*}
\begin{minipage}[b]{0.5\linewidth}
\resizebox{\hsize}{!}{\includegraphics{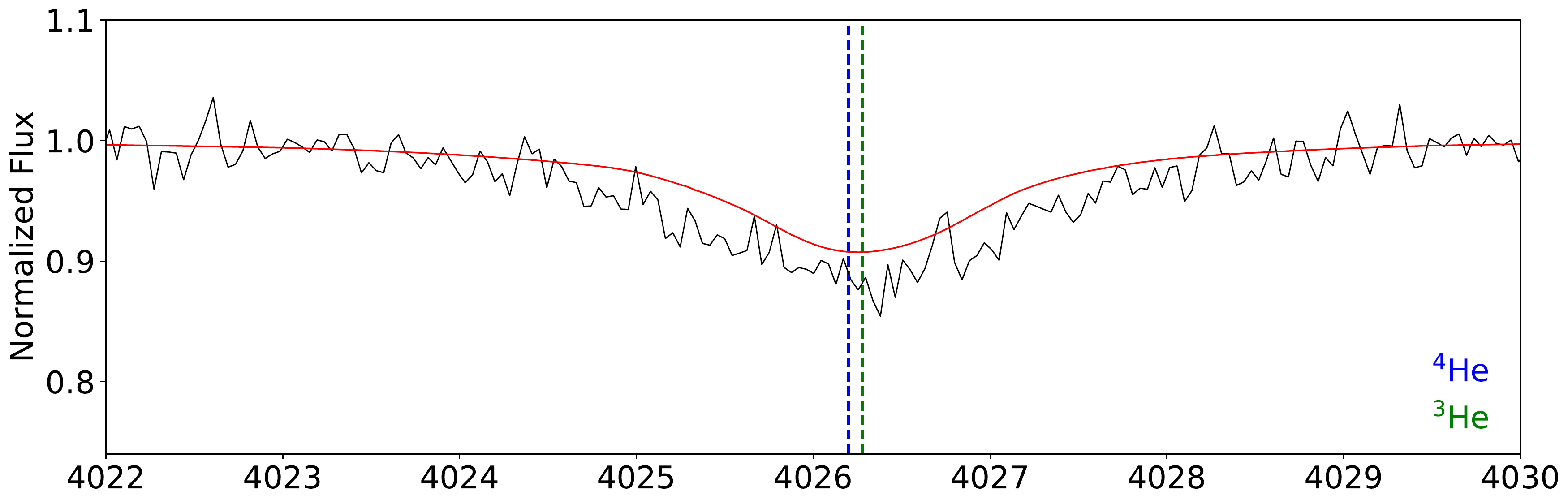}}
\centering
        \end{minipage}\hfill
\begin{minipage}[b]{0.5\linewidth}
\resizebox{\hsize}{!}{\includegraphics{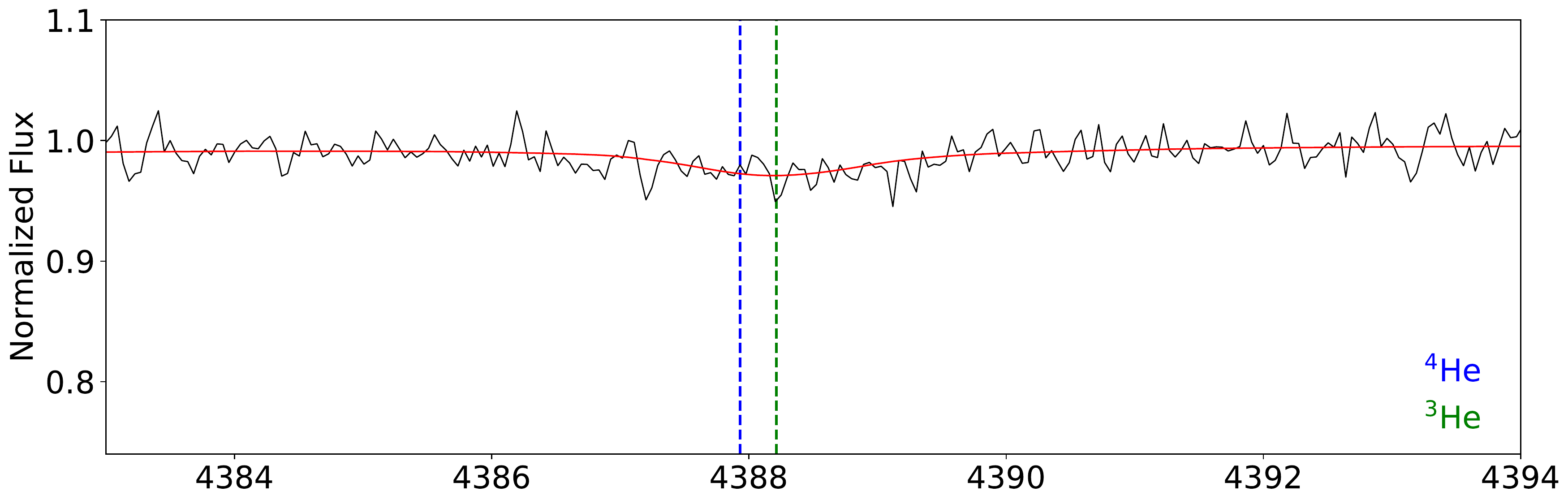}}
        \centering
        \end{minipage}\hfill
\begin{minipage}[b]{0.5\linewidth}      
\resizebox{\hsize}{!}{\includegraphics{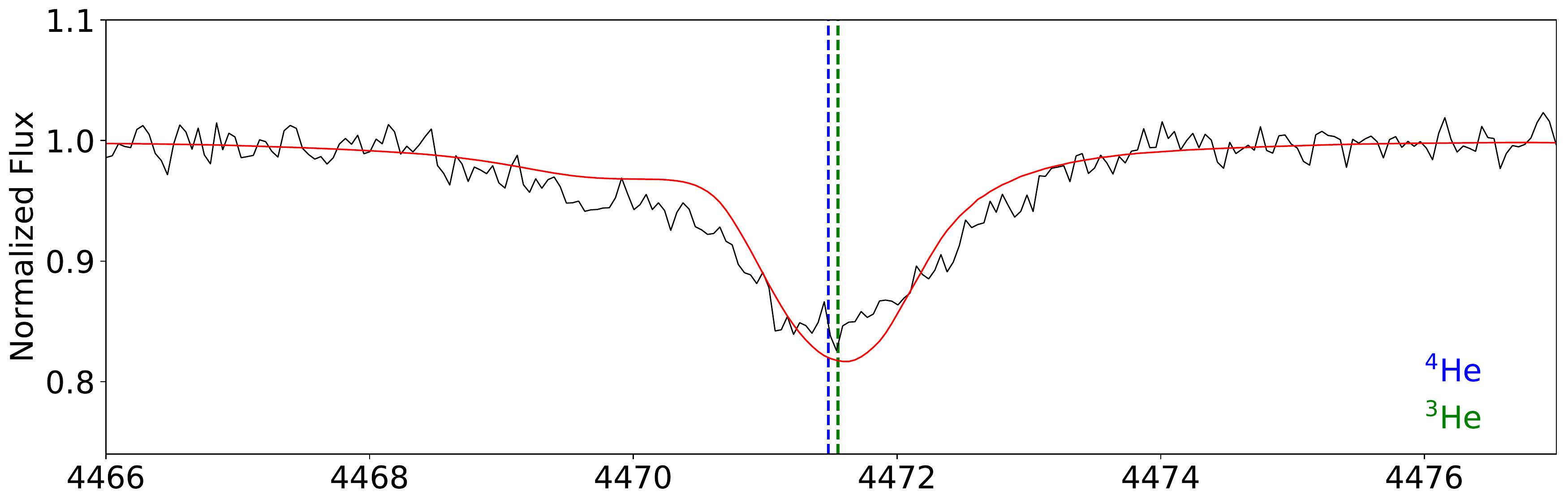}}
        \centering
        \end{minipage}\hfill
\begin{minipage}[b]{0.5\linewidth}      
\resizebox{\hsize}{!}{\includegraphics{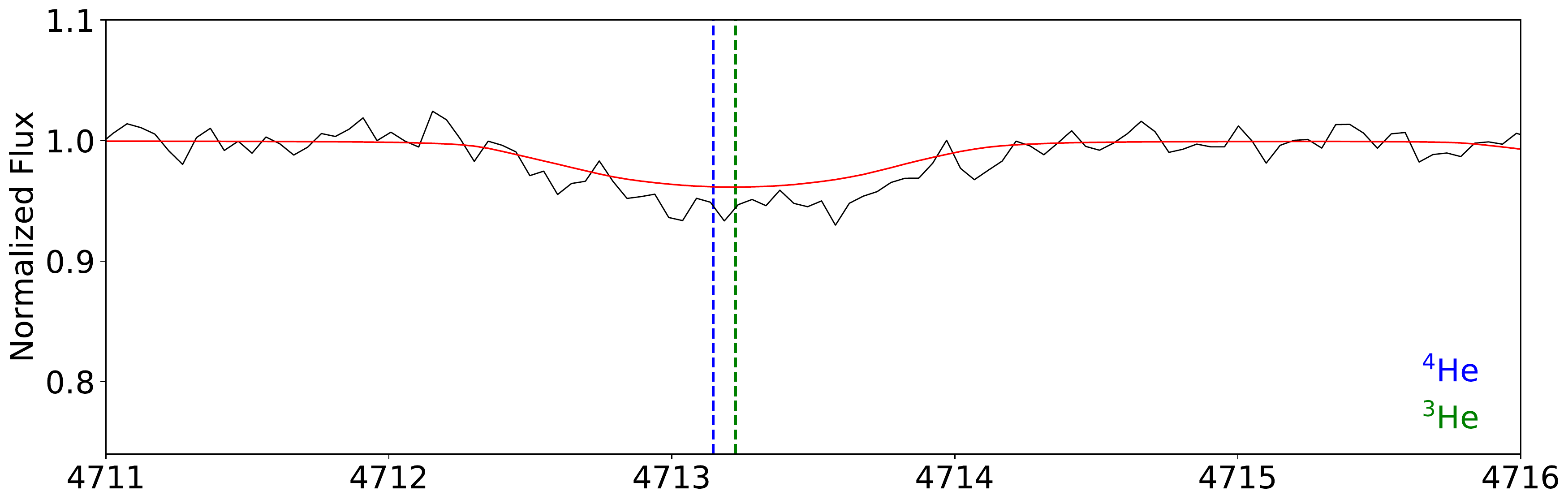}}
        \centering
        \end{minipage}\hfill
\begin{minipage}[b]{0.5\linewidth}      
\resizebox{\hsize}{!}{\includegraphics{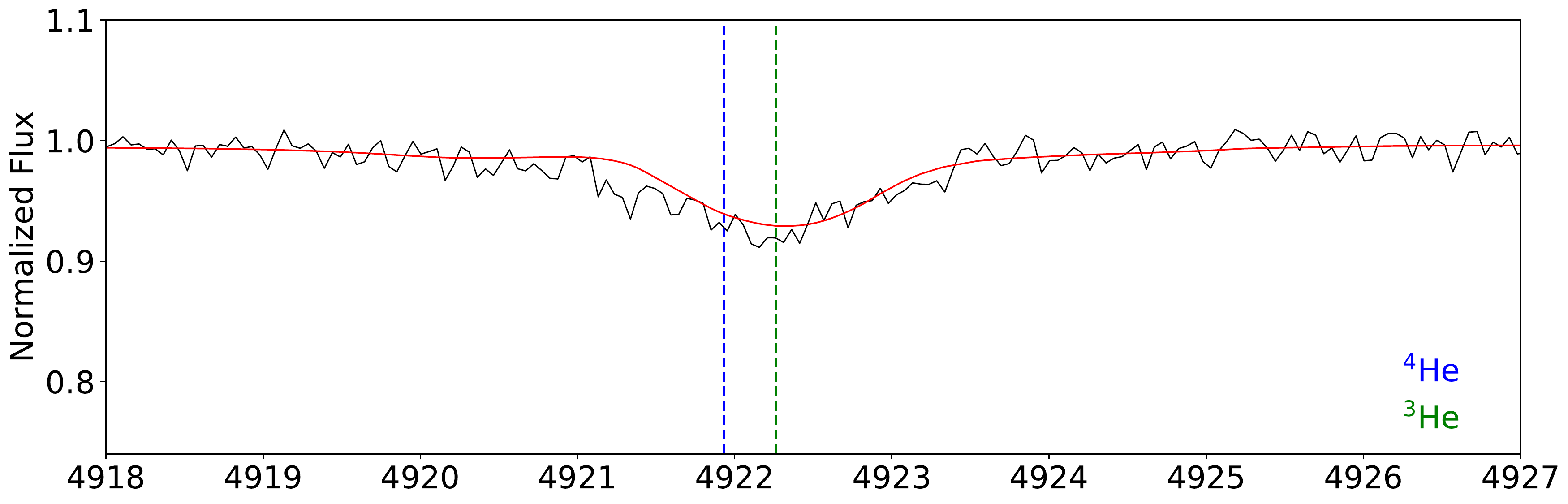}}
        \centering
        \end{minipage}\hfill
\begin{minipage}[b]{0.5\linewidth}      
\resizebox{\hsize}{!}{\includegraphics{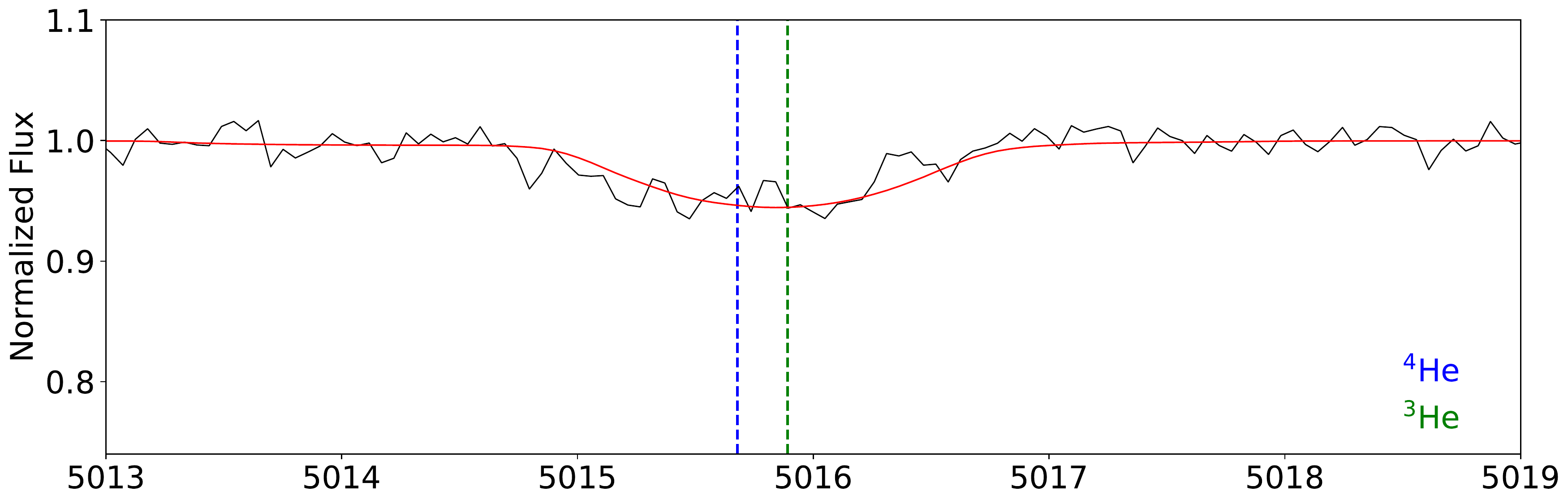}}
        \centering
        \end{minipage}\hfill
\begin{minipage}[b]{0.5\linewidth}      
\resizebox{\hsize}{!}{\includegraphics{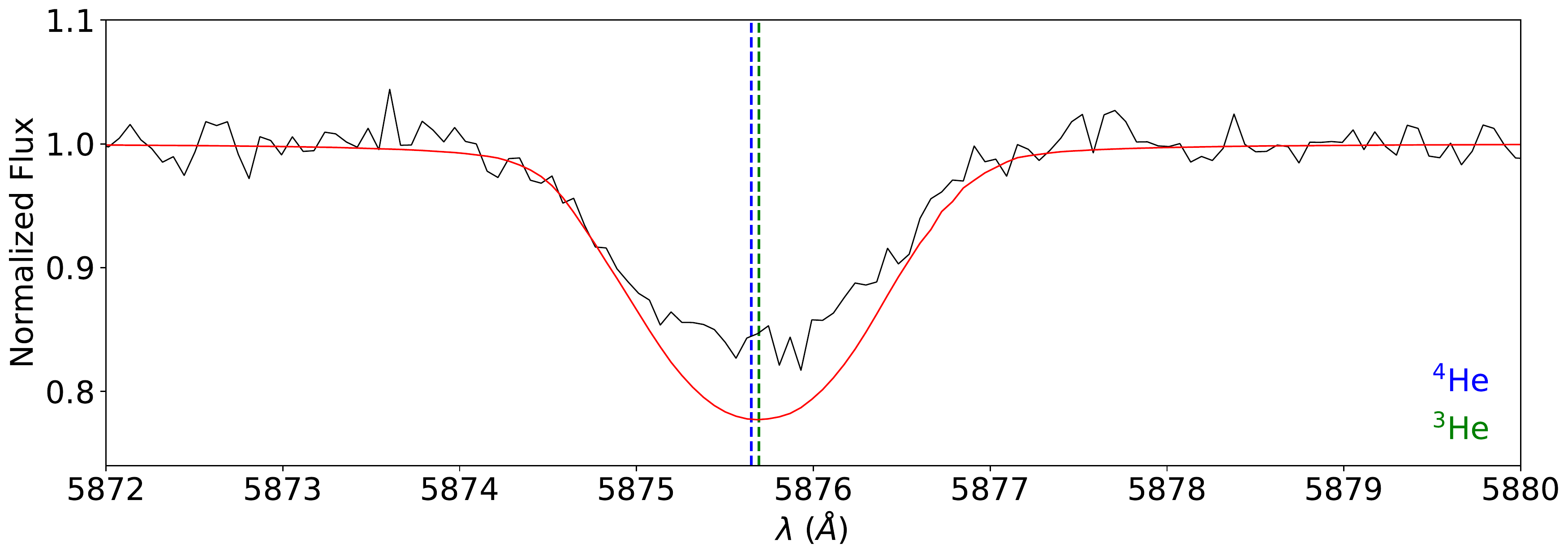}}
        \centering
        \end{minipage}\hfill
\begin{minipage}[b]{0.5\linewidth}      
\resizebox{\hsize}{!}{\includegraphics{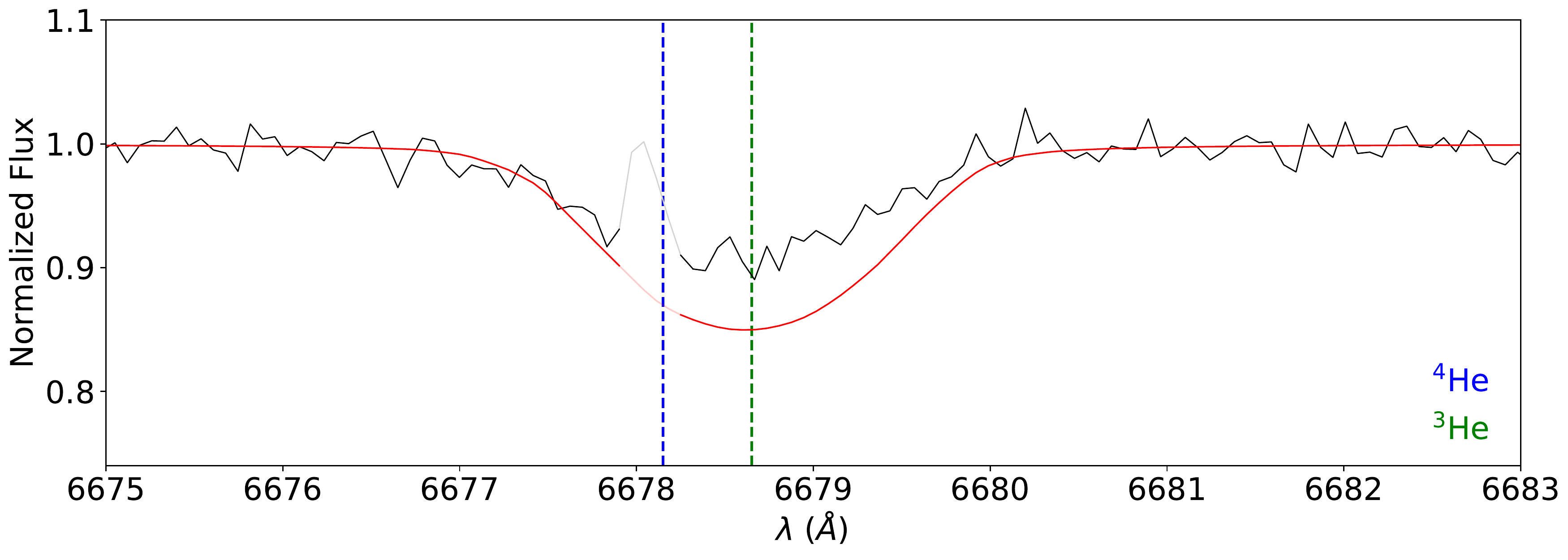}}
        \centering
        \end{minipage}\hfill
\caption{Same as Fig. \ref{Feros HD4539 Hybrid LTE/NLTE Helium Line Fits}, but for the FEROS spectrum of the $\isotope[3]{He}$ star SB 290. The star rotates, but also shows significant mismatches in the strongest $\ion{He}{i}$ lines, which points toward vertical helium stratification (see Sect. \ref{Helium Line Profile Anomalies} for details).}\label{Feros SB290 Hybrid LTE/NLTE Helium Line Fits}
\end{figure*}\clearpage\noindent
\onecolumn
\section{Target sample and data}\label{Target Sample and Data appendix}
\begin{minipage}[b]{2.0\linewidth}
\captionof{table}{Target sample and data of the subluminous B-type program stars.}\label{summary of analyzed spectra}
\begin{tabular}{ccccccc}
\hline\hline
Object & Instrument & No. Spec. & Resolution & S/N & $P$ & References\\
 &  &  &  &  & [\si{\day}] & \\
\hline
HD 4539 & FEROS & 3 & 48\,000 & 71  & - & [1]\\
CD-35$^\circ$ 15910 & FEROS & 2 & 48\,000 & 56 & - & [2]\\
\hline
PHL 25 & HRS & 2 & 60\,000 & 40 & - & [1]\\
PHL 382 & FEROS & 7 & 48\,000 & 88 & - & [1]\\
BD+48$^\circ$ 2721 & FOCES & 1 & 40\,000 & 84 & - & [2,3]\\
\hline
SB 290\tablefootmark{a} & FEROS & 2 & 48\,000 & 68 & - & [1,4]\\
EC 03263-6403 & FEROS & 2 & 48\,000 & 23 & - & [2]\\
EC 03591-3232 & FEROS & 2 & 48\,000 & 65 & - & [2]\\
EC 12234-2607 & FEROS & 4 & 48\,000 & 28 & - & [2]\\
EC 14338-1445 & FEROS & 3 & 48\,000 & 38 & - & [2]\\
Feige 38 & FEROS & 5 & 48\,000 & 69 & - & [2]\\
PG 1710+490 & FOCES & 1 & 40\,000 & 27 & - & [2]\\
Feige 36\tablefootmark{b} & HIRES & 1 & 36\,000 & 116 & $0.35386\pm0.00014$ & [5,6,7,8]\\
PG 1519+640\tablefootmark{b}\tablefootmark{c} & - & - & - & - & $0.539\pm0.003$ & [2,9,10,11]\\
PG 0133+114\tablefootmark{b}\tablefootmark{c} & - & - & - & - & $1.23787\pm0.00003$ & [3,12,13]\\
\hline
HE 0929-0424\tablefootmark{b} & UVES & 2 & 18\,500 & 22 & $0.4400\pm0.0002$ & [14,16,17]\\
HE 1047-0436\tablefootmark{b} & UVES & 4 & 18\,500 & 25 & $1.21325\pm0.00001$ & [15,16,17]\\
HE 2156-3927\tablefootmark{d} & UVES & 3 & 18\,500 & 29 & - & [16,17]\\
HE 2322-0617\tablefootmark{d} & UVES & 2 & 18\,500 & 26 & - & [16,17]\\
\hline
\end{tabular}
\tablefoot{
\tablefoottext{a}{Rotating star.}\tablefoottext{b}{RV-variable star.}\tablefoottext{c}{Available FOCES spectrum not suitable for an analysis.}
\tablefoottext{d}{A cool\\ companion shows $\ion{Mg}{i}$ in the spectrum, and additional features (see Sect. \ref{Two 3He sdB Stars from the ESO SPY Project} for details).}}
\tablebib{
(1) \citet{Heber_1987}; (2) \citet{Geier_2013a}; (3) \citet{Edelmann_2001}; (4) \citet{Geier_SB290_2013};\\ (5) \citet{Saffer_1998}; (6) \citet{Moran_1999}; (7) \citet{Edelmann_1997}; (8) \citet{Edelmann_1999};\\ (9) \citet{Rueda_2003b}; (10) \citet{Edelmann_2004}; (11) \citet{Copperwheat_2011};\\ (12) \citet{Rueda_2003a}; (13) \citet{Edelmann_2005}; (14) \citet{Karl_2006}; (15) \citet{Napiwotzki_2001};\\ (16) \citet{Lisker_2005}; (17) This work.   
}
\end{minipage}\hfill    

\end{appendix}

\end{document}